\documentclass[preprint, tighten, twocolumn]{aastex631}
\usepackage{apjfonts}
\usepackage{amsmath}
\usepackage{multirow}
\usepackage{textcomp}
\usepackage{appendix}
\usepackage{color}
\usepackage{enumitem}
\usepackage{booktabs}

\shorttitle{Stellar Luminosity Models of dust-disk ETGs}
\shortauthors{Davidson et al.}

\usepackage{natbib}
\defcitealias{boizelle17}{Paper I}
\defcitealias{boizelle19}{Paper II}

\newcommand{\kms}{km s\ensuremath{^{-1}}}

\newcommand{\rg}{\ensuremath{r_\mathrm{g}}}
\newcommand{\rb}{\ensuremath{R_\mathrm{b}}}
\newcommand{\re}{\ensuremath{R_\mathrm{e}}}
\newcommand{\mbh}{\ensuremath{M_\mathrm{BH}}}
\newcommand{\msun}{\ensuremath{M_\odot}}
\newcommand{\lsun}{\ensuremath{L_\odot}}

\newcommand{\vc}{\ensuremath{v_\mathrm{c}}}
\newcommand{\per}{\ensuremath{^{-1}}}
\newcommand{\pertwo}{\ensuremath{^{-2}}}

\newcommand{\chisq}{\ensuremath{\chi^2}}

\newcommand{\coone}{CO(1$-$0)}
\newcommand{\cotwo}{CO(2$-$1)}
\newcommand{\cothree}{CO(3$-$2)}

\begin{document}

\title{Circumnuclear Dust in Luminous Early-Type Galaxies -- I. Sample Properties and Stellar Luminosity Models\footnote{Based on observations made with the NASA/ESA Hubble Space Telescope, obtained at the Space Telescope Science Institute, which is operated by the Association of Universities for Research in Astronomy, Inc., under NASA contract NAS5-26555. New observations are associated with GO programs 14920, 15226, and 15909 and the archival data with GO programs 5124, 5446, 5476, 5910, 5927, 5999, 6587, 6791, 6837, 7450, 8686, 9069, 9106, 9401, 9427, 10003, 10217, 11339, 11583, 12220, 14159, and 15444.}}

\author[0000-0003-3900-6189]{Jared R. Davidson}
\affiliation{Department of Physics and Astronomy, N284 ESC, Brigham Young University, Provo, UT, 84602, USA}

\author[0000-0001-6301-570X]{Benjamin D. Boizelle}
\affiliation{Department of Physics and Astronomy, N284 ESC, Brigham Young University, Provo, UT, 84602, USA}
\affiliation{George P. and Cynthia Woods Mitchell Institute for Fundamental Physics and Astronomy, 4242 TAMU, Texas A\&M University, College Station, TX, 77843-4242, USA}
\email{boizellb@byu.edu}

\author[0000-0002-1881-5908]{Jonelle L. Walsh}
\affiliation{George P. and Cynthia Woods Mitchell Institute for Fundamental Physics and Astronomy, 4242 TAMU, Texas A\&M University, College Station, TX, 77843-4242, USA}

\author[0000-0002-3026-0562]{Aaron J. Barth}
\affiliation{Department of Physics and Astronomy, 4129 Frederick Reines Hall, University of California, Irvine, CA, 92697-4575, USA}

\author{Emma Rasmussen}
\affiliation{Department of Physics and Astronomy, N284 ESC, Brigham Young University, Provo, UT, 84602, USA}

\author[0000-0002-7892-396X]{Andrew J. Baker}
\affiliation{Department of Physics and Astronomy, Rutgers, the State University of New Jersey, 136 Frelinghuysen Road, Piscataway, NJ 08854-8019, USA}
\affiliation{Department of Physics and Astronomy, University of the Western Cape, Robert Sobukwe Road, 7535 Bellville, Cape Town, South Africa}

\author[0000-0002-3202-9487]{David A. Buote}
\affiliation{Department of Physics and Astronomy, 4129 Frederick Reines Hall, University of California, Irvine, CA, 92697-4575, USA}

\author[0000-0003-2511-2060]{Jeremy Darling}
\affiliation{Center for Astrophysics and Space Astronomy, Department of Astrophysical and Planetary Sciences, University of Colorado, 389 UCB, Boulder, CO 80309-0389, USA}

\author[0000-0001-6947-5846]{Luis C. Ho}
\affiliation{Kavli Institute for Astronomy and Astrophysics, Peking University, Beijing 100871, China} 
\affiliation{Department of Astronomy, School of Physics, Peking University, Beijing 100871, China}

\author[0000-0003-2632-8875]{Kyle M. Kabasares}
\affiliation{Department of Physics and Astronomy, 4129 Frederick Reines Hall, University of California, Irvine, CA, 92697-4575, USA}
\affiliation{Ames Research Center, National Aeronautics and Space Administration, Moffett Field, CA 94035, USA}
\affiliation{Bay Area Environmental Research Institute, Ames Research Center, Moffett Field, CA 94035, USA}

\author[0000-0003-1420-6037]{Jonathan H. Cohn}
\affiliation{George P. and Cynthia Woods Mitchell Institute for Fundamental Physics and Astronomy, 4242 TAMU, Texas A\&M University, College Station, TX, 77843-4242, USA}
\affiliation{Department of Physics and Astronomy, Dartmouth College, 6127 Wilder Laboratory, Hanover, NH 03755, USA}

\begin{abstract}
Dusty circumnuclear disks (CNDs) in luminous early-type galaxies (ETGs) show regular, dynamically cold molecular gas kinematics. For a growing number of ETGs, Atacama Large Millimeter/sub-millimeter Array (ALMA) CO imaging and detailed gas-dynamical modeling facilitate moderate-to-high precision black hole (BH) mass (\mbh) determinations. From the ALMA archive, we identified a subset of 26 ETGs with estimated $\mbh/\msun \gtrsim 10^8$ to a few $\times$10$^9$ and clean CO kinematics but that previously did not have sufficiently high angular resolution near-IR observations to mitigate dust obscuration when constructing stellar luminosity models. We present new optical and near-IR Hubble Space Telescope (HST) images of this sample to supplement the archival HST data, detailing the sample properties and data analysis techniques. After masking the most apparent dust features, we measure stellar surface brightness profiles and model the luminosities using the multi-Gaussian expansion (MGE) formalism. Some of these MGEs have already been used in CO dynamical modeling efforts to secure quality \mbh\ determinations, and the remaining ETG targets here are expected to significantly improve the high-mass end of the current BH census, facilitating new scrutiny of local BH mass-host galaxy scaling relationships. We also explore stellar isophotal behavior and general dust properties, finding these CNDs generally become optically thick in the near-IR ($A_H \gtrsim 1$ mag). These CNDs are typically well-aligned with the larger-scale stellar photometric axes with a few notable exceptions. Uncertain dust impact on the MGE often dominates the BH mass error budget, so extensions of this work will focus on constraining CND dust attenuation.
\end{abstract}


\section{Introduction}
\label{sec:intro}

Supermassive black holes (BHs), spanning a mass range of $\sim 10^{5}-10^{10}$ \msun, are believed to be at the centers of nearly all large galaxies \citep[i.e., $M_\star>10^{11} \msun$; ][]{mcma13}. Over the past two decades, dynamical modeling techniques have been employed to measure BH masses (\mbh) in well over 100 galaxies \citep[e.g.,][]{korho13,sagl16}. Although BHs gravitationally dominate only the innermost regions of their host galaxies (often not more than the central few pc to few hundreds of parsecs in extreme cases), their masses strongly correlate with several large-scale galaxy properties, especially the stellar bulge velocity dispersion \citep[$\sigma_\star$; e.g.,][]{fermer00,geb00} and luminosity or mass \citep[$L_{\mathrm{bul}}$ or $M_{\mathrm{bul}}$; e.g.,][]{korrich95,mcma13}. Such empirical relations suggest a co-evolution of the central BH and its host galaxy through a series of gas accretion and galaxy merger events across cosmic time. During this galaxy growth, star formation and BH accretion are regulated by both stellar and AGN feedback processes. However, the detailed nature of these scaling relationships -- including the intrinsic scatter and dependence on galaxy morphology -- remain uncertain due to small sample size, some poorly constrained \mbh\ values, and persistent selection biases.

For the most luminous early-type galaxies (ETGs), including several brightest group galaxies (BGGs) and brightest cluster galaxies (BCGs) with cored stellar surface brightness profiles, current data hint at a steeper \mbh$-\sigma_{\star}$ slope \citep{lauer07,bern07,mcma13}. This result suggests that BH growth in high-galaxy density environments follows a different evolutionary path due to the prevalence of dry (gas-poor) mergers in clusters and to a lesser extent in groups \citep[e.g., see][]{bog18}. However, the BH census remains incomplete above $\sim$10$^{8.7}$ \msun, and statistical mass uncertainties in this regime are typically large \citep[of order 25\%; e.g.,][]{sagl16}. In addition, potentially serious (and often unexplored) systematics in both stellar and gas-dynamical models may affect \mbh\ measurements \citep[for more discussion, see][]{korho13}. For example, stellar triaxality is infrequently incorporated into stellar-dynamical modeling \citep[c.f.][]{liepold23}, possibly biasing the best-fitting \mbh\ by a factor of two in some cases \citep{vandb10}. A larger sample of \mbh\ for luminous ETGs, together with greater measurement precision, are necessary for any confident interpretation of BH-host galaxy co-evolution in rich galaxy environments. The most reliable \mbh\ determinations originate from spatially resolved, relaxed tracer kinematics that extend well within the BH sphere of influence approximated by $\rg \equiv G\mbh /\sigma_{\star}^2$, within which the BH's gravitational influence dominates over the galaxy's extended mass contributions.

Dense molecular gas in circumnuclear disks (CNDs) provides an appealing avenue for expanding the high-mass BH census with higher precision than typically possible through other techniques. Because of the small physical extent of these CNDs \citep[dust radii of $\sim$100 pc to a few kpc for most ETGs;][]{tran01}, gas-dynamical modeling processes are less sensitive to uncertainties in large-scale galaxy properties that often hampers stellar-dynamical efforts \citep[e.g.,][]{vandb10,mcma13}. Extended atomic/molecular gas and dust are detected in roughly half of all ETGs \citep[][]{alighieri07,young11,serra12,alatalo13}, with an apparent preference for dense molecular gas to be found in lenticular as opposed to elliptical galaxies \citep[][]{alighieri13}. About 10$-$20\% of all ETGs host morphologically round CNDs that suggest significant molecular gas in dynamically cold rotation that should be amenable to gas-dynamical modeling \citep[e.g.,][]{lauer05,maio08,davis11,alatalo13}. In some CNDs, these tracer kinematics are detected down to (or even well within) \rg, making them appealing targets for constraining BH masses.

Already, resolved low-$J$ CO imaging using the Atacama Large Millimeter/sub-millimeter Array (ALMA) has mapped molecular gas kinematics in a few dozen ETGs with dusty CNDs \citep[e.g.,][]{boizelle17,ruffa19a,zabel19}. This effort is especially valuable for the most luminous dust-disk ETGs, which tend to have large physical \rg\ but are expected to have at least mildly triaxial stellar structures that make global stellar-dynamical modeling challenging. In cases with relaxed gaseous kinematics probing near or within \rg, gas-dynamical modeling of CO cubes has resulted in some of the most precise BH mass measurements to date \citep{barth16a,barth16b,onishi17,davis17,davis18,boizelle19,boizelle21,smith19,north19,smith21,cohn21,cohn23,ruffa23}. In addition to those cases, a greater number of ETGs have ALMA CO imaging with synthesized beam full width at half maximum (FWHM) $\theta_\mathrm{FWHM} \lesssim 2\rg$ that should enable good quality \mbh\ determinations \citep{davis14}.

\begin{deluxetable*}{lcccccccccccc}[!ht]
\tabletypesize{\footnotesize}
\tablecaption{Early-type Galaxy Sample}
\tablewidth{0pt}
\tablehead{
\\[-5mm]
\colhead{Galaxy} & \colhead{RC3} & \colhead{$m-M$} & \colhead{$A_\mathrm{Gal,F160W}$} & \colhead{$z$} & \colhead{$D_{L}$} & \colhead{$D_{A}$} & \colhead{Scale} & \colhead{\re} & \colhead{$\sigma_{c}$} & \colhead{$M_{K}$} & \colhead{$L_{H}$} & \colhead{\rg} \\[-2.5mm]
\colhead{Name} & \colhead{Type} & \colhead{(mag)} & \colhead{(mag)} & \colhead{} & \colhead{(Mpc)} & \colhead{(Mpc)} & \colhead{(kpc arcsec\per)} & \colhead{(kpc)} & \colhead{(km s\per)} & \colhead{(mag)} & \colhead{(10$^{11}$ $L_\sun$)} & \colhead{(arcsec)}\\[-2mm]
\colhead{(1)} & (2) & (3) & (4) & (5) & (6) & (7) & (8) & (9) & (10) & (11) & (12) & (13)
}
\startdata
Hydra A & (R')SA0\scalebox{.8}{\textsuperscript{$\wedge$}}$-$: & $36.89\pm0.15$ & 0.021 & 0.055782 & 238.70 & 214.14 & 1.038 & 17.79 & 341.1 & $-$25.99 & 4.958 & 0.13 \\
\multirow{2}{*}{NGC \phantom{0}612} & SA0\scalebox{.8}{\textsuperscript{$\wedge$}}+ pec & \multirow{2}{*}{$35.46\pm0.15$} & \multirow{2}{*}{0.010} & \multirow{2}{*}{0.029430} & \multirow{2}{*}{123.60} & \multirow{2}{*}{116.63} & \multirow{2}{*}{0.565} & \multirow{2}{*}{5.79} & \multirow{2}{*}{\nodata} & \multirow{2}{*}{$-$25.86} & \multirow{2}{*}{3.591} & \multirow{2}{*}{0.22} \\[-1mm]
 & edge-on &  &  &  &  &  &  &  &  &  &  &  \\
NGC \phantom{0}997 & E & $34.71\pm0.15$ & 0.071 & 0.021015 & 87.70 & 84.13 & 0.408 & 4.96 & \nodata & $-$25.26 & 2.168 & 0.22 \\
\multirow{2}{*}{NGC 1332} & S0\scalebox{.8}{\textsuperscript{$\wedge$}}$-$:(s) & \multirow{2}{*}{$31.80\pm0.18 ^{1}$} & \multirow{2}{*}{0.017} & \multirow{2}{*}{0.005180} & \multirow{2}{*}{22.91} & \multirow{2}{*}{22.67} & \multirow{2}{*}{0.110} & \multirow{2}{*}{2.87} & \multirow{2}{*}{294.6} & \multirow{2}{*}{$-$24.74} & \multirow{2}{*}{1.369} & \multirow{2}{*}{0.57} \\[-1mm]
 & edge-on &  &  &  &  &  &  &  &  &  &  &  \\
NGC 1387 & SAB0\scalebox{.8}{\textsuperscript{$\wedge$}}-(s) & $31.80\pm0.09 ^{5}$ & 0.006 & 0.004079 & 19.32 & 19.16 & 0.093 & 1.40 & 167.3 & $-$23.94 & 0.593 & 0.42 \\
NGC 3245 & SA0\scalebox{.8}{\textsuperscript{$\wedge$}}0(r):? & $31.43\pm0.20 ^{1}$ & 0.013 & 0.005854 & 20.89 & 20.65 & 0.100 & 2.87 & 207.0 & $-$23.70 & 0.658 & 0.33 \\
NGC 3258 & E1 & $32.53\pm0.27 ^{1}$ & 0.041 & 0.009580 & 32.06 & 31.46 & 0.153 & 2.57 & 261.0 & $-$24.34 & 0.788 & 0.32 \\
NGC 3268 & E2 & $32.73\pm0.25 ^{1}$ & 0.053 & 0.009280 & 34.83 & 34.20 & 0.166 & 5.99 & 228.6 & $-$24.54 & 1.342 & 0.34 \\
NGC 3271 & SB0\scalebox{.8}{\textsuperscript{$\wedge$}}0(r) & $33.73\pm0.15$ & 0.056 & 0.013393 & 55.60 & 54.14 & 0.262 & 4.59 & 246.8 & $-$25.54 & 2.765 & 0.40 \\
NGC 3557 & E3 & $33.30\pm0.22 ^{1}$ & 0.052 & 0.009867 & 45.71 & 44.82 & 0.217 & 6.64 & 270.3 & $-$26.06 & 4.844 & 0.65 \\
NGC 3862 & E & $34.95\pm0.15$ & 0.012 & 0.023403 & 97.80 & 93.38 & 0.453 & 15.26 & 265.1 & $-$25.41 & 3.946 & 0.21 \\
NGC 4061 & E: & $35.21\pm0.15$ & 0.018 & 0.026302 & 110.20 & 104.62 & 0.507 & 10.48 & 477.2 & $-$25.32 & 3.116 & 0.18 \\
NGC 4261 & E2$-$3 & $32.34\pm0.19 ^{2}$ & 0.009 & 0.003332 & 29.38 & 29.18 & 0.141 & 5.02 & 296.7 & $-$25.05 & 2.099 & 0.54 \\
\multirow{2}{*}{NGC 4373a} & SA0\scalebox{.8}{\textsuperscript{$\wedge$}}+: & \multirow{2}{*}{$32.60\pm0.52$} & \multirow{2}{*}{0.043} & \multirow{2}{*}{0.008019} & \multirow{2}{*}{33.10} & \multirow{2}{*}{32.58} & \multirow{2}{*}{0.158} & \multirow{2}{*}{2.94} & \multirow{2}{*}{201.6} & \multirow{2}{*}{$-$23.77} & \multirow{2}{*}{0.640} & \multirow{2}{*}{0.23} \\[-1mm]
 & edge-on &  &  &  &  &  &  &  &  &  &  &  \\
NGC 4429 & SA0\scalebox{.8}{\textsuperscript{$\wedge$}}+(r) & $30.73\pm0.15$ & 0.017 & 0.003382 & 13.90 & 13.81 & 0.067 & 2.65 & 173.4 & $-$23.93 & 0.647 & 0.58 \\
NGC 4435 & SB0\scalebox{.8}{\textsuperscript{$\wedge$}}0(s) & $31.12\pm0.05 ^{4}$ & 0.015 & 0.003399 & 16.75 & 16.64 & 0.081 & 1.38 & 155.0 & $-$23.75 & 0.420 & 0.41 \\
NGC 4751 & SA0\scalebox{.8}{\textsuperscript{$\wedge$}}$-$: & $31.86\pm0.16$ & 0.062 & 0.005694 & 23.50 & 23.24 & 0.113 & 2.46 & 350.6 & $-$23.59 & 0.556 & 0.29 \\
NGC 4786 & cD pec & $34.26\pm0.15$ & 0.019 & 0.017115 & 71.20 & 68.82 & 0.334 & 8.49 & 284.7 & $-$25.51 & 3.743 & 0.30 \\
NGC 4797 & S0\scalebox{.8}{\textsuperscript{$\wedge$}}$-$: & $35.35\pm0.15$ & 0.006 & 0.028053 & 117.70 & 111.36 & 0.540 & 6.14 & 201.7 & $-$25.44 & 2.210 & 0.18 \\
NGC 5084 & S0 edge-on & $31.85\pm0.15$ & 0.060 & 0.005664 & 23.40 & 23.14 & 0.112 & 2.50 & 199.8 & $-$24.77 & 1.138 & 0.59 \\
NGC 5193 & E pec: & $33.35\pm0.15 ^{2}$ & 0.029 & 0.010247 & 46.77 & 45.83 & 0.222 & 3.22 & 205.1 & $-$24.66 & 1.187 & 0.24 \\
NGC 5208 & S0 & $35.09\pm0.15$ & 0.018 & 0.024894 & 104.20 & 99.20 & 0.481 & 12.35 & \nodata & $-$25.55 & 4.475 & 0.22 \\
NGC 5838 & SA0\scalebox{.8}{\textsuperscript{$\wedge$}}$-$ & $31.75\pm0.15$ & 0.027 & 0.005420 & 22.40 & 22.16 & 0.107 & 1.74 & 273.6 & $-$24.13 & 0.667 & 0.41 \\
NGC 6861 & SA0\scalebox{.8}{\textsuperscript{$\wedge$}}$-$(s): & $32.24\pm0.36 ^{1}$ & 0.028 & 0.010137 & 28.05 & 27.49 & 0.133 & 2.36 & 387.2 & $-$24.74 & 1.263 & 0.47 \\
NGC 6958 & cD & $33.03\pm0.15$ & 0.023 & 0.009750 & 40.30 & 39.53 & 0.192 & 2.98 & 185.2 & $-$24.59 & 1.246 & 0.30 \\
\enddata
    \tablecomments{Properties of the ETGs observed in these HST programs. Col. (2) gives the galaxy morphology reported by \citet{devaucouleurs91}. Col. (3) lists the adopted distance modulus, with preference for SBF measurements. SBF measurements were drawn from \textsuperscript{1}\citet{tonry01}, \textsuperscript{2}\citet{jensen03}, \textsuperscript{3}\citet{cantiello05}, \textsuperscript{4}\citet{mei07}, and \textsuperscript{5}\citet{blakeslee09}. The remainder were derived from a luminosity distance ($D_L$) that is itself estimated using corrected redshifts and a standard cosmology as described in Section~\ref{sec:intro}. Cols.\ (4) and (5) report Galactic extinction in the WFC3/F160W band and optical redshifts from the NASA/IPAC Extragalactic Database (NED) after correcting for the Virgo + Great Attractor + Shapley inflow model \citep{mould00}. Cols.\ (6) and (7) give $D_L$ and angular size distance ($D_A$) corresponding to the adopted $z$ values and cosmology computed using the \citet{cosmocalc} cosmological calculator, with the corresponding physical scale given in col.\ (8). Col.\ (9) gives an \emph{H}-band effective radius \re\ estimate using the half-light radius from the corresponding multi-Gaussian expansion (MGE) described in Section~\ref{sec:mge}. Col.\ (10) gives the measured central stellar velocity dispersion from the HyperLEDA database \citep{paturel03}; the apparent total $K$-band magnitudes, also from HyperLEDA, are combined with the adopted $D_L$ to estimate $M_K$ for each galaxy in col.\ (11). Col.\ (12) gives the total \emph{H}-band luminosity estimated from MGE models. Lastly, col.\ (13) gives the estimated \rg\ value from the $\mbh - L_K$ scaling relation \citep{korho13}.}
   \label{tbl:galaxy_info}
\end{deluxetable*}

In addition to the quality and coverage of the tracer kinematics, the accuracy of both stellar and gas-dynamical models relies on an accurate model of the galaxy mass as a function of radius. In most cases, the gas masses of CNDs in luminous ETGs are in the range $\sim$10$^5$--10$^9$ \msun\ \citep[e.g.,][]{young11,boizelle17,ruffa19a}, with the result that the enclosed mass profile is dominated by the stellar component from $\sim$\rg\ out to at least a few half-light radii (\re). Models of a galaxy's mass profile are typically derived from two-dimensinoal (2D) optical/near-IR images of the observed stellar surface brightnesses, typically at an angular resolution similar to \rg\ (or at least $\theta_\mathrm{FWHM}$ for the ALMA CO kinematics) to avoid potentially biasing the \mbh\ value \citep{yoon17}.

Unfortunately, previous optical/near-IR imaging did not always have sufficient angular resolution to be fully useful in ALMA CO dynamical modeling efforts for luminous galaxies. At typical distances of 20$-$50 Mpc, these CNDs subtend angular sizes of only a few arcseconds or less, with the typical \rg\ of the host BH on the order of $\sim 0\farcs 05 - 1\farcs 0$. These CND systems are therefore difficult to resolve and study at optical/near-IR wavelengths except with the Hubble Space Telescope (HST), and the James Webb Space Telescope (JWST), or large-aperture ground-based facilities with adaptive optics capabilities. In addition, for certain ETGs the dust accompanying large CO column densities is sufficiently opaque, extended, and/or face-on to limit the usefulness of optical HST imaging when constructing stellar mass models. The intrinsic CND dust extinction is not known \emph{a priori}, but studies have demonstrated peak $A_V \sim 3-5$ mag extinction of the background stellar light in some cases \citep[e.g.,][]{ferr96,viaene17,boizelle19,boizelle21,cohn21,cohn23,kabasares22} that is much higher than foreground screen estimates \citep[e.g.,][]{tran01}. Due in large part to exquisite CO kinematics, select ALMA studies have demonstrated that uncertainty in the dust extinction correction tends to be the dominant term in the BH mass error budget \citep{boizelle19,boizelle21,cohn21,kabasares22}. 

To facilitate more accurate BH mass measurements derived from ALMA CO data, we developed a set of HST programs to obtain new broadband imaging for ETGs with the most promising ALMA CO emission-line data sets. This sample included candidates with large \rg\ and relatively small ALMA $\theta_\mathrm{FWHM}$, together with regular CO kinematics, but having no near-IR (or, at times, even optical) HST imaging. In this paper, we focus on constructing stellar luminosity models that can be employed in ongoing ALMA CO modeling efforts or in future gaseous/stellar-dynamical modeling. We additionally explore analyses of the stellar surface brightness and color behavior of the near-IR data to better place these targets in the context of volume-limited surveys. Our HST programs also include supplementary optical data that are needed to map dust attenuation, but we defer that analysis to a future paper.

This paper is organized as follows. In Sections~\ref{sec:sample} and \ref{sec:imaging}, we introduce the ETG sample and describe the new and archival optical/near-IR observations, respectively. In Section~\ref{sec:sb_behavior}, we explore the isophotal and color behavior of these galaxies. We detail the construction of stellar luminosity models using the multi-Gaussian expansion (MGE) formalism in Section~\ref{sec:mge}. In Section~\ref{sec:discussion}, we discuss these results in the context of past work and analyze the accuracy and consistency of these MGE solutions. In Section~\ref{sec:conclusion}, we preview next steps and discuss conclusions. Throughout this paper, we adopt a standard $\Lambda$CDM cosmology with $\Omega_\mathrm{m} = 0.308$, $\Omega_\mathrm{vac} = 0.692$, and Hubble constant $H_0 = 73$ km s\per\ Mpc\per\ \citep{blakesless21,riess22,kenw22}. Magnitudes are in the Vega system.

\begin{figure*}[!ht]
    \includegraphics[width=\columnwidth]{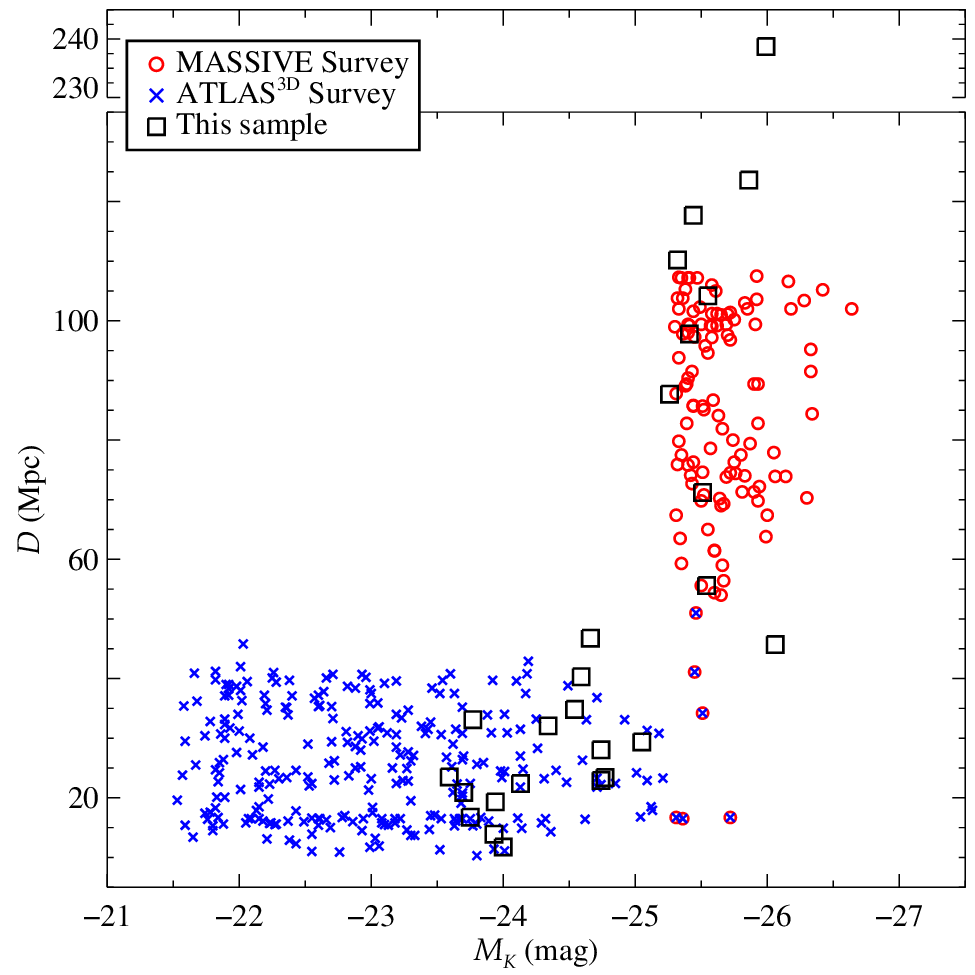}
    \includegraphics[width=\columnwidth]{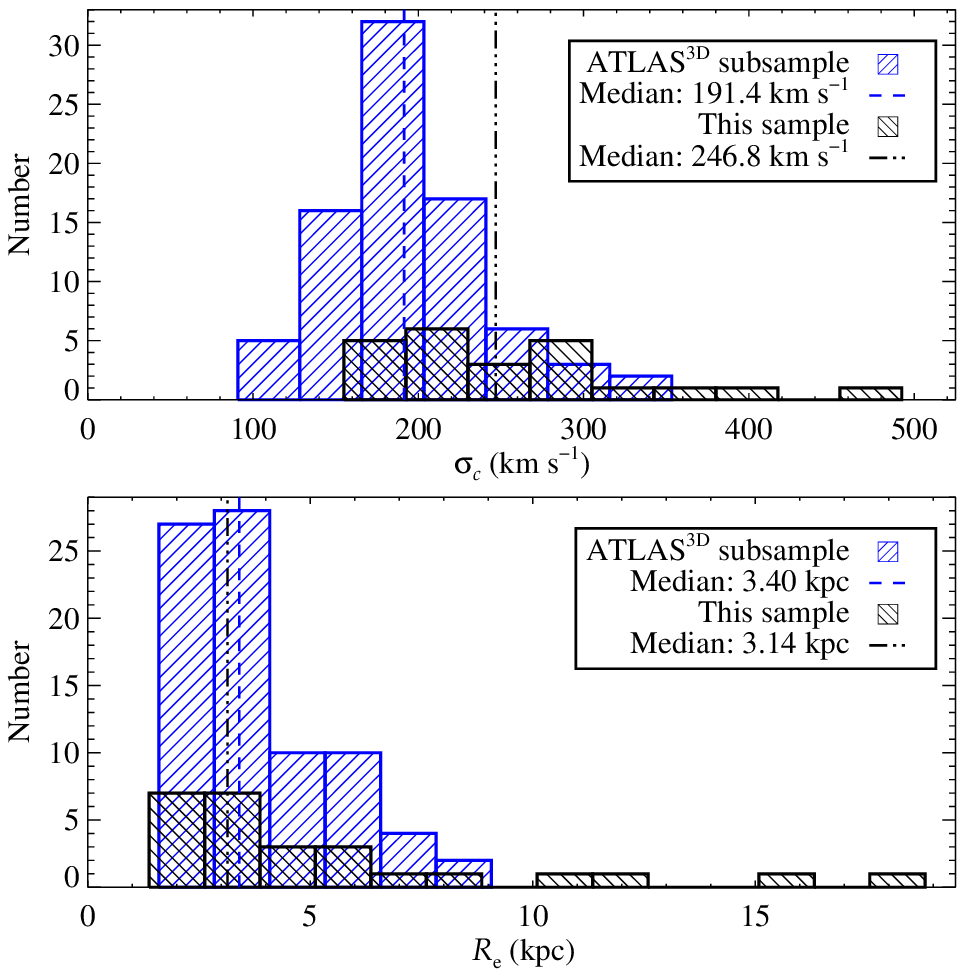}
    \centering
    \caption{Galaxy properties for this sample compared to those from the ATLAS$^\mathrm{3D}$ and MASSIVE surveys. Distances and absolute $K_s$-band magnitudes \citep[\textit{left}; following][]{ma14} demonstrate our sample occupies an intermediate range of luminosities, while tending to have higher central velocity dispersions ($\sigma_c$; \textit{upper right}) but only slightly smaller median half-light radii (\re; \textit{lower right}) when compared to a subsample of ATLAS$^\mathrm{3D}$ targets with $M_K < -23.59$ mag. The \re\ values for the ATLAS$^\mathrm{3D}$ survey were derived from a combination of RC3 and 2MASS determinations, normalized to agree on average with RC3 \citep{capp11}.}
    \label{fig:us_vs_others}
\end{figure*}

\section{ETG Sample} 
\label{sec:sample}

We identified targets for this project based on the existence of \coone, \cotwo, and/or \cothree\ imaging of ETGs in ALMA Cycles 2$-$5. Selection criteria were designed to ensure that current (or future) ALMA CO imaging could enable high-confidence \mbh\ constraints. To that end, we selected systems with very regular CO kinematics and $\theta_\mathrm{FWHM} \sim 0\farcs1 - 0\farcs 6$ to ensure well-resolved disks. This was done by analyzing currently unpublished pipeline-calibrated images from the ALMA archive (Boizelle et al., in prep.) or by looking to literature sources to help identify additional targets with regular CO kinematics \citep{boizelle17,davis17,voort18,babyk19,rose19,ruffa19a,ruffa19b,zabel19,boizelle21,davis22,ruffa23,kabasares24}. Next, we selected ETGs with estimated $\rg \gtrsim 0\farcs 09$ \citep[using measured central $\sigma_\mathrm{c}$ and \mbh\ estimated from the $\mbh - L_K$ scaling relation;][]{korho13} to ensure that \rg\ could be fully resolved, perhaps using a more extended ALMA configuration in a reasonable amount of time. This led to a natural cutoff for distances $D \gtrsim 250$ Mpc or redshifts $z\gtrsim 0.06$. All targets have an expected CND radius $\gtrsim 0\farcs5$ to allow for spatial characterization of the CND dust properties. Lastly, we removed candidates with previous HST wide-field, broadband near-IR (e.g., F110W) imaging to focus on cases that would benefit the most from additional HST data. Previous Near Infrared Camera and Multi-Object Spectrograph \citep[NICMOS;][]{dickinson02} data sets had too narrow a field of view (FOV) to build reliable stellar luminosity models, so we included several HST NICMOS-observed ETGs in our sample.

The final sample containing 26 ETGs is listed in Table~\ref{tbl:galaxy_info} and compared to two surveys of local ETGs in Figure~\ref{fig:us_vs_others}. This sample containing about 42\% elliptical and 58\% lenticular galaxies is not entirely representative of the local population of luminous ETGs, although it does span $K_s$-band absolute magnitudes ($M_K$) of about $-23.5 < M_K < -26$ mag. Depending on the limiting $M_K$, volume-limited surveys return different distributions: ATLAS$^\mathrm{3D}$ \citep[$M_K < -21.5$ mag or $D < 42$ Mpc;][]{capp11} contained 26\% elliptical and 74\% lenticular galaxies, while MASSIVE \citep[$M_K < -23.5$ mag or $D < 108$ Mpc;][]{ma14} contained 68\% and 32\%, respectively. Because of ALMA's declination limit and the abundance of targets in the ALMA archive with $\delta < 0\degr$, we have only six galaxies in common with ATLAS$^{\mathrm{3D}}$ (NGC 3245, NGC 4261, NGC 4429, NGC 4435, NGC 4697, and NGC 5838) and three in common with MASSIVE (NGC 997, NGC 3862, and NGC 5208). If we consider only ATLAS$^{\mathrm{3D}}$ galaxies with $M_K \lesssim -23.6$ mag, the distribution of morphological types becomes identical to that of our sample. The galaxies in our sample have larger $\sigma_\mathrm{c}$ but lie somewhere between ATLAS$^{\mathrm{3D}}$ (with $M_K \lesssim -23.6$ mag and median $\re \sim 3.4$ kpc) and MASSIVE in terms of stellar luminosities (with a median $M_K \sim -24.74$ mag), with similar median \re. However, some of the most luminous ($M_K < -25.3$ mag) and distant ($D \gtrsim 100$ Mpc) ETGs in our sample are very extended ($\re > 10$ kpc) and without analogs in the more local universe. Restricting this comparison to ETGs in our sample that are within the ATLAS$^\mathrm{3D}$ 42 Mpc distance limit, we find that our sample is moderately more compact (median $\re \sim 3.14$ kpc). This selection bias follows a known trend of dynamical BH mass measurements being preferentially pursued in more compact galaxies relative to the distribution at a fixed luminosity, or in systems with larger projected \rg\ \citep{vandb15}. As we determine in Section~\ref{subsec:nuker}, over half of our sample have cored surface brightness profiles, while the remainder exhibit steeper slopes; volume-limited ETG surveys show much greater preference for cuspy central slopes \citep[][]{krajnovic13}. Lastly, nearly every object in our sample is a member of some galaxy group, and five (Hydra A, NGC 3258, NGC 3268, NGC 3557, and NGC 6861) are referenced in the literature as either BGGs or BCGs \citep{garcia93,zabludoff00,sato12,pablocaso13}.

For 12 of these galaxies, we adopted luminosity distances ($D_L$) from surface brightness fluctuation (SBF) distance modulus measurements \citep{tonry01,jensen03,cantiello05,mei07,blakeslee09}. For the remaining 14 galaxies lacking SBF-derived $m-M$ values, we estimated $D_L$ from Hubble flow velocities using the Virgo + Great Attractor + Shapley Supercluster inflow model \citep{mould00} and the \citet{cosmocalc} cosmological calculator, assuming the corrected redshift reported in the NASA/IPAC Extragalactic Database\footnote{\url{https://ned.ipac.caltech.edu/}} (NED) is entirely cosmological in origin. The median uncertainty in SBF-derived $m-M$ corresponds to $\sim$12\% uncertainty in $D_L$, while Hubble flow-derived distance moduli give only lower-bound $\sim$7\% uncertainties from redshift errors alones. The uncertainties in the physical scale are calculated to be $\sim$18\% from the median distance and redshift errors.

We note that nine of the targets in our sample already have published BH mass measurements or estimates \citep{barth16a,barth16b,davis17,davis18,ruffa19a,ruffa23,boizelle19,boizelle21,kabasares22,thater22}. For two of these ETGs (NGC 3258 and NGC 4261), the respective HST data have already been utilized in gas-dynamical modeling \citep{boizelle19,boizelle21}. For uniformity, however, we analyze them again using the approach outlined in Sections~\ref{sec:imaging} and \ref{sec:mge}. In many cases, the stellar luminosity models constructed here will still be useful in any re-analysis of the ALMA CO data, or in more comprehensive explorations of CND dust attenuation.

\begin{deluxetable*}{lcccccl}[!ht]
\tabletypesize{\footnotesize}
\tablecaption{New and Archival HST Observations}
\tablewidth{0pt}
\tablehead{
\\[-5mm]
\multicolumn{1}{c}{\multirow{2}{*}{Galaxy}} & \multicolumn{2}{c}{This Paper$^{1}$} & \multicolumn{3}{c}{Other Programs} & \colhead{GO ID}\\[-1.5mm]
\cmidrule(lr){2-3}\cmidrule(lr){4-6}
\\[-5.5mm]
\multicolumn{1}{c}{Name} & \colhead{WFC3/IR} & \multicolumn{1}{c}{WFC3/UVIS} & \colhead{WFPC2} & \colhead{ACS} & \colhead{WFC3/UVIS} &  }
\startdata
Hydra A & F110W, F160W & F475W & \nodata & F814W & \nodata & 12220 (PI: Mittal) \\
\hline
NGC \phantom{0}612 & F110W, F160W & F475W & \nodata & F814W & \nodata & 15444 (PI: Barth) \\
\hline
NGC \phantom{0}997 & F110W, F160W & F475W, F814W & \nodata & \nodata & \nodata & \multicolumn{1}{c}{\nodata} \\
\hline
NGC 1332 & F110W, F160W & F438W & F814W & \nodata & \nodata & 5999 (PI: Phillips) \\
\hline
\multirow{2}{*}{NGC 1387} & \multirow{2}{*}{F110W, F160W} & \multirow{2}{*}{\nodata} & F606W & \nodata & \nodata & 5446 (PI: Illingworth) \\
 &  &  & \nodata & F475W, F850LP & \nodata & 10217 (PI: Jordan) \\
\hline
NGC 3245 & F110W, F160W & F475W & F547M & \nodata & \nodata & 6837 (PI: Ho) \\
\hline
NGC 3258 & F110W, F160W & \nodata & \nodata & F435W, F814W & \nodata & 9427 (PI: Harris) \\
\hline
NGC 3268 & F110W, F160W & F555W & \nodata & F435W, F814W & \nodata & 9427 (PI: Harris) \\
\hline
NGC 3271 & F110W, F160W & F475W, F814W & \nodata & \nodata & \nodata & \multicolumn{1}{c}{\nodata} \\
\hline
NGC 3557 & F110W, F160W & F438W & F555W & \nodata & \nodata & 6587 (PI: Richstone) \\
\hline
\multirow{3}{*}{NGC 3862} & \multirow{3}{*}{F110W, F160W} & \multirow{3}{*}{\nodata} & F547M, F791W & \nodata & \nodata & 5927 (PI: Ford) \\
 &  &  & F702W & \nodata & \nodata & 9069 (PI: Biretta) \\
 &  &  & \nodata & \nodata & F225W, F475W, F814W & 14159 (PI: Meyer) \\
\hline
NGC 4061 & F110W, F160W & F475W & F555W, F814W & \nodata & \nodata & 9106 (PI: Richstone) \\
\hline
\multirow{3}{*}{NGC 4261} & \multirow{3}{*}{F110W, F160W} & \multirow{3}{*}{\nodata} & F547M, F675W, F791W & \nodata & \nodata & 5124 (PI: Ford) \\
 &  &  & F702W & \nodata & \nodata & 5476 (PI: Sparks) \\
 &  &  & F450W, F606W, F814W & \nodata & \nodata & 11339 (PI: Zezas) \\
\hline
NGC 4373a & F110W, F160W & F475W, F814W & \nodata & \nodata & \nodata & \multicolumn{1}{c}{\nodata} \\
\hline
NGC 4429 & F110W, F160W & F475W, F814W & F606W & \nodata & \nodata & 5446 (PI: Illingworth) \\
\hline
\multirow{2}{*}{NGC 4435} & \multirow{2}{*}{F110W, F160W} & \multirow{2}{*}{\nodata} & F450W, F675W, F814W & \nodata & \nodata & 6791 (PI: Kenney) \\
 &  &  & \nodata & F475W, F850LP & \nodata & 9401 (PI: Cote) \\
\hline
\multirow{2}{*}{NGC 4697} & \multirow{2}{*}{F110W, F160W} & \multirow{2}{*}{F555W} & \nodata & F475W, F850LP & \nodata & 10003 (PI: Sarazin) \\
 &  &  & \nodata & \nodata & F225W, F336W & 11583 (PI: Bregman) \\
\hline
NGC 4751 & F110W, F160W & F475W, F814W & \nodata & \nodata & \nodata & \multicolumn{1}{c}{\nodata} \\
\hline
NGC 4786 & F110W, F160W & F438W & F555W & \nodata & \nodata & 6587 (PI: Richstone) \\
\hline
NGC 4797 & F110W, F160W & F475W, F814W & \nodata & \nodata & \nodata & \multicolumn{1}{c}{\nodata} \\
\hline
NGC 5084 & F110W, F160W & F475W & \nodata & \nodata & \nodata & \multicolumn{1}{c}{\nodata} \\
\hline
NGC 5193 & F110W, F160W & F475W & F814W & \nodata & \nodata & 5910 (PI: Lauer) \\
\hline
NGC 5208 & F110W, F160W & F475W, F814W & \nodata & \nodata & \nodata & \multicolumn{1}{c}{\nodata} \\
\hline
NGC 5838 & F110W, F160W & F555W & F450W, F814W & \nodata & \nodata & 7450 (PI: Peletier) \\
\hline
NGC 6861 & F110W, F160W & F438W & F814W & \nodata & \nodata & 5999 (PI: Phillips) \\
\hline
NGC 6958 & F110W, F160W & F475W & F547M, F814W & \nodata & \nodata & 8686 (PI: Goudfrooij) \\
\enddata
   \tablecomments{Optical and near-IR medium and broadband-filter HST observations that provide good coverage and depth. New WFC3/IR and UVIS observations ($^{1}$GO IDs: 14920, 15226, and 15909; PI: Boizelle) supplement archival data sets that were obtained using the WFPC2, ACS, or WFC3/UVIS instruments, ensuring sufficient coverage and sampling for dust attenuation modeling. The typical \emph{H}-band FOV is 3.6\arcmin$\times$3.6\arcmin with  exposure times ranging from 250 to 400 s. The \emph{J}-band images are typically subarrays with a FOV of $\sim$ 1\arcmin$\times$1\arcmin\ and exposure times ranging from 100 to 250 s. The optical data range of coverage typically fell between the \emph{H} and \emph{J}-band ranges, with an average FOV of 2.5\arcmin$\times$2.5\arcmin\ and exposure times ranging from 150 to 400 s. \label{tbl:hst_obs}}
\end{deluxetable*}


\section{Optical/near-IR Data} 
\label{sec:imaging}

Our sample of 26 ETGs had inconsistent broadband imaging in the optical/near-IR regimes, although all had Spitzer Infrared Array Camera \citep[IRAC;][]{fazio04} channel 1 or 2 imaging that probes far out into the stellar halo. None had high-resolution near-IR data to mitigate the impact of dust attenuation while also covering a sufficiently wide FOV, which is needed to construct stellar luminosity models, and only half had HST imaging in the F814W filter (or similar; see Table~\ref{tbl:hst_obs}). In this section, we detail efforts to calibrate, mosaic, and align HST data across multiple filters. We also discuss near-IR sky subtraction with the help of larger-scale Spitzer data.

\subsection{HST Imaging}
\label{sec:hstimaging}

We observed each ETG in a single HST orbit through programs GO--14920, GO--15226, or GO--15909 (Cycles 24, 25, and 27; PI: Boizelle) using Wide Field Camera 3 \citep[WFC3;][]{dressel22} in the IR channel with a focus on F160W data. For just over half of our targets, we obtained additional WFC3/UVIS imaging to provide either the first or supplemental optical HST imaging, since broad wavelength coverage is crucial in constraining dust attenuation (as a Galactic reddening law has $A_\lambda$ decreasing by a factor of $\sim$8 from $B$ to $H$ band). All these HST data can be found in the Mikulski Archive for Space Telescopes (MAST): \dataset[10.17909/98s2-be33]{http://dx.doi.org/10.17909/98s2-be33} and on Zenodo: \dataset[10.5281/zenodo.11122962]{https://doi.org/10.5281/zenodo.11122962}. In Table~\ref{tbl:hst_obs}, we list these new WFC3 data together with the archival Wide Field Planetary Camera 2 \citep[WFPC2;][]{mcmaster08}, Advanced Camera for Surveys \citep[ACS;][]{ryon22}, and WFC3 observations that were selected for this project to span the desired wavelength range.

\begin{figure*}[!ht]
    \includegraphics[width=\textwidth]{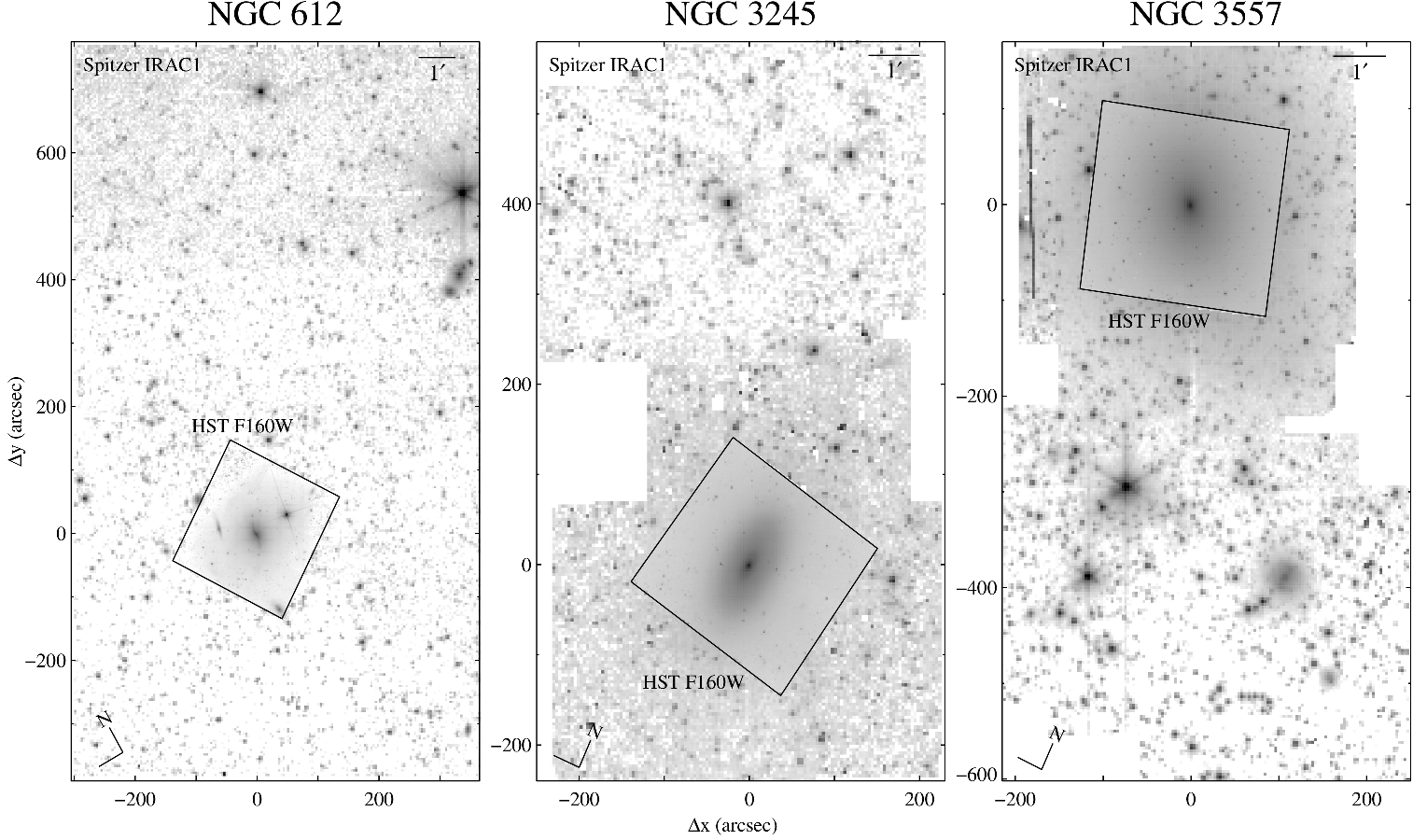}
    \caption{HST WFC3/F160W ($H$-band) drizzled mosaics for three targets overlaid on Spitzer IRAC1 (3.6 $\mu$m) supermosaics that were scaled to match the $H$-band data at the edge of the HST footprint. These targets highlight the diversity of stellar halo extents and the challenges in accurately determining the $H$-band sky background for most of the sample. Images are displayed using a logarithmic intensity scale.}  
    \label{fig:hst_spitz}
\end{figure*}

\subsubsection{New WFC3 Data}

In the IR channel, we obtained both F110W and F160W (hereafter \emph{J} and \emph{H}) imaging for all targets, primarily to construct a near-IR mosaic that adequately samples the CND while also probing well into the stellar halo. We employed a 4-point dither pattern for \emph{H}-band observations, adopting the WFC3-IR-DITHER-BOX-MIN pattern to more optimally sample the \emph{H}-band point-spread function (PSF) for more compact galaxies. For more extended targets, we used a large square dither pattern with offsets of up to 75\arcsec\ and a total coverage of up to 3.6\arcmin$\times$3.6\arcmin. In all cases, we placed the central bright region of each galaxy within the overlap of the four pointings. In most cases, these WFC3/IR mosaics cover out to a few $\times$\re, with a median projected $\re \sim 20\arcsec$ and a maximum of $\sim$40\arcsec. Individual \emph{H}-band exposure times ranged from 250 to 400 s, using various SPARS or STEP sampling sequences to avoid time loss due to buffer dumps. Each MULTIACCUM image had NSTEP = 9 or higher to enable good up-the-ramp calibration and cosmic-ray rejection. Combined exposure times in the overlap regions deliver background-limited sensitivity for this filter, and we estimate a typical $1\sigma$ surface brightness depth of $\sim$25.9 mag arcsec\pertwo\ in a 1\arcsec$\times$1\arcsec\ region measured at the edges of this dither pattern.

For wavelengths bluer than the $H$ band, observations were centered on the CND, with the observing setup adapted to avoid costly buffer dumps and fit each target in a single orbit. We generally obtained the \emph{J}-band data using the IRSUB512-FIX sub-array aperture, employing the 2-point WFC3-IR-DITHER-LINE pattern to better sample the PSF, with a final FOV of about 60\arcsec $\times$60\arcsec. For nearly 80\% of our sample, this $J$-band FOV covers out to at least one (projected) \re\ from the nucleus. In a few cases, the available optical data and orbit constraints allowed for full-aperture \emph{J}-band imaging. Sampling sequences for the \emph{J}-band observations were similar to those employed in acquiring the \emph{H}-band data. Individual exposure times generally ranged between 100$-$250 s, with combined exposure time reaching (or exceeding) the threshold for background-limited sensitivity. Two cases with extended stellar light distributions (NGC 3862 and NGC 4261) host AGN that are more prominent in the optical, and we obtained additional \emph{J} and \emph{H}-band imaging of these galaxies to better characterize the near-IR point sources. These data employed an IRSUB256-FIX aperture with an ideal 4-point dither pattern and the SPARS5 / NSAMP = 15 combination for rapid temporal sampling, with a total exposure time for an individual pointing of 33 s.

For over 80\% of our sample, the existing HST optical data were not sufficient for our eventual goal of constraining dust attenuation arising from the CNDs. We obtained additional WFC3/UVIS data using the F438W or F475W filter ($B$ band) and/or the F814W filter ($I$ band) to ensure broad wavelength coverage. In a few cases, orbit scheduling also allowed for F555W imaging for more complete wavelength sampling. To allow all data for a given object to be scheduled in a single orbit, we chose either the UVIS1-2K2A-SUB or UVIS2-M1K1C-SUB apertures. Total exposure lengths for individual frames ranged from 150 to 400 s depending on the time available. To limit the impact of cosmic rays, observations in a single filter were split into either 2 or 3 frames and dithered using the corresponding WFC3-UVIS-DITHER-LINE pattern.

\begin{figure*}[!ht]
    \includegraphics[width=\textwidth]{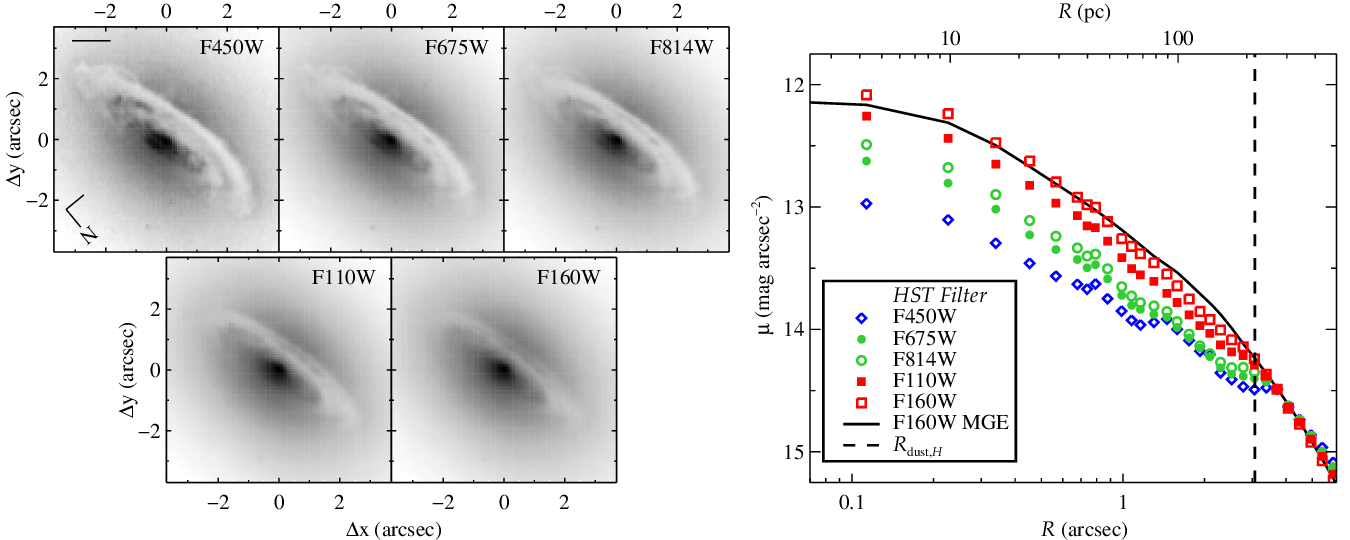} 
    \caption{Aligned HST images (\textit{left}) show the NGC 4435 nucleus and CND across five different optical/near-IR filters and two detectors, displayed using an inverted grayscale. The scale bar corresponds to 100 pc. Surface brightness measurements extracted along the major axis (\textit{right}; without any dust masking and scaled together at $R = 3.5\arcsec$) show the $>$2$\times$ change in attenuation, and the best-fit stellar luminosity model for the $H$-band mosaic (dust-masked; see Section~\ref{sec:mge}) fits the data well. At the disk outskirts between $2\farcs5 < R < 3\farcs5$, the diffuse dust primarily impacts the optical filters, while the highest column density gas lies between $1\farcs 5 < R < 2\farcs 5$. The dust radius $R_{\mathrm{dust},H}$ from Table~\ref{tbl:ell_results} identifies the extent of the near-IR-thick disk.} 
    \label{fig:filters_zoom_in}
\end{figure*}

\subsubsection{Archival Data}

We retrieved optical (and occasionally UV) ACS, WFPC2, and WFC3 images of our target galaxies from MAST\footnote{\url{https://archive.stsci.edu/hst}}. When there were duplicates in a specific wavelength regime, preference was given to data taken with later-generation detectors, sequences that gave better spatial coverage of the dusty CND, and better angular resolution. Additional criteria included good image quality (e.g., low incidence of cosmic rays overlapping with the CND) and an unsaturated nucleus. Narrowband data were not included in the final analysis due to the likelihood of emission-line contamination \citep[e.g.,][]{walsh08}. As mentioned earlier, NICMOS data were not included due to calibration issues towards that instrument's edges and its limited FOV.

\subsubsection{Calibration and Analysis}

After processing these new HST data through the \texttt{CALWF3} pipeline \citep{dressel22}, we created final \emph{H}-band mosaics and \emph{B}, \emph{I}, and \emph{J} subarray products using \texttt{AstroDrizzle} \citep{astrodrizz}. All images were drizzled to the same pixel scale of 0.08\arcsec~pixel\per\ to facilitate the exploration of dust extinction across each CND. For the dithered WFC3/IR data, we adopted a pixel fraction of 0.75 to optimize the point-spread function (PSF) sampling and pixel noise. Since the WFC3/UVIS data had smaller detector pixel sizes but less ideal dithering, we used the same pixel fraction. In general, these WFC3 data obtained in a single orbit remain well aligned after drizzling. In Figure~\ref{fig:hst_spitz}, we show examples of our \emph{H}-band mosaics and the HST footprint overlaid on larger-scale near-IR imaging.

Preliminary mosaicing of the archival HST data did not align well with the new WFC3 data, so we first aligned the pipeline-calibrated files to the \emph{H}-band mosaic using \texttt{TweakReg} \citep{astrodrizz}. We then combined the single-filter data in \texttt{AstroDrizzle} using the same pixel scale. The accuracy of dust attenuation modeling is very sensitive to the relative (sub-pixel) alignment of data across all filters. Slight offsets between different filters were still present, so to further improve the alignment of the HST data for each target, we calculated a luminosity-weighted centroid using a series of stellar isophotes that were measured beyond the dusty features of the CND. Afterwards, we corrected for the offsets of the shorter-wavelength data relative to the \emph{H}-band mosaic. Finally, we confirmed the accuracy of these sub-pixel offsets (or introduced additional fine tuning) by inspecting the resulting color maps. In Figure~\ref{fig:filters_zoom_in}, we show an example of the alignment of these multi-wavelength images for a single target.

Recovering intrinsic stellar luminosities necessitates a prescription for the PSF in a given filter. Following standard practice, we created model \emph{H}-band PSFs for each target by dithering and drizzling copies of the Tiny Tim \citep{tinytim} F160W response as we did for the HST data. This dithering and drizzling process returns PSFs with somewhat narrower full widths at half maximum (average $\mathrm{FWHM}\sim 0.19\arcsec$) than are observed for foreground stars in the \emph{H}-band mosaics (generally $\sim$0.22$-$0.24\arcsec). An alternative approach is to employ an empirical PSF using either stars in each field or an average point-source response near a particular detector location. The $H$-band mosaics of our sample do not contain many suitable PSF stars, and galaxy light often contaminates the PSF wings. Therefore, we employed the composite WFC3/F160W PSF provided by STScI \citep{anderson16}, again dithering and drizzling copies of this frame in the same manner as done for the Tiny Tim files. While the empirical PSF only extends out to $R\sim 1\arcsec$, it produces slightly better agreement ($\mathrm{FWHM}\sim 0.20-0.21\arcsec$) with those measured in the $H$-band mosaics. None of our targets have overly dominant $H$-band point sources, so subsequent stellar luminosity fits are not affected by the limited PSF response range. In Section~\ref{sec:mge}, we compare stellar luminosity models constructed using both a theoretical and an empirical PSF, while for the remainder of this paper we adopt results that employed the Tiny Tim PSF.

\begin{figure*}
    \includegraphics[width=\textwidth]{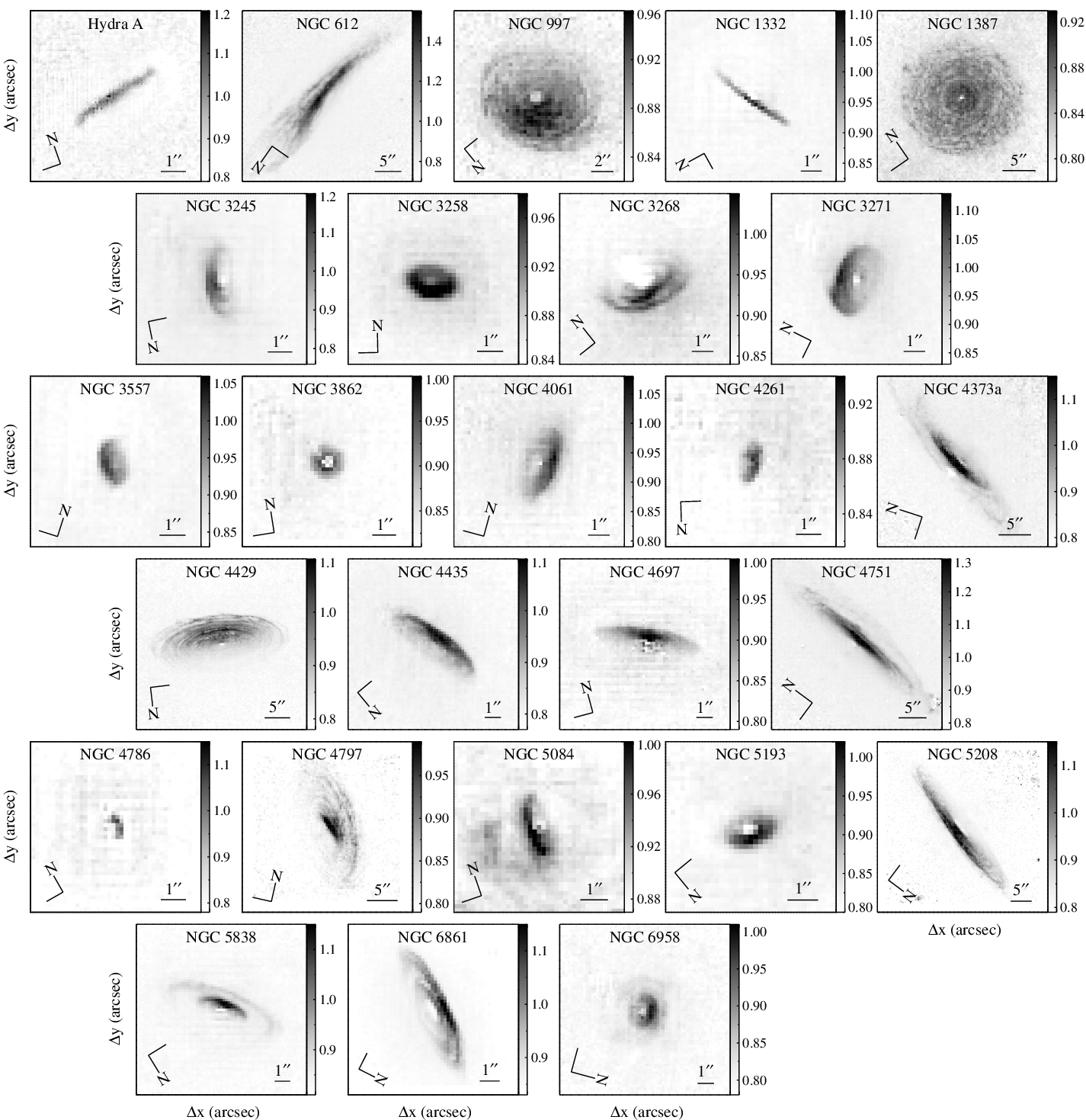} 
    \caption{Central portions of $J-H$ color maps, showing both the CNDs and their environs. The near side of each disk obscures a greater fraction of the stellar distribution, resulting in redder colors. While these CNDs were selected based on their regular dust morphology, filamentary dust features (especially in NGC 5084) or warped disk structures (especially in NGC 612 and NGC 4797) are also detected. $J-H$ colors (in mag) are mapped according to each color bar.}
    \label{fig:dustdisks}
\end{figure*}

\begin{figure*}[!ht]
    \includegraphics[width=\textwidth]{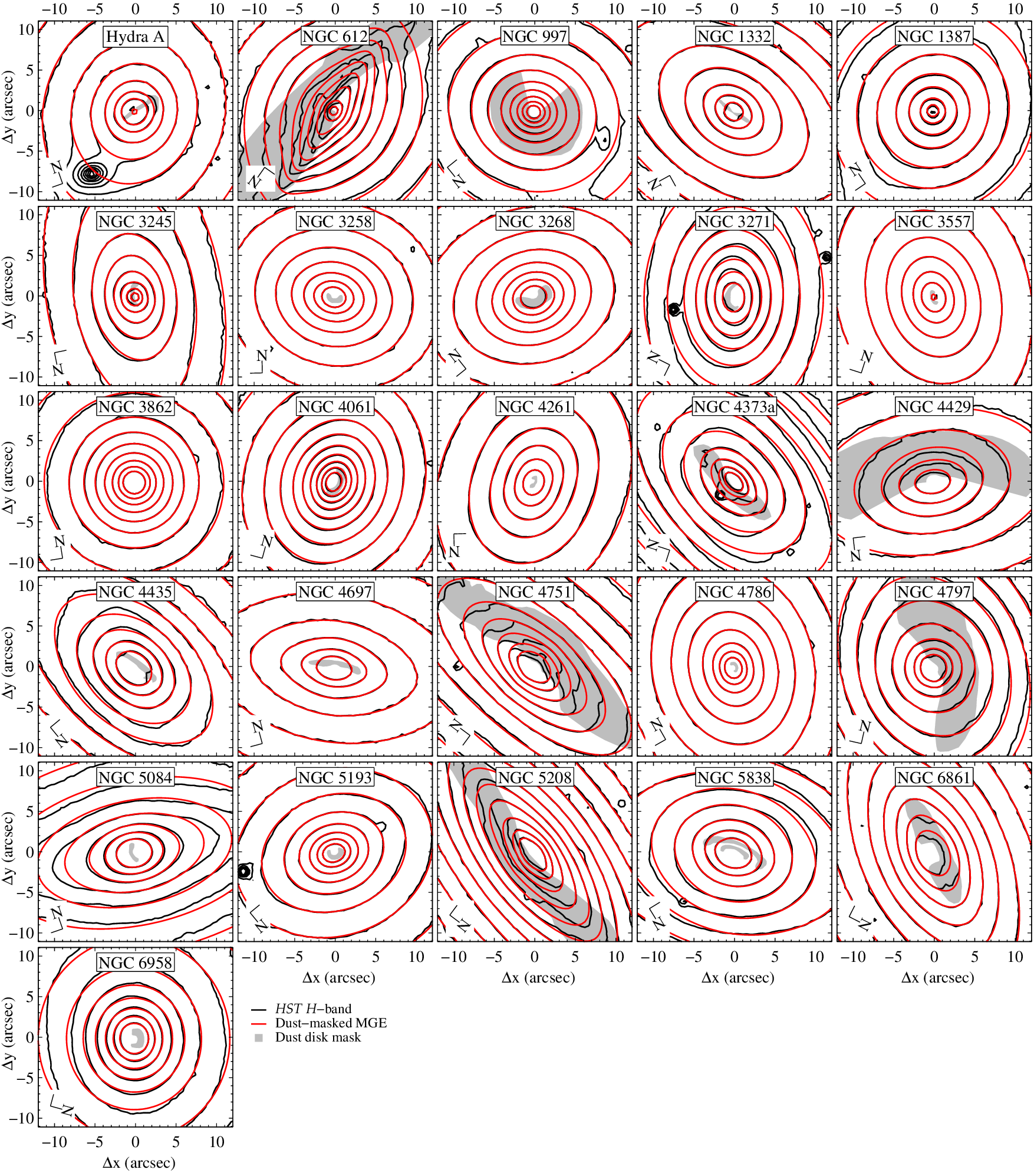}
    \centering
    \caption{Central portions of the $H$-band drizzled mosaics, with shaded regions showing the most dust-obscured portions of each CND based on high $J-H$ colors (see Figure~\ref{fig:dustdisks}) that were masked during the MGE fit (overplotted in red). The CNDs of NGC 1387 and NGC 3862 were left unmasked due to either only marginal evidence for dust attenuation or a more face-on orientation that would require more complete masking of the inner $\sim$1\arcsec. Contours are placed at logarithmic intensity intervals.}
    \label{fig:contours3}
\end{figure*}

\begin{figure*}[!ht]
    \centering
    \includegraphics[width=\textwidth]{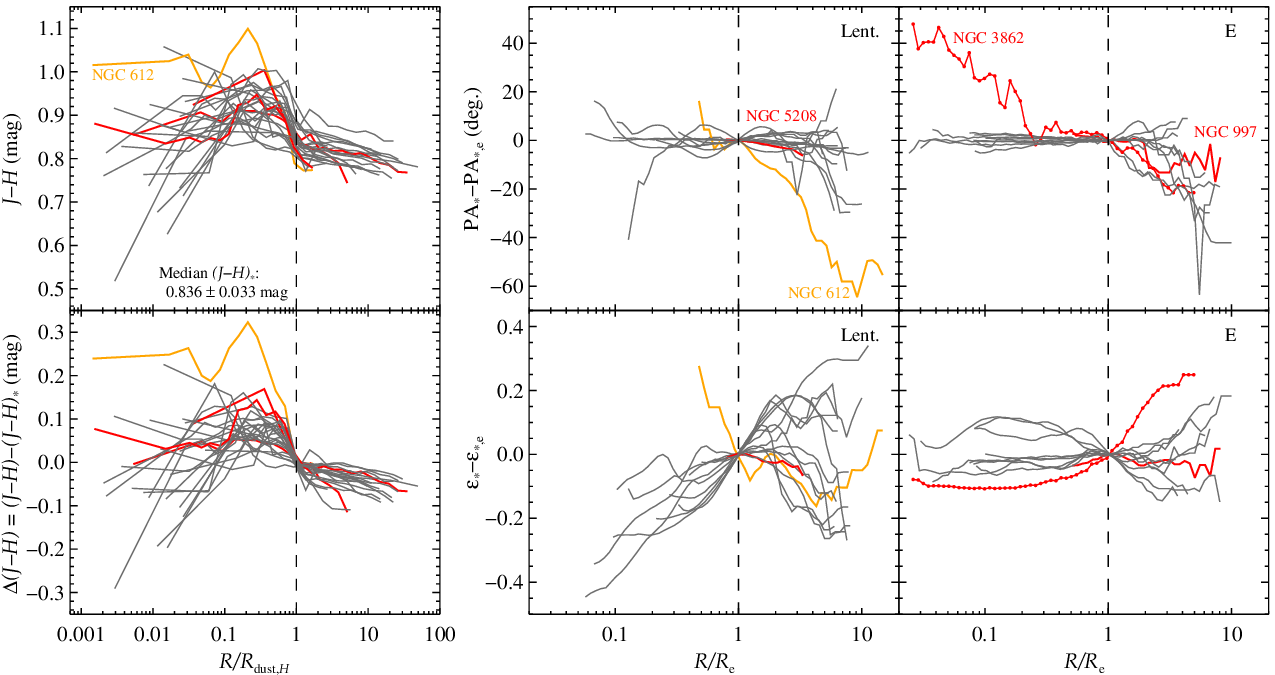} 
    \caption{Radial near-IR color and isophotal behavior of the ETG sample. \textit{Left}: $J-H$ color (\textit{above}) and $\Delta (J-H)$ reddening (\textit{below}) relative to stellar color just outside the CND plotted vs. radius normalized to the CND radius. The median stellar $J-H$ color and its 1$\sigma$ scatter just beyond $R_{\mathrm{dust},H}$ are noted in the upper left panel. \textit{Center and right}: Position angle ($\mathrm{PA}_\star - \mathrm{PA}_{\star,\mathrm{e}}$; \textit{above}) and ellipticity ($\varepsilon_\star - \varepsilon_{\star,\mathrm{e}}$; \textit{below}) plotted vs. radius after removing the values $\mathrm{PA}_{\star,\mathrm{e}}$ and $\varepsilon_{\star,\mathrm{e}}$ measured at the half-light radius \re, normalized to \re, for lenticular (\textit{center}) and elliptical (\textit{right}) galaxies, respectively. The position angle and ellipticity values are plotted starting just beyond the dust disk extent (i.e., for all $R \geq R_{\mathrm{dust},H}$). The only exception is NGC 612 (plotted in orange), for which $R_{\mathrm{dust},H} > \re$. Galaxies labeled and highlighted in red are also contained in the MASSIVE survey, some of which are clear outliers in $\mathrm{PA}_\star - \mathrm{PA}_{\star,\mathrm{e}}$.} 
    \label{fig:color_pa_eps}
\end{figure*}

\subsection{Spitzer Imaging and Sky Subtraction}
\label{sec:spitzerimaging}

Based on an initial analysis, 9 galaxies were sufficiently compact (or distant) to allow for accurate $H$-band sky removal using the edges of the corresponding WFC3 mosaics. For the remaining 17 galaxies (see Table~\ref{tbl:ell_results}), \emph{H}-band stellar light contributions near the edges of the HST footprint were close to the expected background level \citep[primarily zodiacal in origin;][]{pirzkal14}, as estimated using the WFC3/IR Exposure Time Calculator\footnote{\url{https://etc.stsci.edu/etc/input/wfc3ir/imaging/}} (ETC) for the corresponding solar angles. Such a high level of stellar light that persists out to a projected distance of $R\sim 2\arcmin$ precludes accurate sky subtraction using the $H$-band data alone. Following the method outlined by \citet{boizelle19}, we determined those sky values with the aid of larger-scale Spitzer IRAC channel 1 (3.6 $\mu$m) supermosaics from the Spitzer Heritage Archive\footnote{\url{https://irsa.ipac.caltech.edu//onlinehelp/heritage/\#about}}, with the data available at the Infrared Science Archive: \dataset[10.26131/IRSA361]{http://dx.doi.org/10.26131/IRSA361}. After masking galaxies, foreground stars, and noisy regions in these Spitzer data, we extracted \emph{H}-band and IRAC1 surface brightness profiles in the same direction towards the corner of the HST mosaic. Using overlapping measurements between $R\sim 20-70\arcsec$ (median of about 1$-$3 \re), we simultaneously determined both the average $H-\mathrm{IRAC1}$ color and the \emph{H}-band sky background. This radius range avoids PSF blurring effects, and the near-IR color gradients tend to be mild \citep[e.g.,][]{tamura03,iodice19} but increasingly blue with radius. The best-fit \emph{H}-band sky values ($\mu_{\mathrm{sky},H} \sim 20-21$ mag arcsec\pertwo) mostly agree with ETC values, and the median $H-\mathrm{IRAC1} \sim 2.20$ mag at these radii with a standard deviation of $\sim$0.16 mag is consistent with the color of an evolved, metal-rich single-burst stellar population \citep[with total metallicity $\lbrack \mathrm{M/H} \rbrack = 0.22$ with an age of 10 Gyr;][]{vazdekis12,vazdekis16}.

Finally, we removed the sky background from the smaller-FOV \emph{J}-band data. Except for a few cases where the ETG was sufficiently compact, we estimated \emph{J}-band sky values by scaling the measured \emph{H}-band levels by the ratio of $J/H$ zodiacal light from the ETC estimates. Slight adjustments were made to ensure smooth color gradients at the largest radii probed by the WFC3/IR (sub)array mosaics. We note that any uncertainty in the near-IR sky subtraction does not affect the stellar luminosity models described in Section~\ref{sec:mge}. Even if we change the $H$-band sky level by a factor of $\sim$2, the resulting circular velocity curves change by at most a few \kms, and overwhelmingly in the galaxy outskirts.

\begin{deluxetable*}{lcccccccrcccc}[!ht]
\tabletypesize{\footnotesize}
\tablecaption{CND Properties, Isophotal Analysis, and Spitzer Scaling Results}
\tablewidth{0pt}
\tablehead{\\[-3.5mm]
 & \multicolumn{3}{c}{CND Properties} & \multicolumn{7}{c}{Stellar Isophotal Results} & \multicolumn{2}{c}{Near-IR Colors} \\
\cmidrule(lr){2-4}\cmidrule(lr){5-11}\cmidrule(lr){12-13} 
 \\[-5.5mm]
\multicolumn{1}{c}{Galaxy} & \colhead{\textit{R}$_{\mathrm{dust},H}$} & \colhead{$(b,a)_{\mathrm{dust},H}$} & \colhead{PA$_{\mathrm{dust},H}$} & \colhead{$\overline{\mathrm{PA}}_{\star}$} & \colhead{$\Delta$PA$_{\star}$} & \colhead{$\overline{\varepsilon}_{\star}$} & \colhead{$\Delta \varepsilon_\star$} & \colhead{median} & \colhead{min,max} & \colhead{IC} & \colhead{$H-$IRAC1} & \colhead{$\nabla_{J-H}$} \\[-2.5mm]
\multicolumn{1}{c}{Name} & \colhead{(kpc)} & \colhead{(arcsec)} & \colhead{(deg.)} & \colhead{(deg.)} & \colhead{(deg.)} & \colhead{} & \colhead{} & \colhead{$a_4/a$} & \colhead{$a_4/a$} &  & \colhead{(mag)} & \colhead{(mag)} \\[-2mm]
\colhead{(1)} & \colhead{(2)} & \colhead{(3)} & \colhead{(4)} & \colhead{(5)} & \colhead{(6)} & \colhead{(7)} & \colhead{(8)} & \colhead{(9)} & \colhead{(10)} & \colhead{(11)} & \colhead{(12)} & \colhead{(13)}
}
\startdata
Hydra A & 2.03 & 0.27 , 1.96  & $-$75.3 & $-$36.0 & 41.2 & 0.126 & 0.304 & $-$0.006 & $-$0.101 , 0.003 & x0b & \nodata & $-$0.024 \\	
NGC \phantom{0}612 & 10.8 & 2.90 , 19.1 & $-$11.0 & $-$15.4 & 46.8 & 0.265 & 0.198 & 0.014 & $-$0.023 , 0.046 & xd? & 2.26 & $-$0.027 \\
NGC \phantom{0}997 & 2.32 & 4.69 , 5.70 & 32.6 & 29.7 & 11.0 & 0.121 & 0.034 & $-$0.006 & $-$0.043 , $-$0.002 & x0b & \nodata & $-$0.103 \\	
NGC 1332 & 0.24 & 0.17 , 2.17 & 114.9 & 116.7 & 3.4 & 0.321 & 0.445 & $-$0.007 & $-$0.014 , 0.005 & x0b & 2.12 & $-$0.031 \\
NGC 1387 & 0.88 & 8.35 , 9.43 & 52.1 & 108.7 & 54.5 & 0.146 & 0.296 & $-$0.001 & $-$0.015 , 0.006 & x0 & 2.22 & $-$0.024 \\
NGC 3245 & 0.16 & 0.59 , 1.60 & $-$6.9 & $-$3.4 & 19.7 & 0.367 & 0.367 & 0.001 & $-$0.091 , 0.024 & xd0? & 2.65 & $-$0.054 \\
NGC 3258 & 0.15 & 0.58 , 0.99 & 75.0 & 76.3 & 21.3 & 0.176 & 0.188 & $-$0.001 & $-$0.009 , 0.031 & x0d & 2.66 & $-$0.071 \\
NGC 3268 & 0.40 & 1.27 , 2.40 & $-$108.6 & $-$112.1 & 7.4 & 0.196 & 0.097 & 0.001 & $-$0.002 , 0.007 & x0 & 2.20 & $-$0.044 \\
NGC 3271 & 0.46 & 1.04 , 1.74 & $-$86.1 & $-$66.9 & 33.7 & 0.322 & 0.280 & $-$0.003 & $-$0.035 , 0.043 & xd0db & \nodata & $-$0.058 \\	
NGC 3557 & 0.22 & 0.62 , 0.99 & 36.2 & 33.4 & 7.7 & 0.245 & 0.122 & 0.002 & $-$0.006 , 0.011 & x0 & 2.20 & $-$0.025 \\
NGC 3862 & 0.38 & 0.80 , 0.84 & $-$9.0 & $-$16.0 & 62.7 & 0.022 & 0.344 & 0.002 & $-$0.019 , 0.015 & x0d0b & 2.27 & $-$0.046 \\
NGC 4061 & 0.92 & 0.93 , 1.81 & $-$6.2 & $-$5.6 & 28.1 & 0.184 & 0.084 & $-$0.006 & $-$0.076 , 0.027 & x0b & \nodata & $-$0.063 \\	
NGC 4261 & 0.13 & 0.51 , 0.89 & $-$16.4 & $-$22.3 & 11.1 & 0.220 & 0.137 & $-$0.002 & $-$0.014 , 0.004 & x0 & 2.20 & $-$0.024 \\
NGC 4373a & 0.95 & 2.17 , 6.00 & $-$26.0 & $-$32.4 & 9.7 & 0.428 & 0.404 & 0.003 & $-$0.010 , 0.030 & x0d0 & \nodata & $-$0.123 \\
NGC 4429 & 0.90 & 5.62 , 13.5 & 90.3 & 94.4 & 11.5 & 0.439 & 0.248 & 0.004 & $-$0.013 , 0.043 & x0db0 & 2.15 & $-$0.021 \\
NGC 4435 & 0.25 & 0.86 , 3.07 & 13.8 & 6.3 & 28.8 & 0.362 & 0.273 & 0.002 & $-$0.018 , 0.054 & x0bd0d & 2.13 & $-$0.042 \\
NGC 4697 & 0.20 & 0.95 , 3.48 & 65.3 & 66.0 & 2.3 & 0.436 & 0.142 & $-$0.003 & $-$0.010 , 0.003 & x0 & 2.13 & $-$0.034 \\
NGC 4751 & 1.54 & 3.22 , 13.7 & $-$5.1 & $-$4.9 & 5.2 & 0.587 & 0.102 & $-$0.003 & $-$0.031 , 0.008 & x0b & \nodata & $-$0.157 \\
NGC 4786 & 0.19 & 0.32 , 0.57 & $-$13.1 & $-$17.0 & 13.1 & 0.224 & 0.105 & 0.002 & $-$0.011 , 0.033 & x0d & 2.20 & $-$0.025 \\
NGC 4797 & 5.72 & 4.64 , 10.6 & 22.8 & 31.0 & 17.8 & 0.251 & 0.219 & 0.001 & $-$0.013 , 0.024 & x0d & \nodata & $-$0.069 \\
NGC 5084 & 0.13 & 0.44 , 1.18 & $-$2.0 & 82.8 & 5.7 & 0.388 & 0.449 & $-$0.002 & $-$0.016 , 0.006 & x0b? & 2.13 & $-$0.023 \\
NGC 5193 & 0.21 & 0.60 , 0.96 & 64.7 & 70.8 & 46.1 & 0.209 & 0.196 & 0.005 & $-$0.031 , 0.047 & x0d? & \nodata & $-$0.036 \\
NGC 5208 & 8.70 & 3.42 , 18.1 & $-$17.7 & $-$17.1 & 6.2 & 0.611 & 0.050 & $-$0.006 & $-$0.008 , 0.036 & x0? & \nodata & $-$0.083 \\
NGC 5838 & 0.45 & 1.70 , 4.15 & 36.8 & 47.4 & 11.9 & 0.243 & 0.464 & $-$0.004 & $-$0.023 , 0.007 & x0b0 & 2.18 & $-$0.080 \\
NGC 6861 & 1.01 & 1.92 , 7.60 & $-$37.9 & $-$38.0 & 25.7 & 0.458 & 0.224 & 0.002 & $-$0.002 , 0.038 & x0d & 2.19 & $-$0.078 \\
NGC 6958 & 0.25 & 1.15 , 1.29 & 105.5 & 109.7 & 43.7 & 0.125 & 0.225 & $-$0.001 & $-$0.024 , 0.050 & x0bd? & 2.25 & $-$0.059 \\
\enddata
   \tablecomments{General CND properties together with near-IR stellar light behavior. Cols.\ (2), (3), and (4) report the (physical) outer dust-disk radius, the (projected) minor and major axes, and the corresponding position angle, respectively, estimated primarily from the $H$-band data but with some input from bluer data. Cols.\ (5)  and (7) give the average stellar photometric PA and ellipticity, respectively, generally averaged from $R_{\mathrm{dust},H}$ to \re\ in most cases. Cols.\ (6) and (8) give the range of PA$_\star$ and $\varepsilon_\star$ values from stellar isophotal fitting just beyond $R_{\mathrm{dust},H}$ to near the edge of the $H$-band mosaic. Additional details are found in Section~\ref{subsec:isophote}. Cols.\ (9), (10), and (11) report the median and range of $a_4/a$ parameters over this same range, and the isophotal codes \citep{rest01} over the $H$-band surface brightness extent from smallest to largest radii; x, d, b, 0, and ? represents regions undetermined due to dust, disky, boxy, intermediate, and uncertain/inconsistent $a_4/a$ regions. To this table are added the best-fitting $H$--IRAC1 color term in col.\ (12) to scale together these surface brightness profiles. Col.\ (13) reports the logarithmic $J-H$ color gradient, measured from $R_{\mathrm{dust},H}$ to \re\ in most cases.}
   \label{tbl:ell_results}
\end{deluxetable*}

\section{Stellar Surface Brightness Behavior}
\label{sec:sb_behavior}

After WFC3/IR and archival data alignment and sky subtraction, we constructed color maps and extracted radial profiles to explore the CND dust and stellar behavior across the WFC3/IR FOV. Examples of central stellar surface brightness profiles in the available HST filters are shown in Figure~\ref{fig:filters_zoom_in}. To isolate the galaxy light, we masked out all other galaxies, foreground stars and diffraction spikes, detector artifacts, and pixels affected by cosmic rays.

\subsection{Near-IR Colors}
\label{subsec:colors}

HST near-IR color maps shown in Figure~\ref{fig:dustdisks} are expected to isolate the optically-thick dust distribution. As is shown in Figure~\ref{fig:filters_zoom_in}, near-IR colors are not always sensitive to optically-thin and occasionally filamentary features, which are not always detected in ALMA CO imaging \citep[see also][]{boizelle19}. From each $J-H$ map, we constructed an additional mask of the dust (see Figure~\ref{fig:contours3}) that nearly always contains the entire near side of the disk based on a color cutoff of $J-H \gtrsim 0.88$ mag [or an intrinsic color excess $\Delta (J-H) \equiv (J-H) - (J-H)_\star \gtrsim 0.08$ mag after subtracting off the stellar $(J-H)_\star$ that is evaluated just beyond the CND]. From these $J-H$ maps, we visually measured the semi-major and minor axes $a_{\mathrm{dust},H}$ and $b_{\mathrm{dust},H}$ of the optically-thick CND as well as the major-axis position angle PA$_{\mathrm{dust},H}$ (see Table~\ref{tbl:ell_results}). In most cases, the (physical) disk radius $R_{\mathrm{dust},H}$ is a small fraction of \re. In fact, in only five cases (NGC 612, NGC 1387, NGC 4751, NGC 4797, and NGC 5208) is $R_{\mathrm{dust},H} > \re/2$. We note that $R_{\mathrm{dust},H}$ is often 10$-$20\% smaller than $R_\mathrm{dust}$ measured in the $B$ band, but we retain the $H$-band value as it better traces high column density material.

Figure~\ref{fig:color_pa_eps} shows $J-H$ and $\Delta(J-H)$ color profiles extracted along the stellar major axes, with $\Delta(J-H)$ reaching typical 0.06$-$0.10 mag reddening along the major axis at $\sim$(0.4--0.9)$R_{\mathrm{dust},H}$. Adopting a standard Galactic extinction law ($R_V = 3.1$) and assuming that the CND lies in the midplane of each galaxy, the typical major-axis $\Delta(J-H)$ values above correspond to intrinsic $A_V \sim 1.5-2.5$ mag \citep[for details, see equations 1 and 2 from][]{boizelle19}. In some cases, the observed $\Delta(J-H)$ reaches the $\sim$0.15 mag turnover point corresponding to $A_V \sim 5$ mag \citep[see Figure 3 of][]{boizelle19}, suggesting the innermost regions of certain CNDs become optically thick even in the $H$-band. Along the major axis, the rough $\Delta(J-H) \gtrsim 0.08$ mag masking criterion corresponds to $A_H \gtrsim 0.35$ mag, or a drop of background stellar intensity of $\sim$25\%.

Beyond the CND, the stellar $J-H$ colors along the major axis show a gradual blueward trend with increasing radius. These (linear in $\log R$) trends are consistent with inside-out growth for (mostly) relaxed ETGs, with the bluer colors expected to arise from significant dry merger(s) \citep[e.g.,][]{saracco12,kim13}. To better compare these gradients to previous work, \cite[e.g.,][]{labarb10}, we adopt a logarithmic color gradient formalism

\begin{equation}
    \nabla_{J-H} \equiv \frac{\Delta(J-H)}{\Delta(\log_{10} R)}\,\,,
\end{equation}

\noindent where here $\Delta(J-H)$ refers to the difference in $J-H$ color over a large radial range. Inner and outer $R$ were set to slightly beyond $R_{\mathrm{dust},H}$ (to avoid residual dust contamination) and 2\re, respectively, in all cases except NGC 612, where we set the outer $R = 3.5\re$ because of the large $R_{\mathrm{dust},H} \approx 2\re$. The median $\nabla_{J-H} \sim -0.045$ mag and fairly tight standard deviation of 0.034 mag are consistent with established near-IR color gradients of local ETGs \citep[e.g.,][]{aaronson77,labarb10} and do not suggest steeper gradients for ETGs with dust features \citep[c.f.][]{kim13}. Our selection criteria avoided both lower-mass ETGs and those with more diffuse dust, which may explain the relatively tight $\nabla_{J-H}$ distribution.

In general, these CNDs are expected to be relatively thin and flat, and so the (outer) disk kinematic angle $i$ should satisfy $(b/a)_{\mathrm{dust},H} \approx \cos i$ \citep[e.g.,][]{barth16a,davis17}. We note that the observed axis ratio will tend to be more round than the intrinsic one due to beam smearing, especially for the smallest (projected) $a_{\mathrm{dust},H}$. As a result, the inferred $i$ may be systemically smaller than the true value. Estimating $i$ by kinemetric analysis of the observed CO velocity maps \citep{krajnovic06} is similarly fraught, as beam smearing tends to circularize the apparent kinematics except in cases where the outer CO extent $R_\mathrm{CO}$ is much larger than the synthesized beam and the disk is not viewed too edge on \citep[e.g.,][]{boizelle17}. This picture is further complicated by both photometric and kinematic evidence for disk warping. Color maps \textit{can} identify some disk warping signatures (e.g., NGC 612, NGC 3268, and NGC 4797 in Figure~\ref{fig:dustdisks}), although they cannot easily detect twists of $\Delta i \lesssim 10\degr$ or those within the inner couple of resolution elements. The prevalence and degree of disk warping has not yet been explored for a large sample of well-resolved CNDs. For four of our ETGs, however, \citet{boizelle17} find a shift in CO velocity line-of-nodes position angle (PA) of $\Delta\mathrm{PA} \sim 5-10\degr$ across the entire CNDs. \citet{boizelle19} find a central $\Delta i \sim 20\degr$ for NGC 3258 from gas-dynamical modeling, although for most of the disk area, $\Delta i$ is low and $i$ is consistent with the $(b/a)_{\mathrm{dust},H}$ estimate to within about 5\degr.

For the CNDs in our sample, we find a median $i\sim 65\degr$ estimated from $(b/a)_{\mathrm{dust},H}$, with individual values ranging from 18\degr\ to 86\degr. The apparent preference for higher $i$ in this sample likely stems from the greater ease of identifying more inclined disks in optical/near-IR imaging. Highly inclined disks do allow for more straightforward dust masking that removes fewer overall pixels and gives an essentially unobscured view along the minor axis. Despite the benefits when constructing stellar luminosity models, however, disks with $i \gtrsim 75\degr$ are susceptible to substantial modeling degeneracies \citep[e.g.,][]{barth16a,barth16b}. Similar difficulties arise for nearly face-in orientations \citep[$i \lesssim 15\degr$;][]{smith19}. As a result, more intermediate $i$ are preferred for the gas-dynamical approach.

\subsection{Isophotal Analysis}
\label{subsec:isophote}

To better compare to the stellar behavior of other ETGs and to quantify the discrepancies between dust disk and stellar alignment, we analyzed the $H$-band isophotal behavior using the \texttt{IRAF} \citep{tody86,tody93} \texttt{ellipse} task. This task returns the stellar intensity, PA$_\star$, ellipticity ($\varepsilon_\star = 1-b/a$), and deviation from a true ellipse ($a_4$/$a$) at logarithmically spaced radius intervals. Bad pixel maps were passed into the \texttt{IRAF} task using the primary + dust masks. In Figure~\ref{fig:color_pa_eps}, we plot changes in PA$_\star$ and $\varepsilon_\star$ respective to their values at \re. Individual radial PA$_\star$ and $\varepsilon_\star$ results, as well as the $a_4/a$ measurements, are provided in Appendix~\ref{app:sb_isophotal}.

In Table~\ref{tbl:ell_results}, we report the weighted average stellar photometric PA$_\star$, which was calculated as
\begin{equation}
    \overline{\mathrm{PA}}_\star = \frac{\sum w_i\, \mathrm{PA}_i}{\sum w_i}, \quad w_i = 1/\sigma_{\mathrm{PA,i}}^2\,,
    \label{eq:weightedavgpa}
\end{equation}
\noindent between $R_{\mathrm{dust},H}$ and \re. We followed an identical approach to determine the weighted average $\overline{\varepsilon}_{\star}$. The only exceptions are NGC 612 and NGC 4797, whose $R_{\mathrm{dust},H}$ are $\sim$2\re\ and \re, respectively, and for which the radial ranges were extended out to $3\re$ and $1.5\re$. We also report the degree of isophotal twisting ($\Delta\mathrm{PA}_{\star}$) using the method described by \citet{goullaud18}:
\begin{equation}
    \Delta \mathrm{PA}_{\star} = \left(\frac{\mathrm{PA}_{i-1}+\mathrm{PA}_{i}+\mathrm{PA}_{i+1}}{3}\right)-\left(\frac{\mathrm{PA}_{j-1}+\mathrm{PA}_{j}+\mathrm{PA}_{j+1}}{3}\right)\,.
    \label{eq:pa}
\end{equation}
\noindent Here, PA$_i$ and PA$_j$ are the maximum and minimum stellar PA values from just beyond $R_{\mathrm{dust},H}$ to near the edge of the $H$-band mosaic. The range in ellipticity ($\Delta\varepsilon_{\star}$) was computed in an identical manner. Some of the final \texttt{ellipse} results were not included in these $\Delta\mathrm{PA}_{\star}$ and $\Delta\varepsilon_{\star}$ ranges due to unusually large uncertainties or considerable discrepancies between neighboring points, most often near the edge of the $H$-band coverage.

In most cases, photometric twists are small, with over half our targets showing $\Delta\mathrm{PA}_\star <20\degr$ and only about a fifth reaching $\Delta\mathrm{PA}_\star >40\degr$. Figure~\ref{fig:color_pa_eps} separates the PA$_\star$ and $\varepsilon_\star$ responses for different morphological types, with lenticular and elliptical galaxy groups having fairly consistent PA$_\star$ from $R_{\mathrm{dust},H}$ to $\sim$2\re\ and \re, respectively. Some systems' PA$_\star$ and $\varepsilon_\star$ remain flat over the entire WFC3/IR FOV. However, there are some notable exceptions: NGC 3862 shows consistent stellar PA$_\star$ gradients (and increasingly flattened isophotes) with radius, both within and beyond \re; NGC 612 shows similar behavior that is likely due to recent accretion or a major merger \citep{emonts08,asabere16}; and Hydra A shows a consistent PA$_\star$ beyond $\sim$2$R_{\mathrm{dust},H}$ but a rapid $\Delta\mathrm{PA}_\star \sim 40
\degr$ over a radial extent of just 2 kpc. Overall, our sample shows photometric PA behavior similar to that seen in many of the ETGs in the MASSIVE survey \citep{goullaud18}, although the latter include a few cases with more extreme $\Delta\mathrm{PA}_\star$.

Unsurprisingly, the largest $\Delta\varepsilon_\star$ are found in lenticular galaxies, whose $\varepsilon_\star$ typically increases from the nucleus to $\sim$\re\ due to an increasingly dominant disky stellar component. Beyond $\sim$\re, the lenticular population trend bifurcates, with $\varepsilon_\star$ continuing to increase for a few galaxies while the majority show negative trends as the more circular stellar halo begins to dominate the isophotal behavior. Elliptical galaxies have more moderate $\Delta\varepsilon_\star$. To quantify the difference between morphological types, we computed the isophotal axis ratio $1-\varepsilon_\star$ over the entire radial range for each galaxy to measure the ratio of the max/min values. The median and scatter in the ratio of extremal $1-\varepsilon_\star$ values for elliptical galaxies is $1.19\pm 0.13$ while for lenticular galaxies we find a ratio of $1.57\pm 0.42$. The typical lenticular galaxy in our sample shows a factor of $\sim$3 greater change in the isophotal axis ratio than does the typical elliptical galaxy. The half of our sample that have mild isophotal twists ($\Delta\mathrm{PA}_\star < 20\degr$) but have $\Delta\varepsilon_\star \ge 0.2$ correspond to lenticular galaxies, with increasing $\varepsilon_\star$ following the transition from bulge to disk-dominated regions. In the remaining half with more substantial isophotal twists ($\Delta\mathrm{PA}_\star > 20\degr$), every case of high $\Delta \varepsilon_\star \geq 0.2$ is either a lenticular galaxy or, in the cases of NGC 3862 and NGC 6958, an elliptical galaxy that shows $\Delta\mathrm{PA}_\star \gtrsim 45\degr$, which is suggestive of recent merger activity or tidal disruption. In many cases, those lenticular galaxies with high $\Delta \varepsilon_\star$ also show elevated $\Delta\mathrm{PA}_\star$ values.

Most of our targets show $|\overline{\mathrm{PA}}_\star - \mathrm{PA}_{\mathrm{dust},H}| \lesssim 20
\degr$; this good agreement between stellar and CND photometric axes argues either for \textit{in situ} formation of the CNDs or sufficient elapsed time for gas to settle into the galaxy midplane \citep[of order $\sim$1 Gyr; e.g.,][]{tran01,lauer05,davis13b,voort15}. Despite general agreement, four ETGs show more extreme $|\overline{\mathrm{PA}}_\star - \mathrm{PA}_{\mathrm{dust},H}|$. As mentioned earlier, the Hydra A $\overline{\mathrm{PA}}_\star$ only matches $\mathrm{PA}_{\mathrm{dust},H}$ near $R_{\mathrm{dust},H}$; for NGC 1387 and NGC 3271, the stellar isophotes always appear to be misaligned; and, NGC 5084 appears to have a polar-oriented CND that is a strong candidate for external origin. Still other systems show good agreement out to $\sim$(1--3)\re\ followed by high $\mathrm{PA}_\star$ shifts beyond. This group includes four elliptical galaxies (NGC 3258, NGC 4061, NGC 4786, and NGC 6958), and three lenticular galaxies (NGC 612, NGC 4435, and NGC 6861).

In Table~\ref{tbl:ell_results}, we also include the median $a_4/a$ parameter along with its minimum and maximum values over the same radial ranges. Following \citet{rest01}, we computed isophotal codes (ICs) for radial bins to determine the prevalence of boxy ($a_4/a < -0.01$) and disky ($a_4/a > +0.01$) components through the stellar bulges and into the halos of these luminous ETGs. We find roughly equal numbers of ICs that show neither disky/boxy behavior out to the edge of the $H$-band FOV and ICs that show some preference for either disky or boxy isophotes. Nearly one in five systems show both disky and boxy behavior, almost always transitioning quickly between the two. Those that show the strongest disky or boxy isophotes (with $|a_4 /a| > 0.03$; NGC 612, NGC 3271, NGC 4373a, NGC 4429, and NGC 4435) are all lenticular galaxies with a very disky stellar component.

\subsection{Nuker Fits}
\label{subsec:nuker}

To characterize the core vs.\ cusp nature of our sample, we also fit the $H$-band stellar surface brightness profiles with a Nuker profile \citep{lauer95} of the form
\begin{equation}
    I(R) = 2^{(\beta-\gamma)/\alpha} I_\mathrm{b} \left(\frac{\rb}{R}\right)^\gamma \left[1+\left(\frac{R}{\rb}\right)^\alpha\right]^{(\gamma-\beta)/\alpha}\, .
\end{equation}
\noindent This functional form is well-suited to model the light distributions of massive elliptical galaxies, which typically follow log-linear behavior. The Nuker profile connects inner and outer power-law slopes $\gamma$ and $\beta$, respectively, with a transition sharpness $\alpha$ at the break radius \rb.

A 2D approach to fitting the stellar surface brightnesses is challenging due to the observed $\Delta\mathrm{PA}_\star$ and $\Delta\varepsilon_\star$. Instead, we optimized the Nuker parameters using the isophotal intensity curves shown in Appendix~\ref{app:sb_isophotal}. Upon inspection, the $H$-band $I(R)$ of several galaxies show more complicated behavior that would require additional components (or different parameterizations) for adequate global fits. Since we are primarily concerned with the circumnuclear stellar behavior, we restricted these Nuker fits to radial ranges that best matched a broken power law (generally out to $\sim$\re). We followed \citet{lauer95} in not fitting to the centermost data points (for the WFC3/IR data, points with $R\lesssim 0\farcs12$ were removed) due to potential PSF inaccuracies. We also required $\rb > 0\farcs2$ to ensure the solutions are robust against HST PSF effects and do not hinge on a singular central data point. To recover intrinsic $\gamma$ and \rb\ values from the optimization process, we convolved the model surface brightness profile with a 1D approximation of the $H$-band PSF by extracting the Tiny Tim response along the major axis.

For completeness, we also computed the intrinsic Nuker slope

\begin{equation}
    \gamma^\prime = -\frac{d\log I}{d\log R} \biggr\rvert _{R = R^{\prime}}
\end{equation}

\noindent at the resolution limit $R^{\prime}$ of the $H$-band data, i.e., at half the PSF FWHM ($R^{\prime} = 0\farcs12$). Since we have already restricted the Nuker fit to start at this projected distance, it is not surprising that few galaxies in this sample returned $\gamma^{\prime} - \gamma \geq 0.1$. Because of this close agreement in most cases, the statistics discussed below do not depend on the choice of $\gamma$ or $\gamma^{\prime}$. For a broader range of ETGs, however, $\gamma$ and $\gamma^{\prime}$ are more commonly discrepant \citep{lauer05}. Fits to bluer mosaics would allow for more central $\gamma^{\prime}$ measurements, but dust would further complicate the interpretation. 


\begin{deluxetable}{lcccccc}[!t]
\tabletypesize{\footnotesize}
\tablecaption{Nuker Fitting Parameters}
\tablewidth{0pt}
\tablehead{
\\[-5mm]
\colhead{Galaxy} & \colhead{$\mu_{\mathrm{b},H}$} & \colhead{\rb} & \colhead{$\alpha$} & \colhead{$\beta$} & \colhead{$\gamma$} & \colhead{$\gamma^{\prime}$} \\[-2.5mm]
\colhead{Name} & \colhead{(mag arcsec\pertwo)} & \colhead{(arcsec)} & \colhead{} & \colhead{} & \colhead{} & \colhead{} \\[-2mm]
\colhead{(1)} & \colhead{(2)} & \colhead{(3)} & \colhead{(4)} & \colhead{(5)} & \colhead{(6)} & \colhead{(7)}
}
\startdata
Hydra A & 16.28 & 1.16 & 3.73 & 1.31 & 0.04 & 0.04 \\	
NGC 612 & 15.72 & 4.12 & 1.66 & 2.71 & 0.27 & 0.28 \\
NGC 997 & 13.97 & 0.69 & 1.25 & 1.74 & 0.00 & 0.18 \\
NGC 1332 & 13.28 & 1.65 & 0.98 & 1.54 & 0.31 & 0.40 \\
NGC 1387 & 14.35 & 3.25 & 0.98 & 2.23 & 0.37 & 0.44 \\
NGC 3245 & 13.01 & 0.91 & 1.41 & 1.55 & 0.36 & 0.42 \\
NGC 3258 & 14.43 & 1.96 & 1.44 & 1.96 & 0.01 & 0.04 \\
NGC 3268 & 14.09 & 1.43 & 2.67 & 1.59 & 0.05 & 0.05 \\
NGC 3271 & 14.80 & 3.50 & 5.90 & 1.81 & 0.64 & 0.64 \\
NGC 3557 & 13.53 & 1.69 & 1.75 & 1.63 & 0.01 & 0.03 \\
NGC 3862 & 14.56 & 1.00 & 2.29 & 1.76 & 0.00 & 0.01 \\
NGC 4061 & 14.48 & 0.97 & 1.50 & 1.88 & 0.04 & 0.12 \\
NGC 4261 & 13.83 & 1.86 & 2.14 & 1.49 & 0.02 & 0.02 \\
NGC 4373a & 12.94 & 0.39 & 2.15 & 1.24 & 0.25 & 0.32 \\
NGC 4429 & 14.34 & 3.51 & \phantom{$^*$}10.00$^*$ & 1.26 & 0.64 & 0.64 \\
NGC 4435 & 13.63 & 1.76 & 1.11 & 1.66 & 0.23 & 0.30 \\
NGC 4697 & 13.63 & 2.40 & 0.97 & 1.47 & 0.40 & 0.46 \\
NGC 4751 & 12.46 & 0.55 & 2.32 & 1.28 & 0.45 & 0.47 \\
NGC 4786 & 13.58 & 0.65 & 2.99 & 1.24 & 0.15 & 0.16 \\
NGC 4797 & 13.55 & 0.24 & 3.01 & 1.20 & 0.27 & 0.37 \\
NGC 5084 & 12.90 & 1.04 & 2.21& 1.18 & 0.25 & 0.26 \\
NGC 5193 & 14.01 & 1.10 & 2.71 & 1.48 & 0.41 & 0.41 \\
NGC 5208 & 14.84 & 2.47 & \phantom{$^*$}10.00$^*$ & 1.35 & 0.70 & 0.70 \\
NGC 5838 & 14.34 & 3.01 & 1.15 & 1.82 & 0.64 & 0.67 \\
NGC 6861 & 15.07 & 6.64 & 0.73 & 2.94 & 0.13 & 0.27 \\
NGC 6958 & 12.79 & 0.63 & 1.06 & 1.78 & 0.22 & 0.45 \\
\enddata
   \tablecomments{Results of Nuker fits to the $H$-band isophotal surface brightness intensities, which account for telescope resolution effects by blurring the intrinsic Nuker function by the Tiny Tim PSF. The above \rb\ and $\gamma$ represent the intrinsic break radius and inner power-law slope, respectively. The slope $\gamma^{\prime}$ is the slope of the intrinsic Nuker function evaluated at the resolution limit of $\sim$0\farcs12. Cases denoted with $^*$ indicate $\alpha$ was fixed to avoid unphysical solutions.}
   \label{tbl:nuker_results}
\end{deluxetable}

\begin{figure*}[!t]
    \includegraphics[width=0.90\textwidth]{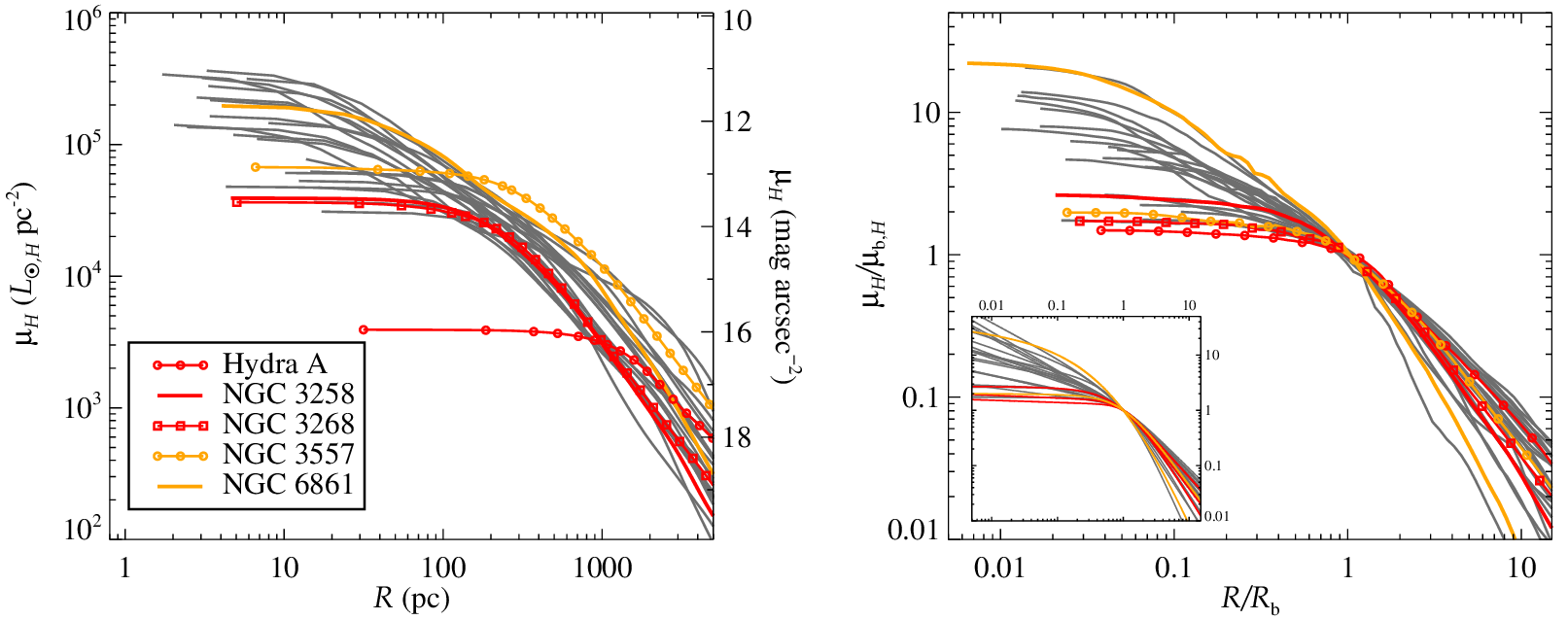}
    \centering
    \caption{\emph{H}-band surface brightness profiles for this ETG sample, showing best-fitting MGE models (\textit{left}) and the isophotal intensities (\textit{right}). The isophotal results are normalized to their respective break radii \rb\ and corresponding surface brightness $\mu_{\mathrm{b},H}$, with the (intrinsic) Nuker profiles shown for comparison (\textit{inset}). Highlighted BCGs/BGGs tend to show lower $\mu_{\mathrm{b},H}$, especially for the only BCG, Hydra A. NGC 6861 is the only exception, largely due to its slightly more cuspy profile.}
    \label{fig:sb+nuker}
\end{figure*}

The best-fitting Nuker parameters in Table~\ref{tbl:nuker_results} show the outer slopes to be fairly steep, with a median and standard deviation $\beta \sim 1.6 \pm 0.4$, although a few cases (especially NGC 612 and NGC 6861) prefer much steeper outer slopes. In every case, we see a distinct break between the inner and outer power-law slopes. Over half our targets have a shallow inner power-law slopes with $\gamma \le 0.3$, often used to identify out centrally cored galaxies \citep{faber97}. Unsurprisingly for the typically higher stellar masses in our sample, only a few systems show inner power-law behavior of $\gamma \ge 0.5$. The remainder (nearly a quarter) of the sample are ``intermediate'' cases with $0.3< \gamma < 0.5$ \citep[for additional examples, see][]{rest01,ravindranath01}. When split into elliptical and lenticular subclassifications, we find distinct medians $\gamma = 0.04$ and 0.31, respectively, but similar overall scatters. This behavior is easily seen in the surface brightness plots of Figure~\ref{fig:sb+nuker}, especially after scaling the isophotal intensities by $\mu_{\mathrm{b},H}$ and \rb. A restricted analysis of the most cored ETGs (with $\gamma < 0.2$) gives a median $M_K \sim -25.29$ mag that is $\sim$60\% more luminous than for the entire sample. These most cored ETGs also show a closer correspondence between the break radius and CND extent, with a median $\rb/a_{\mathrm{dust},H} \approx 1$, although individual ratios range from 0.1$-$2.1.

\section{Stellar Luminosity Models}
\label{sec:mge}

We modeled the sky-subtracted $H$-band surface brightnesses of our galaxies using the multi-Gaussian expansion formalism \citep[MGE;][]{emsell94}, which is convenient for analytical purposes and has been shown to accurately reproduce the stellar profiles of ETGs \citep[see also][]{capp02}. This series expansion recovers the peak surface brightness (in $L_\odot$ pc\pertwo\ units) of each Gaussian component, together with projected dispersions $\sigma^\prime{}$ (in arcsec) and axis ratios $q^\prime{}$. In addition to the benefits of speed and usability of MGEs, this approach also enables simple and efficient convolution with a PSF model to account for blurring effects. In this project, we have employed a 2D concentric MGE and focus on fits that keep the PAs of all Gaussian components tied together during optimization. The ensuing deprojection using the inclination angle $i$ \citep[assuming oblate axisymmetry to determine intrinsic $\sigma$ and $q$;][]{capp02} then results in an intrinsic stellar luminosity density profile. We note that non-parametric approaches allow for more careful deprojection analysis \citep[e.g., reconstruction of intrinsic densities and comparison of the relative likelihood of different deprojections;][]{denicola20}. Since the ALMA CO kinematics do not typically probe beyond $\sim1$ kpc and rarely close to \re\ \citep{boizelle17,sansom19,ruffa19a}, however, uncertainties in the deprojection and the large-scale stellar figure are not expected to be of significant concern for gas-dynamical modeling efforts.

\begin{deluxetable*}{cccc@{\hspace{5mm}}ccc@{\hspace{5mm}}ccc@{\hspace{5mm}}ccc}[!ht]
\tabletypesize{\scriptsize}
\tablecaption{MGE Parameters (with a Uniform PA)}
\tablewidth{0pt}
\tablehead{\\[-5mm]
\colhead{\emph{j}} & \colhead{$\log_{10}~I_{H,j}$} & \colhead{$\sigma^{\prime}_{j}$} & \colhead{\hspace{-3mm}$q^{\prime}_{j}$} & \colhead{$\log_{10}~I_{H,j}$} & \colhead{$\sigma^{\prime}_{j}$} & \colhead{\hspace{-3mm}$q^{\prime}_{j}$} & \colhead{$\log_{10}~I_{H,j}$} & \colhead{$\sigma^{\prime}_{j}$} & \colhead{\hspace{-3mm}$q^{\prime}_{j}$} & \colhead{$\log_{10}~I_{H,j}$} & \colhead{$\sigma^{\prime}_{j}$} & \colhead{\hspace{-3mm}$q^{\prime}_{j}$}\\[-2.5mm]
\colhead{} & \colhead{$(L_{\sun}~\rm{pc}^{-2})$} & \colhead{(arcsec)} & \colhead{} & \colhead{$(L_{\sun}~\rm{pc}^{-2})$} & \colhead{(arcsec)} & \colhead{} & \colhead{$(L_{\sun}~\rm{pc}^{-2})$} & \colhead{(arcsec)} & \colhead{} & \colhead{$(L_{\sun}~\rm{pc}^{-2})$} & \colhead{(arcsec)}\\[-2mm]
\colhead{(1)} & \colhead{(2)} & \colhead{(3)} & \colhead{\hspace{-3mm}(4)} & \colhead{(2)} & \colhead{(3)} & \colhead{\hspace{-3mm}(4)} & \colhead{(2)} & \colhead{(3)} & \colhead{\hspace{-3mm}(4)} & \colhead{(2)} & \colhead{(3)} & \colhead{\hspace{-3mm}(4)}
}
\startdata
& \multicolumn{2}{l}{$\mathrm{PA}=-41.2$} & $q^{\prime}_{min} = 0.23$ & \multicolumn{2}{l}{$\mathrm{PA}=-6.3$} & $q^{\prime}_{min} = 0.24$ & \multicolumn{2}{l}{$\mathrm{PA}=32.3$} & $q^{\prime}_{min} = 0.91$ & 
\multicolumn{2}{l}{$\mathrm{PA}=-63.5$} & $q^{\prime}_{min} = 0.17$\\[-2mm]
& \multicolumn{3}{c}{\hspace{-4mm}\textbf{Hydra A}} & \multicolumn{3}{c}{\hspace{-4mm}\textbf{NGC 612}} & \multicolumn{3}{c}{\hspace{-4mm}\textbf{NGC 997}} & \multicolumn{3}{c}{\hspace{-1mm}\textbf{NGC 1332}}\\
\cline{2-4} \cline{5-7} \cline{8-10} \cline{11-13}
1 & 3.4550 & 1.1649 & 0.9952 & 4.2391 & 0.1104 & 1.0000 & 4.5670 & 0.2489 & 0.9100 & 5.7035 & 0.1397 & 0.3338 \\
2 & 2.9155 & 2.9399 & 0.9060 & 4.1763 & 0.3079 & 0.7770 & 4.2751 & 0.7090 & 0.9118 & 4.2340 & 0.2494 & 0.6765 \\
3 & 2.1662 & 6.4848 & 0.9790 & 3.8616 & 0.5813 & 1.0000 & 3.9212 & 1.4545 & 0.9387 & 4.9782 & 0.4911 & 0.9816 \\
4 & 2.0350 & 7.2705 & 0.6890 & 3.4430 & 0.9568 & 1.0000 & 3.4301 & 3.1261 & 0.9100 & 4.6117 & 1.5227 & 0.7291 \\
5 & 2.1566 & 13.133 & 0.7992 & 3.3977 & 2.3868 & 0.6227 & 2.9587 & 6.5902 & 0.9100 & 4.2075 & 3.4156 & 0.7281 \\
6 & 1.4121 & 26.732 & 0.6940 & 3.4692 & 3.8908 & 0.2553 & 2.4483 & 13.696 & 0.9100 & 3.8332 & 7.3989 & 0.7688 \\
7 & 0.9438 & 58.893 & 0.6676 & 3.2055 & 5.0384 & 0.5201 & 1.8333 & 33.844 & 0.9260 & 3.1562 & 17.528 & 0.3178 \\
8 & \nodata & \nodata & \nodata & 2.9094 & 4.9285 & 0.9815 & \nodata & \nodata & \nodata & 3.0689 & 32.944 & 0.2910 \\
9 & \nodata & \nodata & \nodata & 3.3356 & 4.1166 & 0.2400 & \nodata & \nodata & \nodata & 2.5357 & 52.074 & 0.3259 \\
10 & \nodata & \nodata & \nodata & 2.3213 & 12.502 & 0.6161 & \nodata & \nodata & \nodata & 1.8995 & 78.066 & 0.4190 \\
11 & \nodata & \nodata & \nodata & 2.1723 & 16.296 & 0.9293 & \nodata & \nodata & \nodata & 0.9968 & 165.32 & 0.9901 \\
12 & \nodata & \nodata & \nodata & 1.3514 & 40.481 & 1.0000 & \nodata & \nodata & \nodata & \nodata & \nodata & \nodata \\
13 & \nodata & \nodata & \nodata & 0.2425 & 72.318 & 0.9973 & \nodata & \nodata & \nodata & \nodata & \nodata & \nodata \\
\hline
& \multicolumn{2}{l}{$\mathrm{PA}=-73.2$} & $q^{\prime}_{min} = 0.95$ & \multicolumn{2}{l}{$\mathrm{PA}=-4.3$} & $q^{\prime}_{min} = 0.52$ & \multicolumn{2}{l}{$\mathrm{PA}=74.6$} & $q^{\prime}_{min} = 0.72$ & \multicolumn{2}{l}{$\mathrm{PA}=69.7$} & $q^{\prime}_{min} = 0.67$ \\[-2mm]
\multicolumn{1}{c}{} & \multicolumn{3}{c}{\hspace{-4mm}\textbf{NGC 1387}} & \multicolumn{3}{c}{\hspace{-4mm}\textbf{NGC 3245}} & \multicolumn{3}{c}{\hspace{-4mm}\textbf{NGC 3258}} & \multicolumn{3}{c}{\hspace{-1mm}\textbf{NGC 3268}}\\
\cline{2-4} \cline{5-7} \cline{8-10} \cline{11-13}
1 & 5.3554 & 0.1517 & 0.9973 & 5.5227 & 0.1388 & 0.6686 & 4.1386 & 0.7662 & 0.9671 & 3.6153 & 0.2802 & 0.6700 \\
2 & 4.8519 & 0.5218 & 0.9994 & 4.2977 & 0.2578 & 0.5378 & 4.0292 & 1.1586 & 0.7200 & 3.9109 & 0.9996 & 0.9918 \\
3 & 4.5377 & 1.6441 & 0.9999 & 4.7789 & 0.3163 & 0.5205 & 3.9129 & 2.0040 & 0.7633 & 3.9752 & 1.0794 & 0.7425 \\
4 & 4.0769 & 4.0668 & 0.9500 & 4.7368 & 0.3776 & 0.9647 & 3.6336 & 2.9726 & 0.8064 & 3.8462 & 1.9852 & 0.7188 \\
5 & 3.4977 & 7.1948 & 0.9500 & 4.6770 & 0.6348 & 0.8459 & 3.5087 & 4.8344 & 0.8513 & 3.7069 & 2.2767 & 0.8903 \\
6 & 2.8663 & 14.983 & 0.9500 & 4.5074 & 1.1201 & 0.9631 & 2.6360 & 8.9900 & 0.8158 & 3.4637 & 3.8058 & 0.7805 \\
7 & 2.4841 & 40.343 & 0.9624 & 4.0375 & 2.5263 & 0.6386 & 2.9576 & 11.511 & 0.9348 & 3.2626 & 6.3293 & 0.8090 \\
8 & \nodata & \nodata & \nodata & 3.8275 & 3.2671 & 0.7809 & 2.2148 & 22.281 & 0.9800 & 2.8091 & 12.075 & 0.7841 \\
9 & \nodata & \nodata & \nodata & 3.5080 & 8.9091 & 0.5200 & 2.0012 & 47.787 & 0.7872 & 2.4624 & 21.504 & 0.8048 \\
10 & \nodata & \nodata & \nodata & 2.9232 & 26.569 & 0.5200 & \nodata & \nodata & \nodata & 2.0247 & 51.639 & 0.7253 \\
11 & \nodata & \nodata & \nodata & 1.7579 & 44.971 & 0.5681 & \nodata & \nodata & \nodata & 1.5208 & 87.993 & 0.9289 \\
12 & \nodata & \nodata & \nodata & 1.5658 & 52.299 & 0.6091 & \nodata & \nodata & \nodata & \nodata & \nodata & \nodata \\
13 & \nodata & \nodata & \nodata & 1.1653 & 162.32 & 0.7948 & \nodata & \nodata & \nodata & \nodata & \nodata & \nodata \\
\hline
& \multicolumn{2}{l}{$\mathrm{PA}=-67.9$} & $q^{\prime}_{min} = 0.73$ & \multicolumn{2}{l}{$\mathrm{PA}=33.5$} & $q^{\prime}_{min} = 0.75$ & \multicolumn{2}{l}{$\mathrm{PA}=20.1$} & $q^{\prime}_{min} = 0.99$ & \multicolumn{2}{l}{$\mathrm{PA}=-4.5$} & $q^{\prime}_{min} = 0.66$\\[-2mm]
\multicolumn{1}{c}{}  & \multicolumn{3}{c}{\hspace{-4mm}\textbf{NGC 3271}} & \multicolumn{3}{c}{\hspace{-2.5mm}\textbf{NGC 3557$^*$}} & \multicolumn{3}{c}{\hspace{-2.5mm}\textbf{NGC 3862$^*$}} & \multicolumn{3}{c}{\hspace{-1mm}\textbf{NGC 4061}}\\
\cline{2-4} \cline{5-7} \cline{8-10} \cline{11-13}
1 & 4.9009 & 0.1000 & 0.7376 & 4.2304 & 0.8510 & 0.9866 & 4.1542 & 0.7628 & 0.9900 & 4.1285 & 0.1755 & 0.6600 \\
2 & 5.1274 & 0.1299 & 0.7300 & 4.3214 & 1.2778 & 0.7500 & 3.7966 & 1.4370 & 0.9940 & 4.2827 & 0.5379 & 0.6600 \\
3 & 4.5963 & 0.3489 & 0.8362 & 3.9536 & 2.0599 & 0.7500 & 3.4906 & 2.9353 & 0.9910 & 3.7328 & 1.0829 & 0.9533 \\
4 & 4.2135 & 0.9666 & 0.7307 & 4.0190 & 2.9514 & 0.7500 & 2.7847 & 7.0954 & 0.9933 & 3.6122 & 1.2839 & 0.6600 \\
5 & 4.0445 & 2.4260 & 0.7300 & 3.7178 & 4.5401 & 0.7613 & 2.2149 & 18.472 & 0.9900 & 3.5582 & 2.1409 & 0.8016 \\
6 & 3.7112 & 3.8030 & 0.7300 & 3.4740 & 6.8297 & 0.7535 & 1.7046 & 58.754 & 0.9900 & 3.1855 & 3.7517 & 0.8240 \\
7 & 3.3835 & 6.9061 & 0.7300 & 2.8595 & 11.632 & 0.7500 & \multicolumn{2}{c}{\hspace{-2.5mm}PSF mag = 15.86} & & 2.4901 & 6.1372 & 0.9569 \\
8 & 2.8441 & 18.027 & 0.7300 & 3.0156 & 13.000 & 0.7500 & \nodata & \nodata & \nodata & 2.2961 & 8.1515 & 0.6600 \\
9 & 1.9464 & 26.205 & 0.7300 & 2.6283 & 22.466 & 0.7500 & \nodata & \nodata & \nodata & 2.2072 & 12.643 & 0.8367 \\
10 & 1.8515 & 52.723 & 0.7300 & 2.5442 & 36.827 & 0.7500 & \nodata & \nodata & \nodata & 1.5922 & 20.997 & 0.7836 \\
11 & 1.2997 & 74.622 & 0.7300 & 2.0470 & 77.856 & 0.8731 & \nodata & \nodata & \nodata & 1.5933 & 34.329 & 0.6985 \\
12 & \nodata & \nodata & \nodata & \multicolumn{2}{c}{\hspace{-2.5mm}PSF mag = 18.08} & & \nodata & \nodata & \nodata & 1.0584 & 85.239 & 0.7238 \\
\hline
& \multicolumn{2}{l}{$\mathrm{PA}=-22.4$} & $q^{\prime}_{min} = 0.71$ & \multicolumn{2}{l}{$\mathrm{PA}=-30.8$} & $q^{\prime}_{min} = 0.52$ & \multicolumn{2}{l}{$\mathrm{PA}=-87.0$} & $q^{\prime}_{min} = 0.57$ & 
\multicolumn{2}{l}{$\mathrm{PA}=3.3$} & $q^{\prime}_{min} = 0.41$ \\[-2mm]
\multicolumn{1}{c}{} & \multicolumn{3}{c}{\hspace{-2.5mm}\textbf{NGC 4261$^*$}} & \multicolumn{3}{c}{\hspace{-4mm}\textbf{NGC 4373a}} & \multicolumn{3}{c}{\hspace{-2.5mm}\textbf{NGC 4429$^*$}} & \multicolumn{3}{c}{\hspace{-1mm}\textbf{NGC 4435}} \\
\cline{2-4} \cline{5-7} \cline{8-10} \cline{11-13}
1 & 4.3308 & 1.1185 & 0.8196 & 5.1773 & 0.1233 & 0.5200 & 5.3069 & 0.1099 & 0.8506 & 5.1425 & 0.1893 & 0.4802 \\
2 & 4.0992 & 2.1932 & 0.7100 & 4.8418 & 0.3496 & 0.5200 & 4.8497 & 0.3361 & 0.5700 & 4.6328 & 0.5244 & 0.7973 \\
3 & 3.9663 & 3.7243 & 0.7296 & 4.2347 & 1.0068 & 0.5751 & 4.2484 & 0.7858 & 0.7150 & 4.3501 & 1.2024 & 0.7934 \\
4 & 3.0013 & 6.4422 & 0.7197 & 3.9133 & 2.1370 & 0.5750 & 4.2296 & 2.4506 & 0.5700 & 4.2373 & 2.3678 & 0.6816 \\
5 & 3.3153 & 8.3663 & 0.7143 & 3.4648 & 4.7480 & 0.7467 & 3.5629 & 2.8757 & 0.8448 & 3.9737 & 4.4603 & 0.7654 \\
6 & 3.1800 & 12.611 & 0.8225 & 3.0984 & 12.208 & 0.5200 & 3.8536 & 5.5932 & 0.6529 & 3.2103 & 11.771 & 0.4306 \\
7 & 2.7880 & 19.928 & 0.8384 & 2.2347 & 29.124 & 0.5200 & 3.1773 & 11.419 & 0.5802 & 3.0833 & 16.253 & 0.4100 \\
8 & 1.7008 & 40.664 & 0.8175 & 1.7128 & 55.969 & 0.5200 & 3.2655 & 16.358 & 0.5700 & 2.5444 & 17.346 & 0.9515 \\
9 & 2.3330 & 45.887 & 0.8349 & 0.5813 & 131.04 & 0.9992 & 2.8525 & 48.696 & 0.5700 & 2.4697 & 31.458 & 0.5049 \\
10 & 1.7869 & 94.857 & 0.9587 & \nodata & \nodata & \nodata & 1.9408 & 118.45 & 0.5700 & 1.8732 & 49.121 & 0.9890 \\
11 & \multicolumn{2}{c}{\hspace{-1mm}PSF mag = 20.07} & & \nodata & \nodata & \nodata & \multicolumn{2}{c}{\hspace{-2.5mm}PSF mag = 16.21} & & 1.3577 & 101.44 & 0.5399 \\
\enddata
   \tablecomments{Individual Gaussian components (indexed by $j$) from the best-fitting MGE for each galaxy in this \emph{H}-band sample, after masking out neighboring galaxies, foreground stars, and the most dust-obscured regions of the CND. Projected terms are indicated by a $^\prime{}$. During these fits, the individual $q^\prime$ values were constrained to be equal to or greater than the limit $q^{\prime{}}_\mathrm{min}$ (listed for each galaxy), which ensured the solution could be deprojected for a range of inclination angles. Inner $\sigma^{\prime}$ were constrained to $\gtrsim$0\farcs1 to avoid overly compact stellar distributions. For four of these ETGs (indicated with a $^*$), we included an unresolved point source in the modeling process to remove contamination from known, prominent AGN in the \emph{H} band. The magnitudes of these PSF components are given in such cases.}
   \label{tbl:MGE_params}
\end{deluxetable*}

\setcounter{table}{4}
\begin{deluxetable*}{cccc@{\hspace{5mm}}ccc@{\hspace{5mm}}ccc@{\hspace{5mm}}ccc}[!ht]
\tabletypesize{\scriptsize}
\tablecaption{MGE Parameters (with a Uniform PA; cont.)}
\tablewidth{0pt}
\tablehead{\\[-5mm]
\colhead{\emph{j}} & \colhead{$\log_{10}~I_{H,j}$} & \colhead{$\sigma^{\prime}_{j}$} & \colhead{\hspace{-3mm}$q^{\prime}_{j}$} & \colhead{$\log_{10}~I_{H,j}$} & \colhead{$\sigma^{\prime}_{j}$} & \colhead{\hspace{-3mm}$q^{\prime}_{j}$} & \colhead{$\log_{10}~I_{H,j}$} & \colhead{$\sigma^{\prime}_{j}$} & \colhead{\hspace{-3mm}$q^{\prime}_{j}$} & \colhead{$\log_{10}~I_{H,j}$} & \colhead{$\sigma^{\prime}_{j}$} & \colhead{\hspace{-3mm}$q^{\prime}_{j}$}\\[-2.5mm]
\colhead{} & \colhead{$(L_{\sun}~\rm{pc}^{-2})$} & \colhead{(arcsec)} & \colhead{} & \colhead{$(L_{\sun}~\rm{pc}^{-2})$} & \colhead{(arcsec)} & \colhead{} & \colhead{$(L_{\sun}~\rm{pc}^{-2})$} & \colhead{(arcsec)} & \colhead{} & \colhead{$(L_{\sun}~\rm{pc}^{-2})$} & \colhead{(arcsec)}\\[-2mm]
\colhead{(1)} & \colhead{(2)} & \colhead{(3)} & \colhead{\hspace{-3mm}(4)} & \colhead{(2)} & \colhead{(3)} & \colhead{\hspace{-3mm}(4)} & \colhead{(2)} & \colhead{(3)} & \colhead{\hspace{-3mm}(4)} & \colhead{(2)} & \colhead{(3)} & \colhead{\hspace{-3mm}(4)}
}
\startdata
& \multicolumn{2}{l}{$\mathrm{PA}=66.2$} & $q^{\prime}_{min} = 0.40$ & \multicolumn{2}{l}{$\mathrm{PA}=-5.0$} & $q^{\prime}_{min} = 0.35$ & \multicolumn{2}{l}{$\mathrm{PA}=-16.8$} & $q^{\prime}_{min} = 0.69$ & \multicolumn{2}{l}{$\mathrm{PA}=26.3$} & $q^{\prime}_{min} = 0.59$ \\[-2mm]
\multicolumn{1}{c}{} & \multicolumn{3}{c}{\hspace{-4mm}\textbf{NGC 4697}} & \multicolumn{3}{c}{\hspace{-4mm}\textbf{NGC 4751}} & \multicolumn{3}{c}{\hspace{-4mm}\textbf{NGC 4786}} & \multicolumn{3}{c}{\hspace{-1mm}\textbf{NGC 4797}} \\
\cline{2-4} \cline{5-7} \cline{8-10} \cline{11-13}
1 & 5.5120 & 0.1000 & 0.7453 & 6.0202 & 0.1000 & 0.3500 & 4.3138 & 0.3212 & 0.9949 & 4.9958 & 0.1333 & 0.5900 \\
2 & 5.4715 & 0.1236 & 0.6153 & 5.5755 & 0.2453 & 0.3500 & 4.4351 & 0.5693 & 0.7301 & 4.2163 & 0.4415 & 0.7869 \\
3 & 4.8797 & 0.4738 & 0.7300 & 4.4010 & 0.5827 & 0.9960 & 4.1774 & 1.2630 & 0.8189 & 3.7328 & 0.8758 & 0.9757 \\
4 & 4.5620 & 1.0639 & 0.7217 & 4.4902 & 0.9176 & 0.6279 & 3.5661 & 2.7314 & 0.7362 & 3.3653 & 1.8072 & 0.9995 \\
5 & 3.7315 & 2.2062 & 0.9995 & 4.2328 & 1.8856 & 0.6385 & 3.4393 & 4.7352 & 0.8107 & 2.3105 & 3.5312 & 0.9926 \\
6 & 4.2744 & 2.5638 & 0.4424 & 3.8944 & 4.2760 & 0.4700 & 2.5924 & 5.7134 & 0.8857 & 2.6645 & 4.2037 & 0.9986 \\
7 & 3.8680 & 5.3112 & 0.4314 & 3.3600 & 8.4109 & 0.4177 & 2.5916 & 7.8986 & 0.6900 & 2.8346 & 7.7950 & 0.7230 \\
8 & 3.5558 & 5.7133 & 0.6784 & 3.0034 & 15.594 & 0.4118 & 2.6872 & 12.792 & 0.6900 & 2.2715 & 15.781 & 0.6797 \\
9 & 3.5033 & 9.6667 & 0.6891 & 2.5027 & 34.328 & 0.3974 & 2.3202 & 14.940 & 0.8763 & 1.6517 & 27.516 & 0.9771 \\
10 & 2.9935 & 13.712 & 0.4000 & 1.7134 & 73.467 & 0.5006 & 2.1339 & 23.646 & 0.6900 & \nodata & \nodata & \nodata \\
11 & 2.9662 & 22.580 & 0.5165 & 0.7030 & 146.91 & 0.9050 & 1.7691 & 27.513 & 0.9666 & \nodata & \nodata & \nodata \\
12 & 2.7202 & 24.715 & 0.4000 & \nodata & \nodata & \nodata & 0.5270 & 45.221 & 0.6900 & \nodata & \nodata & \nodata \\
13 & 2.8995 & 34.303 & 0.5967 & \nodata & \nodata & \nodata & 1.3639 & 59.220 & 0.6900 & \nodata & \nodata & \nodata \\
14 & 2.5434 & 55.874 & 0.6750 & \nodata & \nodata & \nodata & 1.0990 & 113.70 & 0.9629 & \nodata & \nodata & \nodata \\
15 & 2.0244 & 124.23 & 0.7377 & \nodata & \nodata & \nodata & \nodata & \nodata & \nodata & \nodata & \nodata & \nodata\\
\hline
& \multicolumn{2}{l}{$\mathrm{PA}=82.3$} & $q^{\prime}_{min} = 0.53$ & \multicolumn{2}{l}{$\mathrm{PA}=71.6$} & $q^{\prime}_{min} = 0.75$ & \multicolumn{2}{l}{$\mathrm{PA}=-16.0$} & $q^{\prime}_{min} = 0.31$ & \multicolumn{2}{l}{$\mathrm{PA}=51.0$} & $q^{\prime}_{min} = 0.56$ \\[-2mm]
\multicolumn{1}{c}{} & \multicolumn{3}{c}{\hspace{-4mm}\textbf{NGC 5084}} & \multicolumn{3}{c}{\hspace{-4mm}\textbf{NGC 5193}} & \multicolumn{3}{c}{\hspace{-4mm}\textbf{NGC 5208}} & \multicolumn{3}{c}{\hspace{-1mm}\textbf{NGC 5838}}\\
\cline{2-4} \cline{5-7} \cline{8-10} \cline{11-13}
1 & 5.1227 & 0.1000 & 0.7858 & 5.2662 & 0.1000 & 0.7500 & 4.7950 & 0.1000 & 0.3100 & 5.5656 & 0.1000 & 0.8742 \\
2 & 4.6764 & 0.4351 & 0.8559 & 4.3592 & 0.4554 & 0.7500 & 4.8307 & 0.2316 & 0.6493 & 5.3226 & 0.1947 & 0.5600 \\
3 & 4.5923 & 0.8891 & 0.8820 & 4.3354 & 0.9520 & 0.7521 & 4.1829 & 0.5332 & 0.6768 & 4.9145 & 0.4656 & 0.9589 \\
4 & 4.4636 & 1.9107 & 0.7429 & 3.9680 & 2.0043 & 0.8089 & 3.8734 & 1.2505 & 0.5766 & 4.4015 & 1.1427 & 0.7839 \\
5 & 4.0177 & 4.6884 & 0.5300 & 3.4391 & 4.6440 & 0.7500 & 3.8380 & 1.6389 & 0.3100 & 4.2892 & 2.1774 & 0.7211 \\
6 & 3.6368 & 8.7730 & 0.5300 & 3.0292 & 9.7925 & 0.8458 & 3.5770 & 2.3291 & 0.5421 & 3.8977 & 4.4992 & 0.8763 \\
7 & 3.1031 & 19.560 & 0.5300 & 2.5007 & 18.614 & 0.9834 & 3.5117 & 5.6546 & 0.3100 & 3.3867 & 9.6607 & 0.5895 \\
8 & 2.4936 & 36.864 & 0.5300 & 1.6639 & 46.271 & 0.9489 & 3.1005 & 6.0425 & 0.5237 & 2.8370 & 16.244 & 0.5721 \\
9 & 2.3514 & 67.627 & 0.5300 & \nodata & \nodata & \nodata & 2.3435 & 11.394 & 0.3100 & 2.6329 & 40.231 & 0.5600 \\
10 & \nodata & \nodata & \nodata & \nodata & \nodata & \nodata & 2.4322 & 14.318 & 0.3100 & 1.0379 & 77.003 & 0.5600 \\
11 & \nodata & \nodata & \nodata & \nodata & \nodata & \nodata & 2.3642 & 16.032 & 0.4188 & \nodata & \nodata & \nodata \\
12 & \nodata & \nodata & \nodata & \nodata & \nodata & \nodata & 1.6524 & 5.3184 & 0.3100 & \nodata & \nodata & \nodata \\
13 & \nodata & \nodata & \nodata & \nodata & \nodata & \nodata & 1.4263 & 31.042 & 0.3904 & \nodata & \nodata & \nodata \\
14 & \nodata & \nodata & \nodata & \nodata & \nodata & \nodata & 1.6982 & 33.695 & 0.4090 & \nodata & \nodata & \nodata \\
15 & \nodata & \nodata & \nodata & \nodata & \nodata & \nodata & 0.6420 & 95.995 & 0.8543 & \nodata & \nodata & \nodata \\
16 & \nodata & \nodata & \nodata & \nodata & \nodata & \nodata & 0.6480 & 155.22 & 0.9660& \nodata & \nodata & \nodata \\
\hline
& \multicolumn{2}{l}{$\mathrm{PA}=-37.8$} & $q^{\prime}_{min} = 0.38$ & \multicolumn{2}{l}{$\mathrm{PA}=-70.0$} & $q^{\prime}_{min} = 0.95$ & \\[-2mm]
\multicolumn{1}{c}{} & \multicolumn{3}{c}{\hspace{-4mm}\textbf{NGC 6861}} & \multicolumn{3}{c}{\hspace{-4mm}\textbf{NGC 6958}} & \multicolumn{1}{c}{} \\
\cline{2-4} \cline{5-7}
1 & 4.8218 & 0.1000 & 0.9551 & 4.1856 & 0.3370 & 0.9522 \\
2 & 4.8781 & 0.1848 & 0.9616 & 5.3515 & 0.1778 & 0.9821 \\
3 & 4.8722 & 0.4997 & 0.3800 & 4.7191 & 0.5959 & 0.9500 \\
4 & 4.5842 & 0.6501 & 0.9987 & 4.3206 & 1.4124 & 0.9500 \\
5 & 4.3006 & 1.5022 & 0.7994 & 3.9184 & 3.0309 & 0.9500 \\
6 & 4.2191 & 3.5366 & 0.4737 & 3.1345 & 7.0509 & 0.9500 \\
7 & 3.6247 & 4.2202 & 0.8086 & 2.8078 & 13.863 & 0.9500 \\
8 & 3.6606 & 7.1797 & 0.4464 & 1.9595 & 32.998 & 0.9500 \\
9 & 3.2430 & 11.910 & 0.4823 & 1.5971 & 41.903 & 0.9500 \\
10 & 2.7818 & 12.513 & 0.6765 & 0.7920 & 101.15 & 0.9500 \\
11 & 2.6121 & 24.214 & 0.4561 & \nodata & \nodata & \nodata \\
12 & 2.3514 & 27.307 & 0.7340 & \nodata & \nodata & \nodata \\
13 & 1.6972 & 52.467 & 0.9997 & \nodata & \nodata & \nodata \\
14 & 1.6257 & 50.880 & 0.4998 & \nodata & \nodata & \nodata \\
15 & 1.1636 & 116.86 & 0.9908 & \nodata & \nodata & \nodata \\
\enddata
   \tablecomments{cont.}
   \label{tbl:MGE_params2}
\end{deluxetable*}

\begin{figure*}[!ht]
    \includegraphics[width=\textwidth]{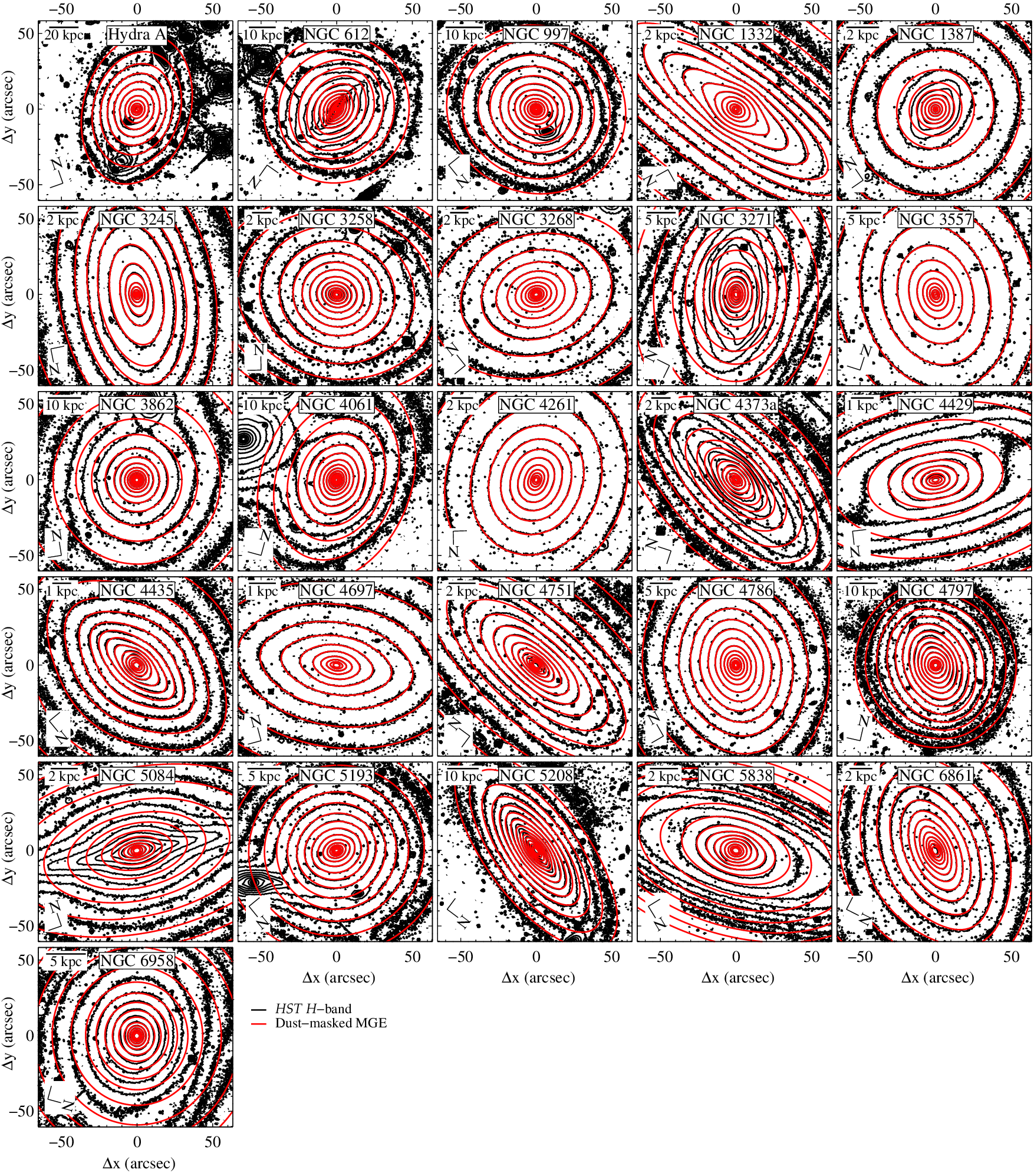}
    \centering
    \caption{Nearly full-frame HST WFC3/F160W mosaics, together with the (dust-masked) \texttt{GALFIT} MGE solutions (overplotted in red) that fit a uniform PA for all components. At larger radii, some galaxies exhibit highly flattened stellar isophotes and/or PA$_\star$ twists, resulting in unavoidable discrepancies. Contours are shown at logarithmic intensity intervals.}
    \label{fig:contours2}
\end{figure*}

We first modeled the 2D surface brightness values using the MGE method presented by \citet{capp02}, using a linear decomposition to determine initial MGE component numbers and Gaussian parameters. However, the \citet{capp02} code requires a symmetric approximation to the PSF shape. For our final $H$-band MGE solutions, we perform the decomposition using the 2D parametric galaxy-fitting algorithm \texttt{GALFIT} \citep{peng10}, including the Tiny Tim F160W PSF to account for blurring effects. The final MGE solutions presented in Table~\ref{tbl:MGE_params} include of between 7 and 16 components. In both MGE approaches, we corrected for foreground Galactic reddening ($A_{\mathrm{Gal},H}$; Table~\ref{tbl:galaxy_info}) and employed the primary and dust masks described in Section~\ref{sec:hstimaging} to mitigate the impact of circumnuclear dust during the optimization. The best-fitting MGEs are overlaid on the observed images in zoom-in plots in Figure~\ref{fig:contours3} and nearly full-frame mosaics in Figure~\ref{fig:contours2}. Comparison between 1D surface brightness profiles extracted from the 2D HST data and MGE models are provided in Appendix~\ref{app:sb_isophotal}. Three ETGs (NGC 3557, NGC 3862, NGC 4261) are Fanaroff-Riley type I \citep[FR-I; i.e., edge-darkened;][]{fr74} radio galaxies, while another (NGC 4429) is sometimes classified as a low-luminosity AGN \citep{ho97,nyland16}. For these targets, preliminary MGE fits preferred the inclusion of nearly unresolved components, which plausibly are due to non-stellar processes. In the final MGEs for these four galaxies, we included PSF components to model and remove possible AGN contributions. Interestingly, only one galaxy (NGC 3862) possesses a prominent $H$-band point source; for another two (NGC 3557 and NGC 4261), the PSF component is entirely negligible, while for one galaxy (NGC 4429), the point source may represent a very compact stellar component.

From these MGEs, we estimated an effective radius by calculating the radius enclosing half the light, so $L(<\re) = L_\mathrm{tot}/2$. Following standard practice \citep[e.g., ATLAS$^\mathrm{3D}$;][]{capp13}, we integrated the light on elliptical annuli to find the half-light radius, which corresponds to the circularized $\re \approx \sqrt{ab}$. We report these \re\ in Table~\ref{tbl:galaxy_info}, and use them to compare PA$_\star$ and $\varepsilon_\star$ trends in Figure~\ref{fig:color_pa_eps}. We note that the $H$-band \re\ for targets also contained in ATLAS$^\mathrm{3D}$ are $\sim$20\% smaller on average than $r$-band estimates, which is expected due to the observed color gradients \citep{ma14}. For the three targets in MASSIVE, we find systemically higher \re\ than those measured from near-IR 2MASS data, although for the two with the largest discrepancies ($\sim$3$\times$ higher) the limiting 2MASS depth may not probe sufficiently far into the stellar halo.

\begin{figure*}[!ht]
    \centering
    \includegraphics[width=\textwidth]{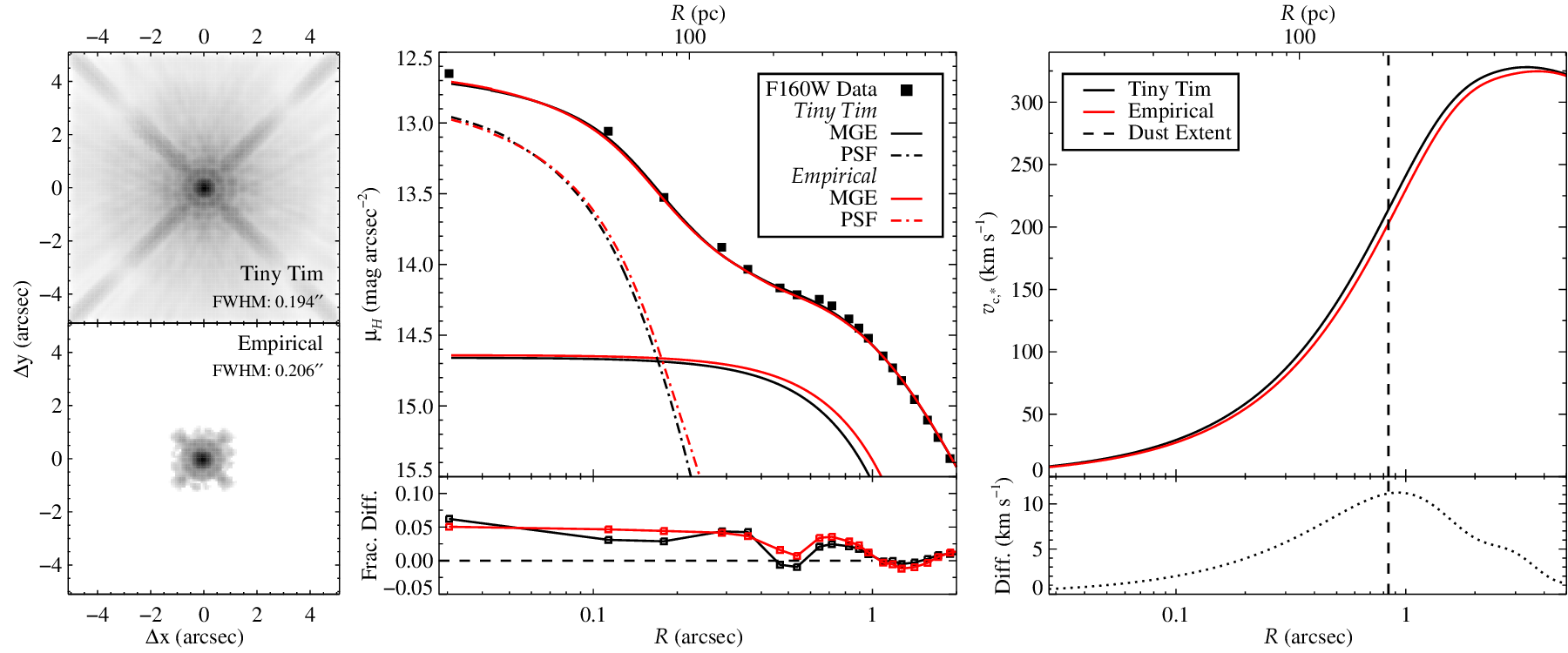} 
    \caption{The \emph{H}-band Tiny Tim PSF response for NGC 3862, showing both drizzled Tiny Tim (\textit{upper left}) and empirical (\textit{lower left}) PSFs that extend to at least an arcsecond. For this target, \texttt{GALFIT} models simultaneously fit both a point source and an MGE using the Tiny Tim and empirical PSFs in turn (\textit{middle}). These show minimal differences in the central PSF and minor changes in the inner couple of MGE components, with generally good agreement (\textit{lower middle}) with the data. The corresponding midplane circular velocity curves (\textit{right}) arising from the different stellar MGE approaches show only small ($\lesssim$10 \kms) discrepancies at all radii.}
    \label{fig:tt_v_emp}
\end{figure*}

Left unconstrained, the MGE optimization process often drives the innermost component(s) to low $\sigma^\prime{}$ values. Some CND regions that fall below the $\Delta (J-H)$ masking criteria may contribute to modestly attenuated stellar light just beyond the nucleus. The result may be the illusion of a centrally compact source \citep[e.g.,][]{bonfini18} that is better fit using MGE component(s) with small dispersions ($\sigma^{\prime} < 0\farcs 1$, typically corresponding to 10$-$30 pc for this sample). When deprojected, these inner component(s) translate to very centrally concentrated stellar luminosity densities. If relatively isolated from adjacent component(s), these compact Gaussian(s) result in peaky circular velocity contributions $v_{\mathrm{c},\star}$ in the inner few $\times$10 pc without clear justification given the PSF FWHM and the inner power-law slope $\gamma$. To avoid complications, we limited $\sigma^\prime{} \gtrsim 0\farcs1$ to ensure these $v_{\mathrm{c,\star}}$ did not translate to stellar mass structure within the $H$-band resolution limit. Other approaches do allow for smaller $\sigma^\prime{}$ \citep[e.g., fitting tightly-spaced MGE components to analytic functions;][]{yildirim17} but are best suited to ETGs without dusty CNDs or with higher $\gamma \gtrsim 1$, for which a truncated $\sigma^{\prime}$ could produce very inaccurate central stellar luminosity densities.

Individual Gaussian components generally do not have physical meaning. However, low $q^\prime{}$ for individual components may fall below $\cos i$ for an assumed $i$, effectively preventing deprojection for oblate axisymmetry. Full gas-dynamical modeling is beyond the scope of this paper, so instead we estimated $i \approx \cos^{-1}[(b/a)_{\mathrm{dust},H}]$. In a limited number of comparisons between $\cos^{-1}[(b/a)_{\mathrm{dust},H}]$ and the average kinematic inclination angle, this proxy has been accurate to within $\sim$4\degr\ \citep[see][]{barth16b,boizelle17,boizelle19}. To ensure that the MGE solutions can be deprojected for a range of reasonable inclination angles about the assumed $i$, we set a limiting $q^\prime{}_\mathrm{min} = \cos [\cos^{-1}(b/a)_{\mathrm{dust},H} - \Delta i]$ where $\Delta i$ increases approximately uniformly from 5\degr\ to 15\degr\ for disks with $i\geq 80\degr$ down to $i \leq 25\degr$, respectively. In most cases, this $q^\prime{}_\mathrm{min}$ constraint does not severely limit the MGE optimization process, although best-fitting MGEs are sometimes noticeably affected as a result, (especially for NGC 997, NGC 3271, NGC 3557, NGC 3862, and NGC 6958). However, this buffer $\Delta i$ is unavoidable and better ensures that gas-dynamical models can explore the full parameter space without deprojection errors. We note that stellar-dynamical modeling typically requires a larger buffer due to the intrinsic galaxy thickness and the optimization of additional intrinsic galaxy viewing angles. Since our primary goal is to aid future ALMA CO work, we restrict ourselves to the $\Delta i$ ranges noted above.



The MGE parameter values change slightly when we use the Tiny Tim rather than the empirical PSF. To quantify the maximal impact of adopting only Tiny Tim PSFs, we also constructed an MGE of the NGC 3862 $H$-band mosaic using the empirical PSF following the outlined method. This target has the most prominent AGN in the sample, and any mismatch of the point-source component is expected to affect the peak surface brightness and FWHM of the centermost MGE component(s). In Figure~\ref{fig:tt_v_emp}, we demonstrate the overall good agreement for the two PSFs. The empirical PSF results in a somewhat better fit to the inner $1\arcsec$, although the MGEs become practically identical much further out. The empirical PSF also leads to $v_{\mathrm{c},\star}(r)$ with slightly smaller velocities ($\lesssim$10 \kms) out to 2.5 kpc or $\sim$6$R_{\mathrm{dust},H}$. These findings are consistent with the conclusion of \citet{zhao21} that an empirical PSF leads to (slightly) better fits. Since our targets do not have strong central point sources at near-IR wavelengths, we expect the Tiny Tim-derived MGEs will be sufficiently accurate for ongoing gas-dynamical modeling.

Photometric PA twists are common in this sample, and accounting for these PA twists results in better overall fits and can be useful for stellar-dynamical efforts that explore triaxality \citep[e.g.,][]{vandb08,krajnovic11,liepold23}. However, allowing for a variable PA between Gaussian components prevents a simple deprojection. In Appendix~\ref{app:free_pa}, we report \texttt{GALFIT} MGE solutions that allow for PAs that differ between components.

\section{Discussion}
\label{sec:discussion}

\subsection{Goodness of Fit}
\label{sec:goodness}

As is shown in Figure~\ref{fig:contours2}, the 2D \texttt{GALFIT} fits to each $H$-band mosaic typically give good global agreement, with formal chi-square per degree of freedom approaching unity. More importantly, fractional residuals along the galaxy major axis are typically smaller than $\sim$10\%. Cases with more serious discrepancies result from either steep rises in $\varepsilon_\star$ or large $\Delta\mathrm{PA}_\star$, and we discuss each in turn. However, we demonstrate in Figure~\ref{fig:contours3} that these large-scale discrepancies have negligible impact on the fidelity of the central MGE fit.

In contrast, $q^{\prime}_\mathrm{min}$ limitations affect nearly every MGE solution here, with two thirds of our sample having two or more components whose $q^{\prime} \rightarrow q^{\prime}_\mathrm{min}$. However, these $q^{\prime}_\mathrm{min}$ are needed to allow for reasonable deprojection in gas-dynamical modeling. For most elliptical galaxies, $q^{\prime}_\mathrm{min}$ does not seem to affect the quality of the MGE fits, while other stellar properties like $\Delta\mathrm{PA}_\star$ and increasing $\varepsilon_\star$ from nearly round isophotes (e.g., NGC 3862) create tension with the CND properties. Large increases in $\varepsilon_\star$ for a third of the lenticular galaxies ($\Delta\varepsilon_\star > 0.2$; NGC 1387, NGC 3271, NGC 4373a, NGC 4429, and NGC 5838) result in poorer overall fits at intermediate (or larger) $R$. MGE fits for two of the three barred lenticular galaxies (NGC 1387 and NGC 3271) are good examples (see also Figure~\ref{fig:ngc1387_ell_mge} and \ref{fig:ngc3271_ell_mge}): over $R\sim 10-35\arcsec$, the poorest-fit regions result from sharp rises in ellipticity (and, for NGC 3271, a PA$_\star$ shift) near where the stellar bar becomes dominant \citep{bettoni97,gadotti05}, with $\varepsilon_\star$ greatly exceeding the maximum allowable $\sim$0.05 level.

Fully half of the ETGs have much smaller $\Delta \mathrm{PA}_\star < 15\degr$, which is more indicative of relaxed systems; just under a quarter have $\Delta \mathrm{PA}_\star \gtrsim 40\degr$, with most of the shift manifesting at large $R$. However, such large $\Delta \mathrm{PA}_\star$ does not always result in visibly worse MGE fits. For example, Hydra A and NGC 3862 show circular inner isophotes with most of the $\Delta\mathrm{PA}_\star$ occurring where $\varepsilon_\star$ remains low. In others (e.g., NGC 4061 and NGC 5193; Figures~\ref{fig:ngc4061_ell_mge} and \ref{fig:ngc5193_ell_mge}) this $\Delta\mathrm{PA}_\star$ is largely confined to the outermost few radial bins.

The large angular extents and high $\Delta(J-H)$ of some CNDs preclude minimal dust masking, possibly leading to less robust central MGE components. The primary example is NGC 612, which contains a large dusty disk ($R_{\mathrm{dust},H}\sim20\arcsec$) that shows evidence for star formation \citep{asabere16} and moderate ($\sim$20\degr) disk warping in both PA and $i$. Even in its \emph{H}-band mosaic, excess light from star formation and opaque dust necessitates masking nearly the entire disk region and nucleus (see Figures~\ref{fig:dustdisks} and \ref{fig:contours3}). As reported in Table~\ref{tbl:MGE_params}, the first 9 MGE components have $\sigma^{\prime{}}$ that are smaller than the semi-major axis extent ($\sim$10\arcsec) of the more regularly-shaped inner disk. Because of the heavy masking, the innermost MGE components are constrained by relatively few data points. At least two other cases have large projected disk sizes ($R_{\mathrm{dust},H}\sim 5-10\arcsec$) and roughly uniform near-IR colors. First, NGC 997 hosts a more face-on ($i \sim 5\degr$) CND for which we masked all of the near side (and much of the far sides) of the disk. Second, NGC 1387 is similarly inclined, and its near-IR colors approach (but do not exceed) the $\Delta (J-H) \gtrsim 0.08$ mag masking criterion. As a result, we do not mask any dust regions for NGC 1387.


Internal dust disk structure gives rise to additional complications when evaluating an MGE. For a galaxy with moderate disk warping, most noticeably within NGC 612 and NGC 4797, it is not clear that $(b/a)_{\mathrm{dust},H}$ measured from near the CND edge is always a good proxy for the galaxy's midplane inclination. Even in beam-smeared ALMA CO moment 1 (velocity) maps, \citet{boizelle17,boizelle19} found that $i(R)\approx \cos^{-1}[q_\mathrm{CO}(R)]$ (where $q_\mathrm{CO}$ is the CO kinematic axis ratio) changes by 5--10\degr\ at larger radii for four CNDs in our sample. Two targets (NGC 3557 and NGC 4261) show CO kinematics that are moderately misaligned with the radio jet orientation \citep[at the 10$-$50\degr\ level;][]{ruffa19a,boizelle21,ruffa20}, at least at the resolution limit, suggesting there may be sharp twists in the PA (and plausibly $i$) of the innermost CO kinematics.

\subsection{Accuracy and Consistency of the MGE Solutions}
\label{subsec:consistency}

It is not trivial to compare the efficacy of MGE solutions that differ significantly in wavelength and/or construction given sparse case studies in dynamical modeling \citep[e.g., see][]{barth16b}. Nor do we attempt to compare the MGE-derived circular velocity profiles to the ALMA CO kinematics. Instead, we use a few MGE solutions and a circular velocity profile from the literature to explore the accuracy and consistency of the MGE solutions in Table~\ref{tbl:MGE_params}.

The same underlying $H$-band data were used by \citet{boizelle19,boizelle21} to construct dust-masked MGEs for two ETGs in this sample (NGC 3258 and NGC 4261), albeit with slight differences in mosaic construction, dust masking, and the number of Gaussian components. We used the different $H$-band MGEs to derive both the stellar-only ($v_{\mathrm{c},\star}$) and total ($v_\mathrm{c}$) circular velocities and the enclosed (spherical) mass distribution $M$($<$\textit{r}) as a function of the physical distance $r$ in the galaxy's midplane. These $v_\mathrm{c}(r)$ and $M$($<$\textit{r}) were constructed using codes in the \texttt{Jeans Anisotropic Models} package \citep[JAM;][]{capp08} after assuming oblate axisymmetry, an inclination angle $i \approx \cos^{-1} [(b/a)_{\mathrm{dust},H}]$, (literature or estimated) \mbh\ values, and a uniform $M/L_H = 1.5$ \msun/\lsun$_{,H}$ based on single stellar population models \citep{vazdekis10}. Dark matter contributions were assumed to be negligible within the central few kpc for such galaxies \citep{debru04}, and gas mass contributions likewise contribute little to \vc\ \citep[e.g.,][]{boizelle19,boizelle21,cohn21,kabasares22}. Within the disk extent, we find the different $v_{\mathrm{c},\star}$ disagree by at most 10\% and become indistinguishable much beyond $R_{\mathrm{dust},H}$, suggesting little impact on gas-dynamical modeling. For NGC 3258, \citet{boizelle21} find that the intrinsic velocity due to all extended mass components is higher than the dust-masked $v_{\mathrm{c},\star}$, as expected, reaching a maximum discrepancy of $\sim$40 \kms but a typical discrepancy closer to 20 \kms\ over most of the disk.

The velocity differences become more serious when comparing near-IR and optically-derived MGEs. Two cases (NGC 3557 and NGC 4429) provide useful case studies. Using the optical MGEs and best-fit $M/L$ (F555W, ignoring the first component: \citealp{ruffa19b}; F606W: \citealp{davis18}), we scaled $M/L_H$ so that the $M$($<$\textit{r}) profiles match a little beyond $R_{\mathrm{dust},H}$, where the CO-bright emission ends. These optical and near-IR MGEs had similar dust masks. We find that the $H$-band MGEs prefer higher stellar mass and $v_{\mathrm{c},\star}$ throughout the CND, reaching excess $\sim$40 \kms\ at the NGC 3557 disk edge and $\sim$90 \kms\ near the NGC 4429 nucleus. For both ETGs, the discrepancy seems largely driven by ($\sim$5$\times$) greater dust attenuation at optical wavelengths in the unmasked regions of the CND, together with some evident $M/L$ gradients. The $H$-band MGEs for both NGC 3557 and NGC 4261 have two Gaussian components with smaller $\sigma^{\prime}$ than those of the optical MGEs.

While these dust-masked MGEs provide good overall fits to the $H$-band mosaics, they may somewhat underestimate the stellar luminosity distributions in the CND regions. Even after careful masking, some residual dust attenuation remains. Based on the $\Delta (J-H) \gtrsim 0.08$ mag criterion and the discussion in Section~\ref{subsec:colors}, unmasked dust should suppress the background $H$-band stellar light by at most $\sim$25\%, and likely much less on average. To explore the maximal possible impact on the stellar luminosity models, we tried increasing the innermost MGE component(s)intensity by up to $\sim25\%$ to compensate for remaining dust obscuration in the \emph{H}-band, with the corresponding $v_{\mathrm{c},\star}$ typically increasing by only 20$-$30 \kms\ within or near $R_{\mathrm{dust},H}$. Relying on a single (dust-masked) stellar luminosity model may therefore bias a BH mass measurement, and including both dust-masked and dust-corrected MGEs provides a broader but more accurate \mbh\ error budget. For a few cored galaxies, the impact has been relatively minor \citep{boizelle19,boizelle21}, with the \mbh\ shifts ranging between $\Delta \mbh \sim 10-20\%$ about a fiducial \mbh\ value and the dust-masked MGEs giving best-fit \mbh\ at the upper end of this mass range. Those particularly cored galaxies are not fully representative of $\Delta \mbh$ for the present sample: they have strong, Keplerian-like CO velocity upturns, insulating these BH mass measurements against changes in the stellar mass model. For a few more cuspy ETGs with CO emission that extends down to $\sim$\rg, different stellar mass models (e.g., MGEs constructed in different filters, or using different dust-masking or correction techniques) returns $\Delta \mbh$ ranging between 30\% and a factor of more than two \citep{barth16a,barth16b,davis18,cohn21,kabasares22}. However, we stress that none of these more cuspy galaxies unambiguously show CO emission arising from deep within \rg, which likely contributes in part to the relatively larger $\Delta\mbh$ shifts. We anticipate the dust-masked MGE presented here will enable \mbh\ determined with accuracies at the 30\% level or slightly better.

Lastly, the near-IR data presented here may not be sufficient to deliver \mbh\ precision even at the 20--30\% level for ETGs with larger angular diameter CNDs \citep[especially when viewed more face-on;][]{davis18,kabasares22} and/or for cuspier surface brightness profiles with higher central $\Delta (J-H)$ \citep[see the discussion by][]{yoon17}. In such cases, extensive masking at the adopted $\Delta (J-H) \gtrsim 0.08$ mag level leads to less secure constraints on inner MGE components. For ETGs with large $R_{\mathrm{dust},H}$ and $(b/a)_{\mathrm{dust},H} \sim 1$, masking difficulties likely result in $H$-band MGEs that underestimate the intrinsic central stellar light contributions. JWST NIRCam imaging at 4.5$-$5 $\mu$m is likely the best avenue to mitigate the impact of dust and allow high-quality MGEs can be constructed from data with an angular resolution $\sim$\rg.

\begin{figure*}
    \centering
    \includegraphics[width=0.85\textwidth]{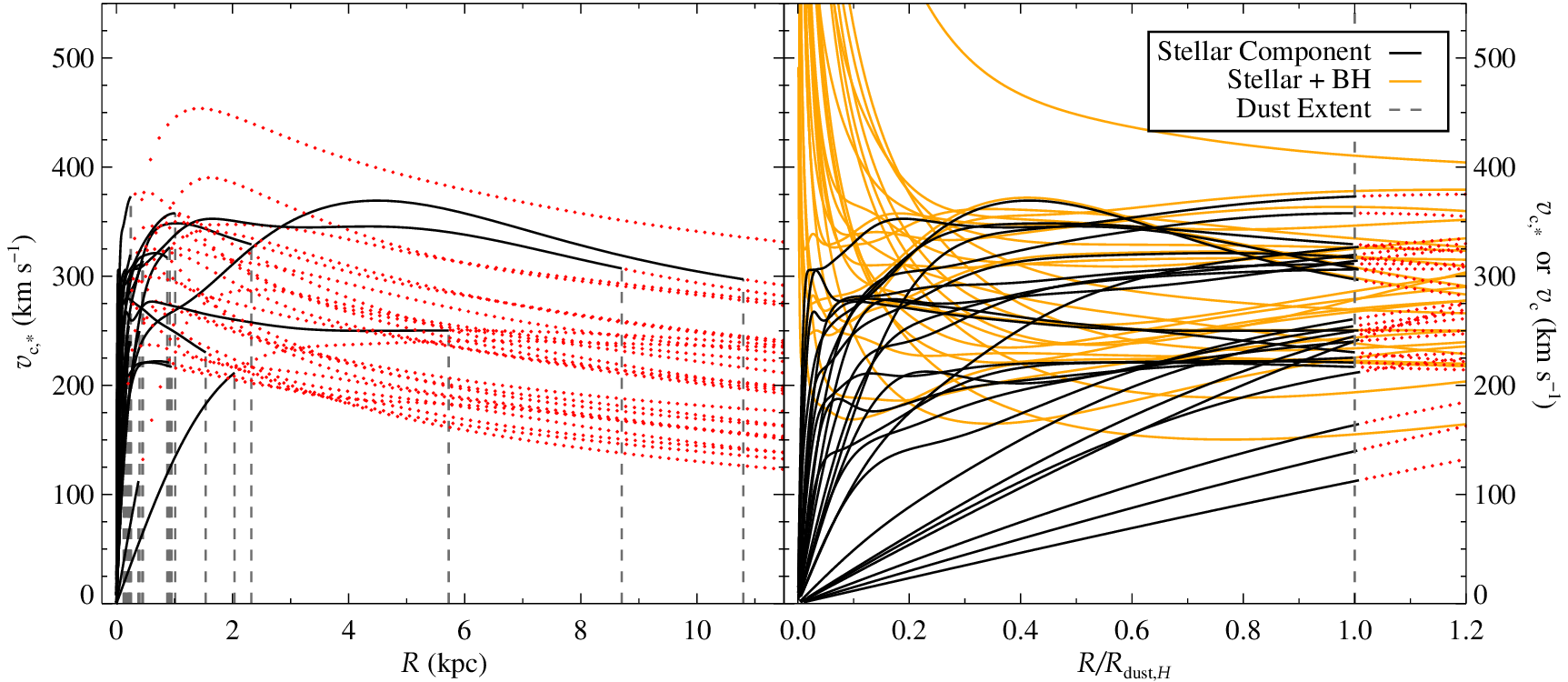} 
    \caption{Circular velocity curves from stellar-only $v_{\mathrm{c},\star}$ and total \vc\ gravitational potentials, constructed by deprojecting the \emph{H}-band MGEs and numerically integrating the stellar luminosity densities. Solutions for $r > R_{\mathrm{dust},H}$ are included (red dotted lines) along with the extents of each dust disk (dashed lines). To better compare the inner curves of these results, the same $v_{\mathrm{c},\star}$ curves are normalized (\textit{right panel}) to their respective dust disk radii, together with \vc\ curves (orange solid lines) that include the influence of the expected BH masses. With the exception of four BH masses already measured by stellar \citep[][]{rusli13} or gas-dynamical modeling \citep{barth16b,davis18,boizelle21}, BH masses were estimated using $M_\mathrm{BH}-\sigma_\star$ or $M_\mathrm{BH}-L_K$ relations.} 
    \label{fig:vcirc}
\end{figure*}

\subsection{Circular Velocities and the CO TF Relationship}
\label{subsec:vcirc}

In addition to providing a kinematic tracer of the innermost gravitational potential, ALMA CO measurements are also being used to explore larger-scale baryonic and dark matter properties in the context of the Tully-Fisher (TF) relation \citep{tulfish77}. Traditionally, a TF analysis employs large-scale gas disks (often from \ion{H}{1} emission at 21 cm) that probe the asymptotic or flat portion of a disk galaxy's rotation curve. Since initial proposals by \citet{dickey92} and \citet{sofue92}, CO emission has also been calibrated as a probe of rotational velocity \citep[e.g., ][]{ho07,davis11,davis16,tiley16,tiley19,topal18}, resulting in the CO TF relation. CO velocity profiles are more irregular in general than traditionally-used \ion{H}{1} profiles \citep{ho07,smith21b}, although morphologically round CNDs typically show the characteristic double-horned, sharp-edged CO profiles \citep[sometimes with high-velocity emission in the wings due to central, Keplerian rotation;][]{boizelle17,boizelle19,boizelle21,ruffa19a}. Previous studies assumed the CO-bright disk probed out to either a maximum circular speed $v_\mathrm{max}$ (followed by a turnover) or an asymptotic circular velocity (occurring beyond the $v_\mathrm{max}$ radius), and therefore the integrated CO line width would be a good proxy for the bulge mass. However, the CO emission in compact CNDs is unlikely to extend far enough to probe $v_\mathrm{max}$, resulting in smaller integrated CO line widths and incorrect estimation of correlated galaxy properties \citep[e.g., $\sigma_\star$, $M_\star$, \mbh;][]{smith21b}. Indeed, the CNDs in our sample have a median $R_{\mathrm{dust},H}\sim 0.4$ kpc that is at least half the median CO extent found in volume-limited surveys \citep{davis13b}. It may be that many of the smaller disks in the volume-limit surveys may have been missed entirely due to sensitivity limitations \citep[e.g., NGC 4261;][]{young11,boizelle21}.

To assess the extent to which CO-bright CNDs can masquerade as extended gas disks in spatially unresolved TF relation studies, we explored the resolved $v_{\mathrm{c},\star}$ and $v_\mathrm{c}$ behavior for our sample following the method outlined in Section~\ref{subsec:consistency}. In Figure~\ref{fig:vcirc}, we plot both $v_{\mathrm{c},\star}$ and $v_\mathrm{c}$ for each ETG as a function of $r$ and after scaling distance by the respective $R_{\mathrm{dust},H}$. While the $v_\mathrm{c}$ velocities upturn within each BH-dominated region ($\lesssim$\rg), this region typically has little to no significant CO emission that could lead to higher integrated CO line widths. In most cases, $\rg < R_{\mathrm{dust},H}$, and the inferred rotation curves increase noticeably between \rg\ and $R_{\mathrm{dust},H}$. Even in cases where CO emission extends well within \rg\ \citep[e.g., NGC 3258;][]{boizelle17,boizelle19}, this Keplerian signature provides at most faint wings to the integrated line profiles and little -- if any -- change to the measured line widths. Still, the (literature or estimated) \mbh\ help to flatten $v_\mathrm{c}$ within $R_{\mathrm{dust},H}$. About half of these $v_\mathrm{c}$ reach $v_\mathrm{max}$ within the CND (but only for those with $R_{\mathrm{dust},H} \geq 0.9$ kpc) while only $\sim$20\% of the sample reach an asymptotic velocity at these radii. The more cored ETGs ($\gamma < 0.2$) typically have sufficiently small $R_{\mathrm{dust},H} \sim 0.1-1$ kpc such that most reach $v_\mathrm{max}$ beyond the CND extent. Cuspy ETGs tend to reach asymptotic velocities within (or just beyond) the disk edge. Unsurprisingly for the fairly compact molecular gas disks that are contained in our sample, we find that the CNDs in at least a third do not probe sufficiently far out to reach a maximum (or asymptotic) circular velocity. Future CO TF studies will benefit from a similar $v_\mathrm{c}$ analysis or resolved CO kinematics to confirm the suitability of a target within a broader sample.


\section{Conclusion}
\label{sec:conclusion}

ALMA CO imaging of dynamically cold CO kinematics in ETGs provides an appealing avenue to more completely and securely populating the high-mass end of \mbh-galaxy correlations. For many ETGs with dusty CNDs, the CO-bright emission is coincident with optically thick dust, limiting the accuracy of stellar luminosity models derived from optical data alone. In this paper, we presented new near-IR HST data for 26 ETGs whose ALMA CO measurements show clean kinematics and good prospects for eventual \mbh\ determination. We detailed masking efforts and constructed stellar luminosity models using the MGE formalism. When fully utilized, these stellar luminosity models will help to expand the number of accurate \mbh\ measurements using ALMA by up to a factor of $\sim$3 from the present number. When considering all methods to determine BH masses, these dust-masked MGEs could increase the number of E/S0 galaxies with accurate \mbh\ by up to $\sim$25\%. BH mass error budgets are often dominated by uncertainties in the central stellar surface brightness slope due to CND dust attenuation, and few studies have explored the effects that these uncertainties have on stellar luminosity models and final \mbh\ measurements. Future work will explore dust attenuation modeling to create corrected MGEs.

Of course, more complicated CNDs and stellar light distributions may provide challenges to simple interpretation and application of these MGEs. The CNDs in this sample were selected because of an expectation of dynamically relaxed CO kinematics. Most gas disks do appear to have settled into their respective galaxy's midplanes, and the corresponding MGEs can be applied in a straightforward manner in dynamical modeling efforts. However, disk warping is evident in some dust features as well as CO kinematics, both of which can only reveal warping on scales larger than the angular resolution of the data. CO kinematics suggest warping at the 5$-$10\degr\ level throughout the CND is typical, although moderate-to-high disk warping has been seen. The constraints on MGE component axis ratio $q^{\prime}$ were based on the axis ratio of the outer dust features, so larger changes in the disk inclination angle or misalignments of the outer CND disk structure with the galaxy midplane may impact the MGE solutions presented here. Users of these dust-masked MGEs should evaluate their appropriateness given the observed gas and/or stellar kinematics. Lastly, dust attenuation still influences the MGE goodness-of-fit in the central regions despite our best masking efforts, leading to inner components that may slightly underestimate the intrinsic stellar distribution.

Comparing these stellar luminosity models to other dust-masked MGEs using same-filter data, we find minimal ($\lesssim$10\%) differences between the derived circular velocity profiles $v_{\mathrm{c}}$ and the corresponding enclosed mass profiles $M$($<$\textit{r}). When we compare $H$-band to optically-derived MGEs with nearly identical dust masks, however, the $v_{\mathrm{c}}$ and $M$($<$\textit{r}) profiles show more significant differences that may impact gas-dynamical modeling of the ALMA CO data. The $H$-band MGEs suggest greater stellar mass within the CNDs, as near-IR data better recover the intrinsic light in unmasked regions.

The HST data files and products arising from this project will prove useful beyond their primary goal of supporting the existing ALMA data sets. These dust-masked MGEs will facilitate other dynamical modeling efforts, including those using stellar kinematic data or those that will pursue a re-analysis of past ionized gas-dynamical modeling \citep{beifiori09}. Because of their depth and wavelength coverage, the optical and near-IR HST data and ongoing analysis will have additional legacy value in studies of stellar population gradients, central star formation, and globular cluster populations in ETGs. These multi-wavelength HST data will be key to constraining dust attenuation across the disk, and forthcoming dust-corrected MGEs will ensure robust exploration of BH mass measurement errors.

   
\begin{acknowledgments}
     This work is based on observations made with the NASA/ESA Hubble Space Telescope, obtained at the Space Telescope Science Institute, which is operated by the Association of Universities for Research in Astronomy, Inc., under NASA contract NAS5-26555. New observations are associated with GO programs 14920, 15226, and 15909. Support for HST GO program number 15909 was provided through a grant from the STScI under NASA contract NAS5-26555. This work has made use of archival data obtained with the Spitzer Space Telescope, which was operated by the Jet Propulsion Laboratory, California Institute of Technology under a contract with NASA. This work as made use of observations made with the NASA/ESA Hubble Space Telescope, and obtained from the Hubble Legacy Archive, which is a collaboration between the Space Telescope Science Institute (STScI/NASA), the Space Telescope European Coordinating Facility (ST-ECF/ESA) and the Canadian Astronomy Data Centre (CADC/NRC/CSA). This research has made use of the HyperLEDA database and the NASA/IPAC Extragalactic Database (NED) which is operated by the Jet Propulsion Laboratory, California Institute of Technology, under contract with the National Aeronautics and Space Administration. JRD thanks the Brigham Young University Department of Physics and Astronomy for their Graduate Assistance Awards. JLW is supported by NSF AST-2206219. AJBak acknowledges support from the Radcliffe Institute for Advanced Study at Harvard University. LCH was supported by the National Science Foundation of China (11721303, 11991052, 12011540375, 12233001) and the China Manned Space Project (CMS-CSST-2021-A04, CMS-CSST-2021-A06).
\end{acknowledgments}

\facilities{HST (WFPC2,ACS,WFC3), IRSA, MAST}

\software{AstroDrizzle \citep{astrodrizz}, GALFIT \citep{peng10}, IRAF \citep{tody86,tody93}, PyRAF \citep{pyraf}, Astropy \citep{astropy}, MgeFit \citep{capp02}, Tiny Tim \citep{tinytim}.}

\bibliography{main}

\begin{thebibliography}{}
\expandafter\ifx\csname natexlab\endcsname\relax\def\natexlab#1{#1}\fi
\providecommand{\url}[1]{\href{#1}{#1}}
\providecommand{\dodoi}[1]{doi:~\href{http://doi.org/#1}{\nolinkurl{#1}}}
\providecommand{\doeprint}[1]{\href{http://ascl.net/#1}{\nolinkurl{http://ascl.net/#1}}}
\providecommand{\doarXiv}[1]{\href{https://arxiv.org/abs/#1}{\nolinkurl{https://arxiv.org/abs/#1}}}

\bibitem[{{Aaronson}(1977)}]{aaronson77}
{Aaronson}, M. 1977, PhD thesis, Harvard University, Massachusetts

\bibitem[{{Alatalo} {et~al.}(2013){Alatalo}, {Davis}, {Bureau}, {Young},
  {Blitz}, {Crocker}, {Bayet}, {Bois}, {Bournaud}, {Cappellari}, {Davies}, {de
  Zeeuw}, {Duc}, {Emsellem}, {Khochfar}, {Krajnovi{\'c}}, {Kuntschner},
  {Lablanche}, {Morganti}, {McDermid}, {Naab}, {Oosterloo}, {Sarzi}, {Scott},
  {Serra}, \& {Weijmans}}]{alatalo13}
{Alatalo}, K., {Davis}, T.~A., {Bureau}, M., {et~al.} 2013, \mnras, 432, 1796,
  \dodoi{10.1093/mnras/sts299}

\bibitem[{{Anderson}(2016)}]{anderson16}
{Anderson}, J. 2016, {Empirical Models for the WFC3/IR PSF}, Instrument Science
  Report WFC3 2016-12, 42 pages

\bibitem[{{Astropy Collaboration} {et~al.}(2018){Astropy Collaboration},
  {Price-Whelan}, {Sip{\H{o}}cz}, {G{\"u}nther}, {Lim}, {Crawford}, {Conseil},
  {Shupe}, {Craig}, {Dencheva}, {Ginsburg}, {VanderPlas}, {Bradley},
  {P{\'e}rez-Su{\'a}rez}, {de Val-Borro}, {Aldcroft}, {Cruz}, {Robitaille},
  {Tollerud}, {Ardelean}, {Babej}, {Bach}, {Bachetti}, {Bakanov}, {Bamford},
  {Barentsen}, {Barmby}, {Baumbach}, {Berry}, {Biscani}, {Boquien}, {Bostroem},
  {Bouma}, {Brammer}, {Bray}, {Breytenbach}, {Buddelmeijer}, {Burke},
  {Calderone}, {Cano Rodr{\'\i}guez}, {Cara}, {Cardoso}, {Cheedella}, {Copin},
  {Corrales}, {Crichton}, {D'Avella}, {Deil}, {Depagne}, {Dietrich}, {Donath},
  {Droettboom}, {Earl}, {Erben}, {Fabbro}, {Ferreira}, {Finethy}, {Fox},
  {Garrison}, {Gibbons}, {Goldstein}, {Gommers}, {Greco}, {Greenfield},
  {Groener}, {Grollier}, {Hagen}, {Hirst}, {Homeier}, {Horton}, {Hosseinzadeh},
  {Hu}, {Hunkeler}, {Ivezi{\'c}}, {Jain}, {Jenness}, {Kanarek}, {Kendrew},
  {Kern}, {Kerzendorf}, {Khvalko}, {King}, {Kirkby}, {Kulkarni}, {Kumar},
  {Lee}, {Lenz}, {Littlefair}, {Ma}, {Macleod}, {Mastropietro}, {McCully},
  {Montagnac}, {Morris}, {Mueller}, {Mumford}, {Muna}, {Murphy}, {Nelson},
  {Nguyen}, {Ninan}, {N{\"o}the}, {Ogaz}, {Oh}, {Parejko}, {Parley}, {Pascual},
  {Patil}, {Patil}, {Plunkett}, {Prochaska}, {Rastogi}, {Reddy Janga},
  {Sabater}, {Sakurikar}, {Seifert}, {Sherbert}, {Sherwood-Taylor}, {Shih},
  {Sick}, {Silbiger}, {Singanamalla}, {Singer}, {Sladen}, {Sooley},
  {Sornarajah}, {Streicher}, {Teuben}, {Thomas}, {Tremblay}, {Turner},
  {Terr{\'o}n}, {van Kerkwijk}, {de la Vega}, {Watkins}, {Weaver}, {Whitmore},
  {Woillez}, {Zabalza}, \& {Astropy Contributors}}]{astropy}
{Astropy Collaboration}, {Price-Whelan}, A.~M., {Sip{\H{o}}cz}, B.~M., {et~al.}
  2018, \aj, 156, 123, \dodoi{10.3847/1538-3881/aabc4f}

\bibitem[{{Babyk} {et~al.}(2019){Babyk}, {McNamara}, {Tamhane}, {Nulsen},
  {Russell}, \& {Edge}}]{babyk19}
{Babyk}, I.~V., {McNamara}, B.~R., {Tamhane}, P.~D., {et~al.} 2019, \apj, 887,
  149, \dodoi{10.3847/1538-4357/ab54ce}

\bibitem[{{Barth} {et~al.}(2016{\natexlab{a}}){Barth}, {Boizelle}, {Darling},
  {Baker}, {Buote}, {Ho}, \& {Walsh}}]{barth16a}
{Barth}, A.~J., {Boizelle}, B.~D., {Darling}, J., {et~al.} 2016{\natexlab{a}},
  \apjl, 822, L28, \dodoi{10.3847/2041-8205/822/2/L28}

\bibitem[{{Barth} {et~al.}(2016{\natexlab{b}}){Barth}, {Darling}, {Baker},
  {Boizelle}, {Buote}, {Ho}, \& {Walsh}}]{barth16b}
{Barth}, A.~J., {Darling}, J., {Baker}, A.~J., {et~al.} 2016{\natexlab{b}},
  \apj, 823, 51, \dodoi{10.3847/0004-637X/823/1/51}

\bibitem[{{Beifiori} {et~al.}(2009){Beifiori}, {Sarzi}, {Corsini}, {Dalla
  Bont{\`a}}, {Pizzella}, {Coccato}, \& {Bertola}}]{beifiori09}
{Beifiori}, A., {Sarzi}, M., {Corsini}, E.~M., {et~al.} 2009, \apj, 692, 856,
  \dodoi{10.1088/0004-637X/692/1/856}

\bibitem[{{Bernardi} {et~al.}(2007){Bernardi}, {Hyde}, {Sheth}, {Miller}, \&
  {Nichol}}]{bern07}
{Bernardi}, M., {Hyde}, J.~B., {Sheth}, R.~K., {Miller}, C.~J., \& {Nichol},
  R.~C. 2007, \aj, 133, 1741, \dodoi{10.1086/511783}

\bibitem[{{Bettoni} \& {Galletta}(1997)}]{bettoni97}
{Bettoni}, D., \& {Galletta}, G. 1997, \aaps, 124, 61,
  \dodoi{10.1051/aas:1997180}

\bibitem[{{Blakeslee} {et~al.}(2021){Blakeslee}, {Jensen}, {Ma}, {Milne}, \&
  {Greene}}]{blakesless21}
{Blakeslee}, J.~P., {Jensen}, J.~B., {Ma}, C.-P., {Milne}, P.~A., \& {Greene},
  J.~E. 2021, \apj, 911, 65, \dodoi{10.3847/1538-4357/abe86a}

\bibitem[{{Blakeslee} {et~al.}(2009){Blakeslee}, {Jord{\'a}n}, {Mei},
  {C{\^o}t{\'e}}, {Ferrarese}, {Infante}, {Peng}, {Tonry}, \&
  {West}}]{blakeslee09}
{Blakeslee}, J.~P., {Jord{\'a}n}, A., {Mei}, S., {et~al.} 2009, \apj, 694, 556,
  \dodoi{10.1088/0004-637X/694/1/556}

\bibitem[{{Bogd{\'a}n} {et~al.}(2018){Bogd{\'a}n}, {Lovisari}, {Volonteri}, \&
  {Dubois}}]{bog18}
{Bogd{\'a}n}, {\'A}., {Lovisari}, L., {Volonteri}, M., \& {Dubois}, Y. 2018,
  \apj, 852, 131, \dodoi{10.3847/1538-4357/aa9ab5}

\bibitem[{{Boizelle} {et~al.}(2017){Boizelle}, {Barth}, {Darling}, {Baker},
  {Buote}, {Ho}, \& {Walsh}}]{boizelle17}
{Boizelle}, B.~D., {Barth}, A.~J., {Darling}, J., {et~al.} 2017, \apj, 845,
  170, \dodoi{10.3847/1538-4357/aa8266}

\bibitem[{{Boizelle} {et~al.}(2019){Boizelle}, {Barth}, {Walsh}, {Buote},
  {Baker}, {Darling}, \& {Ho}}]{boizelle19}
{Boizelle}, B.~D., {Barth}, A.~J., {Walsh}, J.~L., {et~al.} 2019, \apj, 881,
  10, \dodoi{10.3847/1538-4357/ab2a0a}

\bibitem[{{Boizelle} {et~al.}(2021){Boizelle}, {Walsh}, {Barth}, {Buote},
  {Baker}, {Darling}, {Ho}, {Cohn}, \& {Kabasares}}]{boizelle21}
{Boizelle}, B.~D., {Walsh}, J.~L., {Barth}, A.~J., {et~al.} 2021, \apj, 908,
  19, \dodoi{10.3847/1538-4357/abd24d}

\bibitem[{{Bonfini} {et~al.}(2018){Bonfini}, {Gonz{\'a}lez-Mart{\'\i}n},
  {Fritz}, {Bitsakis}, {Bruzual}, \& {Cervantes Sodi}}]{bonfini18}
{Bonfini}, P., {Gonz{\'a}lez-Mart{\'\i}n}, O., {Fritz}, J., {et~al.} 2018,
  \mnras, 478, 1161, \dodoi{10.1093/mnras/sty1087}

\bibitem[{{Cantiello} {et~al.}(2005){Cantiello}, {Blakeslee}, {Raimondo},
  {Mei}, {Brocato}, \& {Capaccioli}}]{cantiello05}
{Cantiello}, M., {Blakeslee}, J.~P., {Raimondo}, G., {et~al.} 2005, \apj, 634,
  239, \dodoi{10.1086/491694}

\bibitem[{{Cappellari}(2002)}]{capp02}
{Cappellari}, M. 2002, \mnras, 333, 400,
  \dodoi{10.1046/j.1365-8711.2002.05412.x}

\bibitem[{{Cappellari}(2008)}]{capp08}
---. 2008, \mnras, 390, 71, \dodoi{10.1111/j.1365-2966.2008.13754.x}

\bibitem[{{Cappellari} {et~al.}(2011){Cappellari}, {Emsellem}, {Krajnovi{\'c}},
  {McDermid}, {Scott}, {Verdoes Kleijn}, {Young}, {Alatalo}, {Bacon}, {Blitz},
  {Bois}, {Bournaud}, {Bureau}, {Davies}, {Davis}, {de Zeeuw}, {Duc},
  {Khochfar}, {Kuntschner}, {Lablanche}, {Morganti}, {Naab}, {Oosterloo},
  {Sarzi}, {Serra}, \& {Weijmans}}]{capp11}
{Cappellari}, M., {Emsellem}, E., {Krajnovi{\'c}}, D., {et~al.} 2011, \mnras,
  413, 813, \dodoi{10.1111/j.1365-2966.2010.18174.x}

\bibitem[{{Cappellari} {et~al.}(2013){Cappellari}, {Scott}, {Alatalo}, {Blitz},
  {Bois}, {Bournaud}, {Bureau}, {Crocker}, {Davies}, {Davis}, {de Zeeuw},
  {Duc}, {Emsellem}, {Khochfar}, {Krajnovi{\'c}}, {Kuntschner}, {McDermid},
  {Morganti}, {Naab}, {Oosterloo}, {Sarzi}, {Serra}, {Weijmans}, \&
  {Young}}]{capp13}
{Cappellari}, M., {Scott}, N., {Alatalo}, K., {et~al.} 2013, \mnras, 432, 1709,
  \dodoi{10.1093/mnras/stt562}

\bibitem[{{Caso} {et~al.}(2013){Caso}, {Bassino}, {Richtler}, {Smith Castelli},
  \& {Faifer}}]{pablocaso13}
{Caso}, J.~P., {Bassino}, L.~P., {Richtler}, T., {Smith Castelli}, A.~V., \&
  {Faifer}, F.~R. 2013, \mnras, 430, 1088, \dodoi{10.1093/mnras/sts687}

\bibitem[{{Cohn} {et~al.}(2021){Cohn}, {Walsh}, {Boizelle}, {Barth},
  {Gebhardt}, {G{\"u}ltekin}, {Y{\i}ld{\i}r{\i}m}, {Buote}, {Darling}, {Baker},
  {Ho}, \& {Kabasares}}]{cohn21}
{Cohn}, J.~H., {Walsh}, J.~L., {Boizelle}, B.~D., {et~al.} 2021, \apj, 919, 77,
  \dodoi{10.3847/1538-4357/ac0f78}

\bibitem[{{Cohn} {et~al.}(2023){Cohn}, {Curliss}, {Walsh}, {Kabasares},
  {Boizelle}, {Barth}, {Gebhardt}, {G{\"u}ltekin}, {Y{\i}ld{\i}r{\i}m},
  {Buote}, {Darling}, {Baker}, \& {Ho}}]{cohn23}
{Cohn}, J.~H., {Curliss}, M., {Walsh}, J.~L., {et~al.} 2023, arXiv e-prints,
  arXiv:2310.11296, \dodoi{10.48550/arXiv.2310.11296}

\bibitem[{{Davis}(2014)}]{davis14}
{Davis}, T.~A. 2014, \mnras, 443, 911, \dodoi{10.1093/mnras/stu1163}

\bibitem[{{Davis} {et~al.}(2017){Davis}, {Bureau}, {Onishi}, {Cappellari},
  {Iguchi}, \& {Sarzi}}]{davis17}
{Davis}, T.~A., {Bureau}, M., {Onishi}, K., {et~al.} 2017, \mnras, 468, 4675,
  \dodoi{10.1093/mnras/stw3217}

\bibitem[{{Davis} {et~al.}(2016){Davis}, {Greene}, {Ma}, {Pandya}, {Blakeslee},
  {McConnell}, \& {Thomas}}]{davis16}
{Davis}, T.~A., {Greene}, J., {Ma}, C.-P., {et~al.} 2016, \mnras, 455, 214,
  \dodoi{10.1093/mnras/stv2313}

\bibitem[{{Davis} {et~al.}(2011){Davis}, {Bureau}, {Young}, {Alatalo}, {Blitz},
  {Cappellari}, {Scott}, {Bois}, {Bournaud}, {Davies}, {de Zeeuw}, {Emsellem},
  {Khochfar}, {Krajnovi{\'c}}, {Kuntschner}, {Lablanche}, {McDermid},
  {Morganti}, {Naab}, {Oosterloo}, {Sarzi}, {Serra}, \& {Weijmans}}]{davis11}
{Davis}, T.~A., {Bureau}, M., {Young}, L.~M., {et~al.} 2011, \mnras, 414, 968,
  \dodoi{10.1111/j.1365-2966.2011.18284.x}

\bibitem[{{Davis} {et~al.}(2013){Davis}, {Alatalo}, {Bureau}, {Cappellari},
  {Scott}, {Young}, {Blitz}, {Crocker}, {Bayet}, {Bois}, {Bournaud}, {Davies},
  {de Zeeuw}, {Duc}, {Emsellem}, {Khochfar}, {Krajnovi{\'c}}, {Kuntschner},
  {Lablanche}, {McDermid}, {Morganti}, {Naab}, {Oosterloo}, {Sarzi}, {Serra},
  \& {Weijmans}}]{davis13b}
{Davis}, T.~A., {Alatalo}, K., {Bureau}, M., {et~al.} 2013, \mnras, 429, 534,
  \dodoi{10.1093/mnras/sts353}

\bibitem[{{Davis} {et~al.}(2018){Davis}, {Bureau}, {Onishi}, {van de Voort},
  {Cappellari}, {Iguchi}, {Liu}, {North}, {Sarzi}, \& {Smith}}]{davis18}
{Davis}, T.~A., {Bureau}, M., {Onishi}, K., {et~al.} 2018, \mnras, 473, 3818,
  \dodoi{10.1093/mnras/stx2600}

\bibitem[{{Davis} {et~al.}(2022){Davis}, {Gensior}, {Bureau}, {Cappellari},
  {Choi}, {Elford}, {Kruijssen}, {Lelli}, {Liang}, {Liu}, {Ruffa}, {Saito},
  {Sarzi}, {Schruba}, \& {Williams}}]{davis22}
{Davis}, T.~A., {Gensior}, J., {Bureau}, M., {et~al.} 2022, \mnras, 512, 1522,
  \dodoi{10.1093/mnras/stac600}

\bibitem[{{De Bruyne} {et~al.}(2004){De Bruyne}, {De Rijcke}, {Dejonghe}, \&
  {Zeilinger}}]{debru04}
{De Bruyne}, V., {De Rijcke}, S., {Dejonghe}, H., \& {Zeilinger}, W.~W. 2004,
  \mnras, 349, 440, \dodoi{10.1111/j.1365-2966.2004.07495.x}

\bibitem[{{de Nicola} {et~al.}(2020){de Nicola}, {Saglia}, {Thomas}, {Dehnen},
  \& {Bender}}]{denicola20}
{de Nicola}, S., {Saglia}, R.~P., {Thomas}, J., {Dehnen}, W., \& {Bender}, R.
  2020, \mnras, 496, 3076, \dodoi{10.1093/mnras/staa1703}

\bibitem[{{de Vaucouleurs} {et~al.}(1991){de Vaucouleurs}, {de Vaucouleurs},
  {Corwin}, {Buta}, {Paturel}, \& {Fouque}}]{devaucouleurs91}
{de Vaucouleurs}, G., {de Vaucouleurs}, A., {Corwin}, Herold~G., J., {et~al.}
  1991, {Third Reference Catalogue of Bright Galaxies}

\bibitem[{{di Serego Alighieri} {et~al.}(2007){di Serego Alighieri}, {Gavazzi},
  {Giovanardi}, {Giovanelli}, {Grossi}, {Haynes}, {Kent}, {Koopmann},
  {Pellegrini}, {Scodeggio}, \& {Trinchieri}}]{alighieri07}
{di Serego Alighieri}, S., {Gavazzi}, G., {Giovanardi}, C., {et~al.} 2007,
  \aap, 474, 851, \dodoi{10.1051/0004-6361:20078205}

\bibitem[{{di Serego Alighieri} {et~al.}(2013){di Serego Alighieri}, {Bianchi},
  {Pappalardo}, {Zibetti}, {Auld}, {Baes}, {Bendo}, {Corbelli}, {Davies},
  {Davis}, {De Looze}, {Fritz}, {Gavazzi}, {Giovanardi}, {Grossi}, {Hunt},
  {Magrini}, {Pierini}, \& {Xilouris}}]{alighieri13}
{di Serego Alighieri}, S., {Bianchi}, S., {Pappalardo}, C., {et~al.} 2013,
  \aap, 552, A8, \dodoi{10.1051/0004-6361/201220551}

\bibitem[{{Dickey} \& {Kazes}(1992)}]{dickey92}
{Dickey}, J.~M., \& {Kazes}, I. 1992, \apj, 393, 530, \dodoi{10.1086/171526}

\bibitem[{{Dickinson} {et~al.}(2002){Dickinson}, {Arribas}, {Bergeron},
  {Boeker}, {Calzetti}, {Holfeltz}, {Mobasher}, {Monroe}, {Noll}, {Roye},
  {Schultz}, {Sosey}, \& {Xu}}]{dickinson02}
{Dickinson}, M.~E., {Arribas}, S., {Bergeron}, L.~E., {et~al.} 2002, in HST
  NICMOS Data Handbook v. 5, Vol.~5, 300

\bibitem[{{Dominiak} {et~al.}(2024){Dominiak}, {Bureau}, {Davis}, {Ma},
  {Greene}, \& {Gu}}]{dom24}
{Dominiak}, P., {Bureau}, M., {Davis}, T.~A., {et~al.} 2024, \mnras,
  \dodoi{10.1093/mnras/stae314}

\bibitem[{{Dressel}(2022)}]{dressel22}
{Dressel}, L. 2022, in WFC3 Instrument Handbook for Cycle 30 v. 14, Vol.~14, 14

\bibitem[{{Duah Asabere} {et~al.}(2016){Duah Asabere}, {Horellou}, {Jarrett},
  \& {Winkler}}]{asabere16}
{Duah Asabere}, B., {Horellou}, C., {Jarrett}, T.~H., \& {Winkler}, H. 2016,
  \aap, 592, A20, \dodoi{10.1051/0004-6361/201528047}

\bibitem[{{Emonts} {et~al.}(2008){Emonts}, {Morganti}, {Oosterloo}, {Holt},
  {Tadhunter}, {van der Hulst}, {Ojha}, \& {Sadler}}]{emonts08}
{Emonts}, B.~H.~C., {Morganti}, R., {Oosterloo}, T.~A., {et~al.} 2008, \mnras,
  387, 197, \dodoi{10.1111/j.1365-2966.2008.13142.x}

\bibitem[{{Emsellem} {et~al.}(1994){Emsellem}, {Monnet}, \& {Bacon}}]{emsell94}
{Emsellem}, E., {Monnet}, G., \& {Bacon}, R. 1994, \aap, 285, 723

\bibitem[{{Faber} {et~al.}(1997){Faber}, {Tremaine}, {Ajhar}, {Byun},
  {Dressler}, {Gebhardt}, {Grillmair}, {Kormendy}, {Lauer}, \&
  {Richstone}}]{faber97}
{Faber}, S.~M., {Tremaine}, S., {Ajhar}, E.~A., {et~al.} 1997, \aj, 114, 1771,
  \dodoi{10.1086/118606}

\bibitem[{{Fanaroff} \& {Riley}(1974)}]{fr74}
{Fanaroff}, B.~L., \& {Riley}, J.~M. 1974, \mnras, 167, 31P,
  \dodoi{10.1093/mnras/167.1.31P}

\bibitem[{{Fazio} {et~al.}(2004){Fazio}, {Hora}, {Allen}, {Ashby}, {Barmby},
  {Deutsch}, {Huang}, {Kleiner}, {Marengo}, {Megeath}, {Melnick}, {Pahre},
  {Patten}, {Polizotti}, {Smith}, {Taylor}, {Wang}, {Willner}, {Hoffmann},
  {Pipher}, {Forrest}, {McMurty}, {McCreight}, {McKelvey}, {McMurray}, {Koch},
  {Moseley}, {Arendt}, {Mentzell}, {Marx}, {Losch}, {Mayman}, {Eichhorn},
  {Krebs}, {Jhabvala}, {Gezari}, {Fixsen}, {Flores}, {Shakoorzadeh}, {Jungo},
  {Hakun}, {Workman}, {Karpati}, {Kichak}, {Whitley}, {Mann}, {Tollestrup},
  {Eisenhardt}, {Stern}, {Gorjian}, {Bhattacharya}, {Carey}, {Nelson},
  {Glaccum}, {Lacy}, {Lowrance}, {Laine}, {Reach}, {Stauffer}, {Surace},
  {Wilson}, {Wright}, {Hoffman}, {Domingo}, \& {Cohen}}]{fazio04}
{Fazio}, G.~G., {Hora}, J.~L., {Allen}, L.~E., {et~al.} 2004, \apjs, 154, 10,
  \dodoi{10.1086/422843}

\bibitem[{{Ferrarese} {et~al.}(1996){Ferrarese}, {Ford}, \& {Jaffe}}]{ferr96}
{Ferrarese}, L., {Ford}, H.~C., \& {Jaffe}, W. 1996, \apj, 470, 444,
  \dodoi{10.1086/177876}

\bibitem[{{Ferrarese} \& {Merritt}(2000)}]{fermer00}
{Ferrarese}, L., \& {Merritt}, D. 2000, \apjl, 539, L9, \dodoi{10.1086/312838}

\bibitem[{{Gadotti} \& {de Souza}(2005)}]{gadotti05}
{Gadotti}, D.~A., \& {de Souza}, R.~E. 2005, \apj, 629, 797,
  \dodoi{10.1086/431717}

\bibitem[{{Garcia}(1993)}]{garcia93}
{Garcia}, A.~M. 1993, \aaps, 100, 47

\bibitem[{{Gebhardt} {et~al.}(2000){Gebhardt}, {Bender}, {Bower}, {Dressler},
  {Faber}, {Filippenko}, {Green}, {Grillmair}, {Ho}, {Kormendy}, {Lauer},
  {Magorrian}, {Pinkney}, {Richstone}, \& {Tremaine}}]{geb00}
{Gebhardt}, K., {Bender}, R., {Bower}, G., {et~al.} 2000, \apjl, 539, L13,
  \dodoi{10.1086/312840}

\bibitem[{{Gonzaga} {et~al.}(2012){Gonzaga}, {Hack}, {Fruchter}, \&
  {Mack}}]{astrodrizz}
{Gonzaga}, S., {Hack}, W., {Fruchter}, A., \& {Mack}, J. 2012, {The DrizzlePac
  Handbook}

\bibitem[{{Goullaud} {et~al.}(2018){Goullaud}, {Jensen}, {Blakeslee}, {Ma},
  {Greene}, \& {Thomas}}]{goullaud18}
{Goullaud}, C.~F., {Jensen}, J.~B., {Blakeslee}, J.~P., {et~al.} 2018, \apj,
  856, 11, \dodoi{10.3847/1538-4357/aab1f3}

\bibitem[{{Ho}(2007)}]{ho07}
{Ho}, L.~C. 2007, \apj, 669, 821, \dodoi{10.1086/521917}

\bibitem[{{Ho} {et~al.}(1997){Ho}, {Filippenko}, \& {Sargent}}]{ho97}
{Ho}, L.~C., {Filippenko}, A.~V., \& {Sargent}, W. L.~W. 1997, \apjs, 112, 315,
  \dodoi{10.1086/313041}

\bibitem[{{Iodice} {et~al.}(2019){Iodice}, {Spavone}, {Capaccioli}, {Peletier},
  {van de Ven}, {Napolitano}, {Hilker}, {Mieske}, {Smith}, {Pasquali},
  {Limatola}, {Grado}, {Venhola}, {Cantiello}, {Paolillo}, {Falcon-Barroso},
  {D'Abrusco}, \& {Schipani}}]{iodice19}
{Iodice}, E., {Spavone}, M., {Capaccioli}, M., {et~al.} 2019, \aap, 623, A1,
  \dodoi{10.1051/0004-6361/201833741}

\bibitem[{{Jensen} {et~al.}(2003){Jensen}, {Tonry}, {Barris}, {Thompson},
  {Liu}, {Rieke}, {Ajhar}, \& {Blakeslee}}]{jensen03}
{Jensen}, J.~B., {Tonry}, J.~L., {Barris}, B.~J., {et~al.} 2003, \apj, 583,
  712, \dodoi{10.1086/345430}

\bibitem[{{Kabasares} {et~al.}(2022){Kabasares}, {Barth}, {Buote}, {Boizelle},
  {Walsh}, {Baker}, {Darling}, {Ho}, \& {Cohn}}]{kabasares22}
{Kabasares}, K.~M., {Barth}, A.~J., {Buote}, D.~A., {et~al.} 2022, \apj, 934,
  162, \dodoi{10.3847/1538-4357/ac7a38}

\bibitem[{{Kabasares} {et~al.}(2024){Kabasares}, {Cohn}, {Barth}, {Boizelle},
  {Davidson}, {Sy}, {Flores-Vel{\'a}zquez}, {Delgado Andrade}, {Buote},
  {Walsh}, {Baker}, {Darling}, \& {Ho}}]{kabasares24}
{Kabasares}, K.~M., {Cohn}, J.~H., {Barth}, A.~J., {et~al.} 2024, \apj, 966,
  132, \dodoi{10.3847/1538-4357/ad2f36}

\bibitem[{{Kenworthy} {et~al.}(2022){Kenworthy}, {Riess}, {Scolnic}, {Yuan},
  {Bernal}, {Brout}, {Casertano}, {Jones}, {Macri}, \& {Peterson}}]{kenw22}
{Kenworthy}, W.~D., {Riess}, A.~G., {Scolnic}, D., {et~al.} 2022, \apj, 935,
  83, \dodoi{10.3847/1538-4357/ac80bd}

\bibitem[{{Kim} \& {Im}(2013)}]{kim13}
{Kim}, D., \& {Im}, M. 2013, \apj, 766, 109,
  \dodoi{10.1088/0004-637X/766/2/109}

\bibitem[{{Kormendy} \& {Ho}(2013)}]{korho13}
{Kormendy}, J., \& {Ho}, L.~C. 2013, \araa, 51, 511,
  \dodoi{10.1146/annurev-astro-082708-101811}

\bibitem[{{Kormendy} \& {Richstone}(1995)}]{korrich95}
{Kormendy}, J., \& {Richstone}, D. 1995, \araa, 33, 581,
  \dodoi{10.1146/annurev.aa.33.090195.003053}

\bibitem[{{Krajnovi{\'c}} {et~al.}(2006){Krajnovi{\'c}}, {Cappellari}, {de
  Zeeuw}, \& {Copin}}]{krajnovic06}
{Krajnovi{\'c}}, D., {Cappellari}, M., {de Zeeuw}, P.~T., \& {Copin}, Y. 2006,
  \mnras, 366, 787, \dodoi{10.1111/j.1365-2966.2005.09902.x}

\bibitem[{{Krajnovi{\'c}} {et~al.}(2011){Krajnovi{\'c}}, {Emsellem},
  {Cappellari}, {Alatalo}, {Blitz}, {Bois}, {Bournaud}, {Bureau}, {Davies},
  {Davis}, {de Zeeuw}, {Khochfar}, {Kuntschner}, {Lablanche}, {McDermid},
  {Morganti}, {Naab}, {Oosterloo}, {Sarzi}, {Scott}, {Serra}, {Weijmans}, \&
  {Young}}]{krajnovic11}
{Krajnovi{\'c}}, D., {Emsellem}, E., {Cappellari}, M., {et~al.} 2011, \mnras,
  414, 2923, \dodoi{10.1111/j.1365-2966.2011.18560.x}

\bibitem[{{Krajnovi{\'c}} {et~al.}(2013){Krajnovi{\'c}}, {Karick}, {Davies},
  {Naab}, {Sarzi}, {Emsellem}, {Cappellari}, {Serra}, {de Zeeuw}, {Scott},
  {McDermid}, {Weijmans}, {Davis}, {Alatalo}, {Blitz}, {Bois}, {Bureau},
  {Bournaud}, {Crocker}, {Duc}, {Khochfar}, {Kuntschner}, {Morganti},
  {Oosterloo}, \& {Young}}]{krajnovic13}
{Krajnovi{\'c}}, D., {Karick}, A.~M., {Davies}, R.~L., {et~al.} 2013, \mnras,
  433, 2812, \dodoi{10.1093/mnras/stt905}

\bibitem[{{Krist} \& {Hook}(2004)}]{tinytim}
{Krist}, J., \& {Hook}, R. 2004, The Tiny Tim User’s Guide,
  \url{https://www.stsci.edu/files/live/sites/www/files/home/hst/instrumentation/focus-and-pointing/documentation/_documents/tinytim.pdf},
  Baltimore: STScI

\bibitem[{{La Barbera} {et~al.}(2010){La Barbera}, {De Carvalho}, {De La Rosa},
  {Gal}, {Swindle}, \& {Lopes}}]{labarb10}
{La Barbera}, F., {De Carvalho}, R.~R., {De La Rosa}, I.~G., {et~al.} 2010,
  \aj, 140, 1528, \dodoi{10.1088/0004-6256/140/5/1528}

\bibitem[{{Lauer} {et~al.}(1995){Lauer}, {Ajhar}, {Byun}, {Dressler}, {Faber},
  {Grillmair}, {Kormendy}, {Richstone}, \& {Tremaine}}]{lauer95}
{Lauer}, T.~R., {Ajhar}, E.~A., {Byun}, Y.~I., {et~al.} 1995, \aj, 110, 2622,
  \dodoi{10.1086/117719}

\bibitem[{{Lauer} {et~al.}(2005){Lauer}, {Faber}, {Gebhardt}, {Richstone},
  {Tremaine}, {Ajhar}, {Aller}, {Bender}, {Dressler}, {Filippenko}, {Green},
  {Grillmair}, {Ho}, {Kormendy}, {Magorrian}, {Pinkney}, \& {Siopis}}]{lauer05}
{Lauer}, T.~R., {Faber}, S.~M., {Gebhardt}, K., {et~al.} 2005, \aj, 129, 2138,
  \dodoi{10.1086/429565}

\bibitem[{{Lauer} {et~al.}(2007){Lauer}, {Faber}, {Richstone}, {Gebhardt},
  {Tremaine}, {Postman}, {Dressler}, {Aller}, {Filippenko}, {Green}, {Ho},
  {Kormendy}, {Magorrian}, \& {Pinkney}}]{lauer07}
{Lauer}, T.~R., {Faber}, S.~M., {Richstone}, D., {et~al.} 2007, \apj, 662, 808,
  \dodoi{10.1086/518223}

\bibitem[{{Liepold} {et~al.}(2023){Liepold}, {Ma}, \& {Walsh}}]{liepold23}
{Liepold}, E.~R., {Ma}, C.-P., \& {Walsh}, J.~L. 2023, \apjl, 945, L35,
  \dodoi{10.3847/2041-8213/acbbcf}

\bibitem[{{Ma} {et~al.}(2014){Ma}, {Greene}, {McConnell}, {Janish},
  {Blakeslee}, {Thomas}, \& {Murphy}}]{ma14}
{Ma}, C.-P., {Greene}, J.~E., {McConnell}, N., {et~al.} 2014, \apj, 795, 158,
  \dodoi{10.1088/0004-637X/795/2/158}

\bibitem[{{Maiolino}(2008)}]{maio08}
{Maiolino}, R. 2008, \nar, 52, 339, \dodoi{10.1016/j.newar.2008.06.012}

\bibitem[{{McConnell} \& {Ma}(2013)}]{mcma13}
{McConnell}, N.~J., \& {Ma}, C.-P. 2013, \apj, 764, 184,
  \dodoi{10.1088/0004-637X/764/2/184}

\bibitem[{{McMaster} \& {et al.}(2008)}]{mcmaster08}
{McMaster}, M., \& {et al.} 2008, {Wide Field and Planetary Camera 2 Instrument
  Handbook v. 10.0}

\bibitem[{{Mei} {et~al.}(2007){Mei}, {Blakeslee}, {C{\^o}t{\'e}}, {Tonry},
  {West}, {Ferrarese}, {Jord{\'a}n}, {Peng}, {Anthony}, \& {Merritt}}]{mei07}
{Mei}, S., {Blakeslee}, J.~P., {C{\^o}t{\'e}}, P., {et~al.} 2007, \apj, 655,
  144, \dodoi{10.1086/509598}

\bibitem[{{Mould} {et~al.}(2000){Mould}, {Huchra}, {Freedman}, {Kennicutt},
  {Ferrarese}, {Ford}, {Gibson}, {Graham}, {Hughes}, {Illingworth}, {Kelson},
  {Macri}, {Madore}, {Sakai}, {Sebo}, {Silbermann}, \& {Stetson}}]{mould00}
{Mould}, J.~R., {Huchra}, J.~P., {Freedman}, W.~L., {et~al.} 2000, \apj, 529,
  786, \dodoi{10.1086/308304}

\bibitem[{{North} {et~al.}(2019){North}, {Davis}, {Bureau}, {Cappellari},
  {Iguchi}, {Liu}, {Onishi}, {Sarzi}, {Smith}, \& {Williams}}]{north19}
{North}, E.~V., {Davis}, T.~A., {Bureau}, M., {et~al.} 2019, \mnras, 490, 319,
  \dodoi{10.1093/mnras/stz2598}

\bibitem[{{Nyland} {et~al.}(2016){Nyland}, {Young}, {Wrobel}, {Sarzi},
  {Morganti}, {Alatalo}, {Blitz}, {Bournaud}, {Bureau}, {Cappellari},
  {Crocker}, {Davies}, {Davis}, {de Zeeuw}, {Duc}, {Emsellem}, {Khochfar},
  {Krajnovi{\'c}}, {Kuntschner}, {McDermid}, {Naab}, {Oosterloo}, {Scott},
  {Serra}, \& {Weijmans}}]{nyland16}
{Nyland}, K., {Young}, L.~M., {Wrobel}, J.~M., {et~al.} 2016, \mnras, 458,
  2221, \dodoi{10.1093/mnras/stw391}

\bibitem[{{Onishi} {et~al.}(2017){Onishi}, {Iguchi}, {Davis}, {Bureau},
  {Cappellari}, {Sarzi}, \& {Blitz}}]{onishi17}
{Onishi}, K., {Iguchi}, S., {Davis}, T.~A., {et~al.} 2017, \mnras, 468, 4663,
  \dodoi{10.1093/mnras/stx631}

\bibitem[{{Paturel} {et~al.}(2003){Paturel}, {Petit}, {Prugniel}, {Theureau},
  {Rousseau}, {Brouty}, {Dubois}, \& {Cambr{\'e}sy}}]{paturel03}
{Paturel}, G., {Petit}, C., {Prugniel}, P., {et~al.} 2003, \aap, 412, 45,
  \dodoi{10.1051/0004-6361:20031411}

\bibitem[{{Peng} {et~al.}(2010){Peng}, {Ho}, {Impey}, \& {Rix}}]{peng10}
{Peng}, C.~Y., {Ho}, L.~C., {Impey}, C.~D., \& {Rix}, H.-W. 2010, \aj, 139,
  2097, \dodoi{10.1088/0004-6256/139/6/2097}

\bibitem[{{Pirzkal}(2014)}]{pirzkal14}
{Pirzkal}, N. 2014, {The Near Infrared Sky Background}, Instrument Science
  Report WFC3 2014-11, 11 pages

\bibitem[{{Ravindranath} {et~al.}(2001){Ravindranath}, {Ho}, {Peng},
  {Filippenko}, \& {Sargent}}]{ravindranath01}
{Ravindranath}, S., {Ho}, L.~C., {Peng}, C.~Y., {Filippenko}, A.~V., \&
  {Sargent}, W. L.~W. 2001, arXiv e-prints, astro,
  \dodoi{10.48550/arXiv.astro-ph/0102505}

\bibitem[{{Rest} {et~al.}(2001){Rest}, {van den Bosch}, {Jaffe}, {Tran},
  {Tsvetanov}, {Ford}, {Davies}, \& {Schafer}}]{rest01}
{Rest}, A., {van den Bosch}, F.~C., {Jaffe}, W., {et~al.} 2001, \aj, 121, 2431,
  \dodoi{10.1086/320370}

\bibitem[{{Riess} {et~al.}(2022){Riess}, {Yuan}, {Macri}, {Scolnic}, {Brout},
  {Casertano}, {Jones}, {Murakami}, {Anand}, {Breuval}, {Brink}, {Filippenko},
  {Hoffmann}, {Jha}, {D'arcy Kenworthy}, {Mackenty}, {Stahl}, \&
  {Zheng}}]{riess22}
{Riess}, A.~G., {Yuan}, W., {Macri}, L.~M., {et~al.} 2022, \apjl, 934, L7,
  \dodoi{10.3847/2041-8213/ac5c5b}

\bibitem[{{Rose} {et~al.}(2019){Rose}, {Edge}, {Combes}, {Gaspari}, {Hamer},
  {Nesvadba}, {Russell}, {Tremblay}, {Baum}, {O'Dea}, {Peck}, {Sarazin},
  {Vantyghem}, {Bremer}, {Donahue}, {Fabian}, {Ferland}, {McNamara}, {Mittal},
  {Oonk}, {Salom{\'e}}, {Swinbank}, \& {Voit}}]{rose19}
{Rose}, T., {Edge}, A.~C., {Combes}, F., {et~al.} 2019, \mnras, 485, 229,
  \dodoi{10.1093/mnras/stz406}

\bibitem[{{Ruffa} {et~al.}(2020){Ruffa}, {Laing}, {Prandoni}, {Paladino},
  {Parma}, {Davis}, \& {Bureau}}]{ruffa20}
{Ruffa}, I., {Laing}, R.~A., {Prandoni}, I., {et~al.} 2020, \mnras, 499, 5719,
  \dodoi{10.1093/mnras/staa3166}

\bibitem[{{Ruffa} {et~al.}(2019{\natexlab{a}}){Ruffa}, {Prandoni}, {Laing},
  {Paladino}, {Parma}, {de Ruiter}, {Mignano}, {Davis}, {Bureau}, \&
  {Warren}}]{ruffa19a}
{Ruffa}, I., {Prandoni}, I., {Laing}, R.~A., {et~al.} 2019{\natexlab{a}},
  \mnras, 484, 4239, \dodoi{10.1093/mnras/stz255}

\bibitem[{{Ruffa} {et~al.}(2019{\natexlab{b}}){Ruffa}, {Davis}, {Prandoni},
  {Laing}, {Paladino}, {Parma}, {de Ruiter}, {Casasola}, {Bureau}, \&
  {Warren}}]{ruffa19b}
{Ruffa}, I., {Davis}, T.~A., {Prandoni}, I., {et~al.} 2019{\natexlab{b}},
  \mnras, 489, 3739, \dodoi{10.1093/mnras/stz2368}

\bibitem[{{Ruffa} {et~al.}(2023){Ruffa}, {Davis}, {Cappellari}, {Bureau},
  {Elford}, {Iguchi}, {Lelli}, {Liang}, {Liu}, {Lu}, {Sarzi}, \&
  {Williams}}]{ruffa23}
{Ruffa}, I., {Davis}, T.~A., {Cappellari}, M., {et~al.} 2023, \mnras, 522,
  6170, \dodoi{10.1093/mnras/stad1119}

\bibitem[{{Rusli} {et~al.}(2013){Rusli}, {Thomas}, {Saglia}, {Fabricius},
  {Erwin}, {Bender}, {Nowak}, {Lee}, {Riffeser}, \& {Sharp}}]{rusli13}
{Rusli}, S.~P., {Thomas}, J., {Saglia}, R.~P., {et~al.} 2013, \aj, 146, 45,
  \dodoi{10.1088/0004-6256/146/3/45}

\bibitem[{{Ryon}(2022)}]{ryon22}
{Ryon}, J.~E. 2022, in ACS Instrument Handbook for Cycle 30 v. 21.0, Vol.~21,
  21

\bibitem[{{Saglia} {et~al.}(2016){Saglia}, {Opitsch}, {Erwin}, {Thomas},
  {Beifiori}, {Fabricius}, {Mazzalay}, {Nowak}, {Rusli}, \& {Bender}}]{sagl16}
{Saglia}, R.~P., {Opitsch}, M., {Erwin}, P., {et~al.} 2016, \apj, 818, 47,
  \dodoi{10.3847/0004-637X/818/1/47}

\bibitem[{{Sansom} {et~al.}(2019){Sansom}, {Glass}, {Bendo}, {Davis},
  {Rowlands}, {Bourne}, {Dunne}, {Eales}, {Kaviraj}, {Popescu}, {Smith}, \&
  {Viaene}}]{sansom19}
{Sansom}, A.~E., {Glass}, D.~H.~W., {Bendo}, G.~J., {et~al.} 2019, \mnras, 482,
  4617, \dodoi{10.1093/mnras/sty3021}

\bibitem[{{Saracco} {et~al.}(2012){Saracco}, {Gargiulo}, \&
  {Longhetti}}]{saracco12}
{Saracco}, P., {Gargiulo}, A., \& {Longhetti}, M. 2012, \mnras, 422, 3107,
  \dodoi{10.1111/j.1365-2966.2012.20830.x}

\bibitem[{{Sato} {et~al.}(2012){Sato}, {Sasaki}, {Matsushita}, {Sakuma},
  {Sato}, {Fujita}, {Okabe}, {Fukazawa}, {Ichikawa}, {Kawaharada}, {Nakazawa},
  {Ohashi}, {Ota}, {Takizawa}, \& {Tamura}}]{sato12}
{Sato}, T., {Sasaki}, T., {Matsushita}, K., {et~al.} 2012, \pasj, 64, 95,
  \dodoi{10.1093/pasj/64.5.95}

\bibitem[{{Science Software Branch at STScI}(2012)}]{pyraf}
{Science Software Branch at STScI}. 2012, {PyRAF: Python alternative for IRAF},
  Astrophysics Source Code Library, record ascl:1207.011.
\newblock \doeprint{1207.011}

\bibitem[{{Serra} {et~al.}(2012){Serra}, {Oosterloo}, {Morganti}, {Alatalo},
  {Blitz}, {Bois}, {Bournaud}, {Bureau}, {Cappellari}, {Crocker}, {Davies},
  {Davis}, {de Zeeuw}, {Duc}, {Emsellem}, {Khochfar}, {Krajnovi{\'c}},
  {Kuntschner}, {Lablanche}, {McDermid}, {Naab}, {Sarzi}, {Scott}, {Trager},
  {Weijmans}, \& {Young}}]{serra12}
{Serra}, P., {Oosterloo}, T., {Morganti}, R., {et~al.} 2012, \mnras, 422, 1835,
  \dodoi{10.1111/j.1365-2966.2012.20219.x}

\bibitem[{{Smith} {et~al.}(2019){Smith}, {Bureau}, {Davis}, {Cappellari},
  {Liu}, {North}, {Onishi}, {Iguchi}, \& {Sarzi}}]{smith19}
{Smith}, M.~D., {Bureau}, M., {Davis}, T.~A., {et~al.} 2019, \mnras, 485, 4359,
  \dodoi{10.1093/mnras/stz625}

\bibitem[{{Smith} {et~al.}(2021{\natexlab{a}}){Smith}, {Bureau}, {Davis},
  {Cappellari}, {Liu}, {Onishi}, {Iguchi}, {North}, {Sarzi}, \&
  {Williams}}]{smith21}
---. 2021{\natexlab{a}}, \mnras, 503, 5984, \dodoi{10.1093/mnras/stab791}

\bibitem[{{Smith} {et~al.}(2021{\natexlab{b}}){Smith}, {Bureau}, {Davis},
  {Cappellari}, {Liu}, {Onishi}, {Iguchi}, {North}, \& {Sarzi}}]{smith21b}
---. 2021{\natexlab{b}}, \mnras, 500, 1933, \dodoi{10.1093/mnras/staa3274}

\bibitem[{{Sofue}(1992)}]{sofue92}
{Sofue}, Y. 1992, \pasj, 44, L231

\bibitem[{{Spiniello} {et~al.}(2015){Spiniello}, {Napolitano}, {Coccato},
  {Pota}, {Romanowsky}, {Tortora}, {Covone}, \& {Capaccioli}}]{spiniello15}
{Spiniello}, C., {Napolitano}, N.~R., {Coccato}, L., {et~al.} 2015, \mnras,
  452, 99, \dodoi{10.1093/mnras/stv1260}

\bibitem[{{Tamura} \& {Ohta}(2003)}]{tamura03}
{Tamura}, N., \& {Ohta}, K. 2003, \aj, 126, 596, \dodoi{10.1086/376469}

\bibitem[{{Thater} {et~al.}(2022){Thater}, {Krajnovi{\'c}}, {Weilbacher},
  {Nguyen}, {Bureau}, {Cappellari}, {Davis}, {Iguchi}, {McDermid}, {Onishi},
  {Sarzi}, \& {van de Ven}}]{thater22}
{Thater}, S., {Krajnovi{\'c}}, D., {Weilbacher}, P.~M., {et~al.} 2022, \mnras,
  509, 5416, \dodoi{10.1093/mnras/stab3210}

\bibitem[{{Tiley} {et~al.}(2016){Tiley}, {Bureau}, {Saintonge}, {Topal},
  {Davis}, \& {Torii}}]{tiley16}
{Tiley}, A.~L., {Bureau}, M., {Saintonge}, A., {et~al.} 2016, \mnras, 461,
  3494, \dodoi{10.1093/mnras/stw1545}

\bibitem[{{Tiley} {et~al.}(2019){Tiley}, {Bureau}, {Cortese}, {Harrison},
  {Johnson}, {Stott}, {Swinbank}, {Smail}, {Sobral}, {Bunker}, {Glazebrook},
  {Bower}, {Obreschkow}, {Bryant}, {Jarvis}, {Bland-Hawthorn}, {Magdis},
  {Medling}, {Sweet}, {Tonini}, {Turner}, {Sharples}, {Croom}, {Goodwin},
  {Konstantopoulos}, {Lorente}, {Lawrence}, {Mould}, {Owers}, \&
  {Richards}}]{tiley19}
{Tiley}, A.~L., {Bureau}, M., {Cortese}, L., {et~al.} 2019, \mnras, 482, 2166,
  \dodoi{10.1093/mnras/sty2794}

\bibitem[{{Tody}(1986)}]{tody86}
{Tody}, D. 1986, in Society of Photo-Optical Instrumentation Engineers (SPIE)
  Conference Series, Vol. 627, Instrumentation in astronomy VI, ed. D.~L.
  {Crawford}, 733, \dodoi{10.1117/12.968154}

\bibitem[{{Tody}(1993)}]{tody93}
{Tody}, D. 1993, in Astronomical Society of the Pacific Conference Series,
  Vol.~52, Astronomical Data Analysis Software and Systems II, ed. R.~J.
  {Hanisch}, R.~J.~V. {Brissenden}, \& J.~{Barnes}, 173

\bibitem[{{Tonry} {et~al.}(2001){Tonry}, {Dressler}, {Blakeslee}, {Ajhar},
  {Fletcher}, {Luppino}, {Metzger}, \& {Moore}}]{tonry01}
{Tonry}, J.~L., {Dressler}, A., {Blakeslee}, J.~P., {et~al.} 2001, \apj, 546,
  681, \dodoi{10.1086/318301}

\bibitem[{{Topal} {et~al.}(2018){Topal}, {Bureau}, {Tiley}, {Davis}, \&
  {Torii}}]{topal18}
{Topal}, S., {Bureau}, M., {Tiley}, A.~L., {Davis}, T.~A., \& {Torii}, K. 2018,
  \mnras, 479, 3319, \dodoi{10.1093/mnras/sty1617}

\bibitem[{{Tran} {et~al.}(2001){Tran}, {Tsvetanov}, {Ford}, {Davies}, {Jaffe},
  {van den Bosch}, \& {Rest}}]{tran01}
{Tran}, H.~D., {Tsvetanov}, Z., {Ford}, H.~C., {et~al.} 2001, \aj, 121, 2928,
  \dodoi{10.1086/321072}

\bibitem[{{Tully} \& {Fisher}(1977)}]{tulfish77}
{Tully}, R.~B., \& {Fisher}, J.~R. 1977, \aap, 54, 661

\bibitem[{{van de Voort} {et~al.}(2015){van de Voort}, {Davis}, {Kere{\v{s}}},
  {Quataert}, {Faucher-Gigu{\`e}re}, \& {Hopkins}}]{voort15}
{van de Voort}, F., {Davis}, T.~A., {Kere{\v{s}}}, D., {et~al.} 2015, \mnras,
  451, 3269, \dodoi{10.1093/mnras/stv1217}

\bibitem[{{van de Voort} {et~al.}(2018){van de Voort}, {Davis}, {Matsushita},
  {Rowlands}, {Shabala}, {Allison}, {Ting}, {Sansom}, \& {van der
  Werf}}]{voort18}
{van de Voort}, F., {Davis}, T.~A., {Matsushita}, S., {et~al.} 2018, \mnras,
  476, 122, \dodoi{10.1093/mnras/sty228}

\bibitem[{{van den Bosch} \& {de Zeeuw}(2010)}]{vandb10}
{van den Bosch}, R. C.~E., \& {de Zeeuw}, P.~T. 2010, \mnras, 401, 1770,
  \dodoi{10.1111/j.1365-2966.2009.15832.x}

\bibitem[{{van den Bosch} {et~al.}(2015){van den Bosch}, {Gebhardt},
  {G{\"u}ltekin}, {Y{\i}ld{\i}r{\i}m}, \& {Walsh}}]{vandb15}
{van den Bosch}, R. C.~E., {Gebhardt}, K., {G{\"u}ltekin}, K.,
  {Y{\i}ld{\i}r{\i}m}, A., \& {Walsh}, J.~L. 2015, \apjs, 218, 10,
  \dodoi{10.1088/0067-0049/218/1/10}

\bibitem[{{van den Bosch} {et~al.}(2008){van den Bosch}, {van de Ven},
  {Verolme}, {Cappellari}, \& {de Zeeuw}}]{vandb08}
{van den Bosch}, R.~C.~E., {van de Ven}, G., {Verolme}, E.~K., {Cappellari},
  M., \& {de Zeeuw}, P.~T. 2008, \mnras, 385, 647,
  \dodoi{10.1111/j.1365-2966.2008.12874.x}

\bibitem[{{Vazdekis} {et~al.}(2016){Vazdekis}, {Koleva}, {Ricciardelli},
  {R{\"o}ck}, \& {Falc{\'o}n-Barroso}}]{vazdekis16}
{Vazdekis}, A., {Koleva}, M., {Ricciardelli}, E., {R{\"o}ck}, B., \&
  {Falc{\'o}n-Barroso}, J. 2016, \mnras, 463, 3409,
  \dodoi{10.1093/mnras/stw2231}

\bibitem[{{Vazdekis} {et~al.}(2012){Vazdekis}, {Ricciardelli}, {Cenarro},
  {Rivero-Gonz{\'a}lez}, {D{\'\i}az-Garc{\'\i}a}, \&
  {Falc{\'o}n-Barroso}}]{vazdekis12}
{Vazdekis}, A., {Ricciardelli}, E., {Cenarro}, A.~J., {et~al.} 2012, \mnras,
  424, 157, \dodoi{10.1111/j.1365-2966.2012.21179.x}

\bibitem[{{Vazdekis} {et~al.}(2010){Vazdekis}, {S{\'a}nchez-Bl{\'a}zquez},
  {Falc{\'o}n-Barroso}, {Cenarro}, {Beasley}, {Cardiel}, {Gorgas}, \&
  {Peletier}}]{vazdekis10}
{Vazdekis}, A., {S{\'a}nchez-Bl{\'a}zquez}, P., {Falc{\'o}n-Barroso}, J.,
  {et~al.} 2010, \mnras, 404, 1639, \dodoi{10.1111/j.1365-2966.2010.16407.x}

\bibitem[{{Viaene} {et~al.}(2017){Viaene}, {Sarzi}, {Baes}, {Fritz}, \&
  {Puerari}}]{viaene17}
{Viaene}, S., {Sarzi}, M., {Baes}, M., {Fritz}, J., \& {Puerari}, I. 2017,
  \mnras, 472, 1286, \dodoi{10.1093/mnras/stx1781}

\bibitem[{{Walsh} {et~al.}(2008){Walsh}, {Barth}, {Ho}, {Filippenko}, {Rix},
  {Shields}, {Sarzi}, \& {Sargent}}]{walsh08}
{Walsh}, J.~L., {Barth}, A.~J., {Ho}, L.~C., {et~al.} 2008, \aj, 136, 1677,
  \dodoi{10.1088/0004-6256/136/4/1677}

\bibitem[{{Wright}(2006)}]{cosmocalc}
{Wright}, E.~L. 2006, \pasp, 118, 1711, \dodoi{10.1086/510102}

\bibitem[{{Y{\i}ld{\i}r{\i}m} {et~al.}(2017){Y{\i}ld{\i}r{\i}m}, {van den
  Bosch}, {van de Ven}, {Mart{\'\i}n-Navarro}, {Walsh}, {Husemann},
  {G{\"u}ltekin}, \& {Gebhardt}}]{yildirim17}
{Y{\i}ld{\i}r{\i}m}, A., {van den Bosch}, R. C.~E., {van de Ven}, G., {et~al.}
  2017, \mnras, 468, 4216, \dodoi{10.1093/mnras/stx732}

\bibitem[{{Yoon}(2017)}]{yoon17}
{Yoon}, I. 2017, \mnras, 466, 1987, \dodoi{10.1093/mnras/stw3171}

\bibitem[{{Young} {et~al.}(2011){Young}, {Bureau}, {Davis}, {Combes},
  {McDermid}, {Alatalo}, {Blitz}, {Bois}, {Bournaud}, {Cappellari}, {Davies},
  {de Zeeuw}, {Emsellem}, {Khochfar}, {Krajnovi{\'c}}, {Kuntschner},
  {Lablanche}, {Morganti}, {Naab}, {Oosterloo}, {Sarzi}, {Scott}, {Serra}, \&
  {Weijmans}}]{young11}
{Young}, L.~M., {Bureau}, M., {Davis}, T.~A., {et~al.} 2011, \mnras, 414, 940,
  \dodoi{10.1111/j.1365-2966.2011.18561.x}

\bibitem[{{Zabel} {et~al.}(2019){Zabel}, {Davis}, {Smith}, {Maddox}, {Bendo},
  {Peletier}, {Iodice}, {Venhola}, {Baes}, {Davies}, {de Looze}, {Gomez},
  {Grossi}, {Kenney}, {Serra}, {van de Voort}, {Vlahakis}, \&
  {Young}}]{zabel19}
{Zabel}, N., {Davis}, T.~A., {Smith}, M. W.~L., {et~al.} 2019, \mnras, 483,
  2251, \dodoi{10.1093/mnras/sty3234}

\bibitem[{{Zabludoff} \& {Mulchaey}(2000)}]{zabludoff00}
{Zabludoff}, A.~I., \& {Mulchaey}, J.~S. 2000, \apj, 539, 136,
  \dodoi{10.1086/309191}

\bibitem[{{Zhao} {et~al.}(2021){Zhao}, {Ho}, {Shangguan}, {Kim}, {Zhao}, \&
  {Gao}}]{zhao21}
{Zhao}, Y., {Ho}, L.~C., {Shangguan}, J., {et~al.} 2021, \apj, 911, 94,
  \dodoi{10.3847/1538-4357/abe8d4}

\end{thebibliography}

\pagebreak

\appendix

\section{Surface Brightness Profiles and Isophotal Analysis Results}
\label{app:sb_isophotal}

\begin{figure*}[!h]
\includegraphics[width=0.9\textwidth]{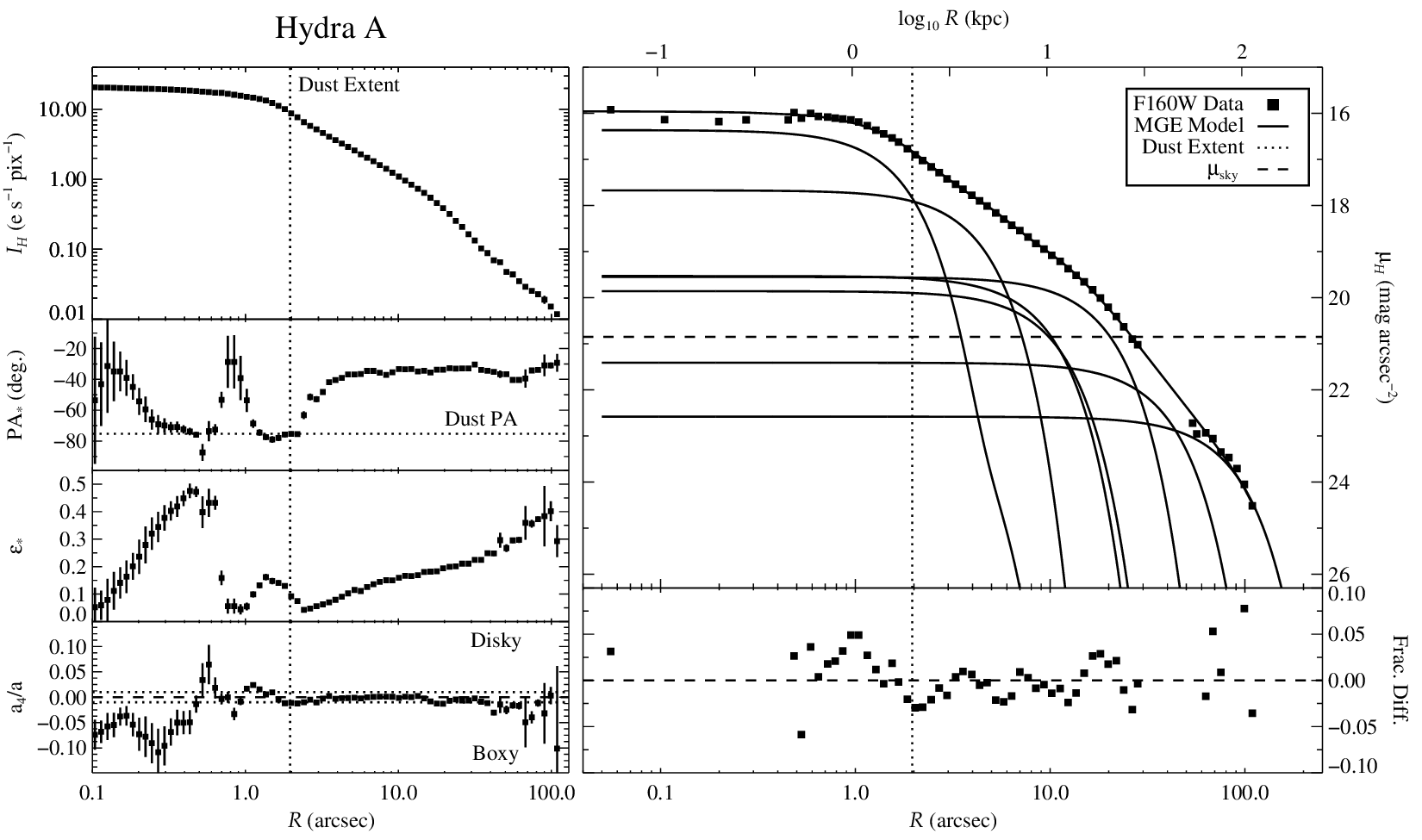}
\centering
\caption{Isophotal analysis (\textit{left panels}) and MGE (shown along the major axis; \textit{right panels}) for Hydra A after masking the most dust-obscured regions of the CND. Each panel gives the dust disk extent (vertical dotted line), beyond which the fitted PA$_\star$ and ellipticity $\varepsilon_\star$ from the \texttt{ellipse} task are reliable and show generally smooth variations. The CND PA$_{\mathrm{dust},H}$ is also shown for reference (horizontal dotted line). The $a_4 / a$ deviations from a perfect ellipse are likewise generally small. The full MGE shows generally good agreement (\textit{lower right panel}) to the major-axis surface brightness measurements.}
\label{fig:hydraa_ell_mge}
\end{figure*}

\begin{figure*}
\includegraphics[width=0.9\textwidth]{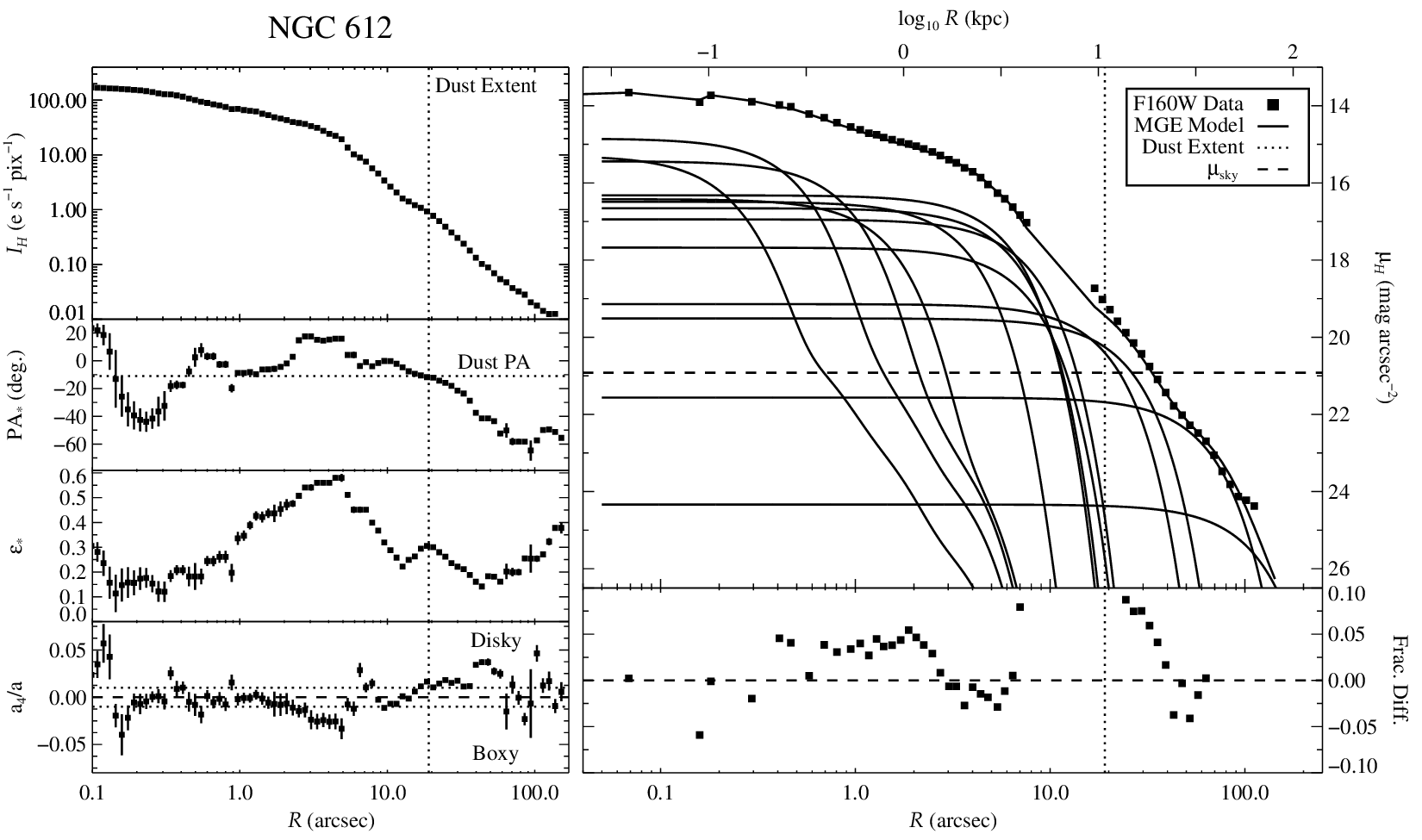}
\centering
\caption{Same as for Figure~\ref{fig:hydraa_ell_mge} but for NGC 612.}
\label{fig:ngc612_ell_mge}
\end{figure*}

\begin{figure*}
\includegraphics[width=0.9\textwidth]{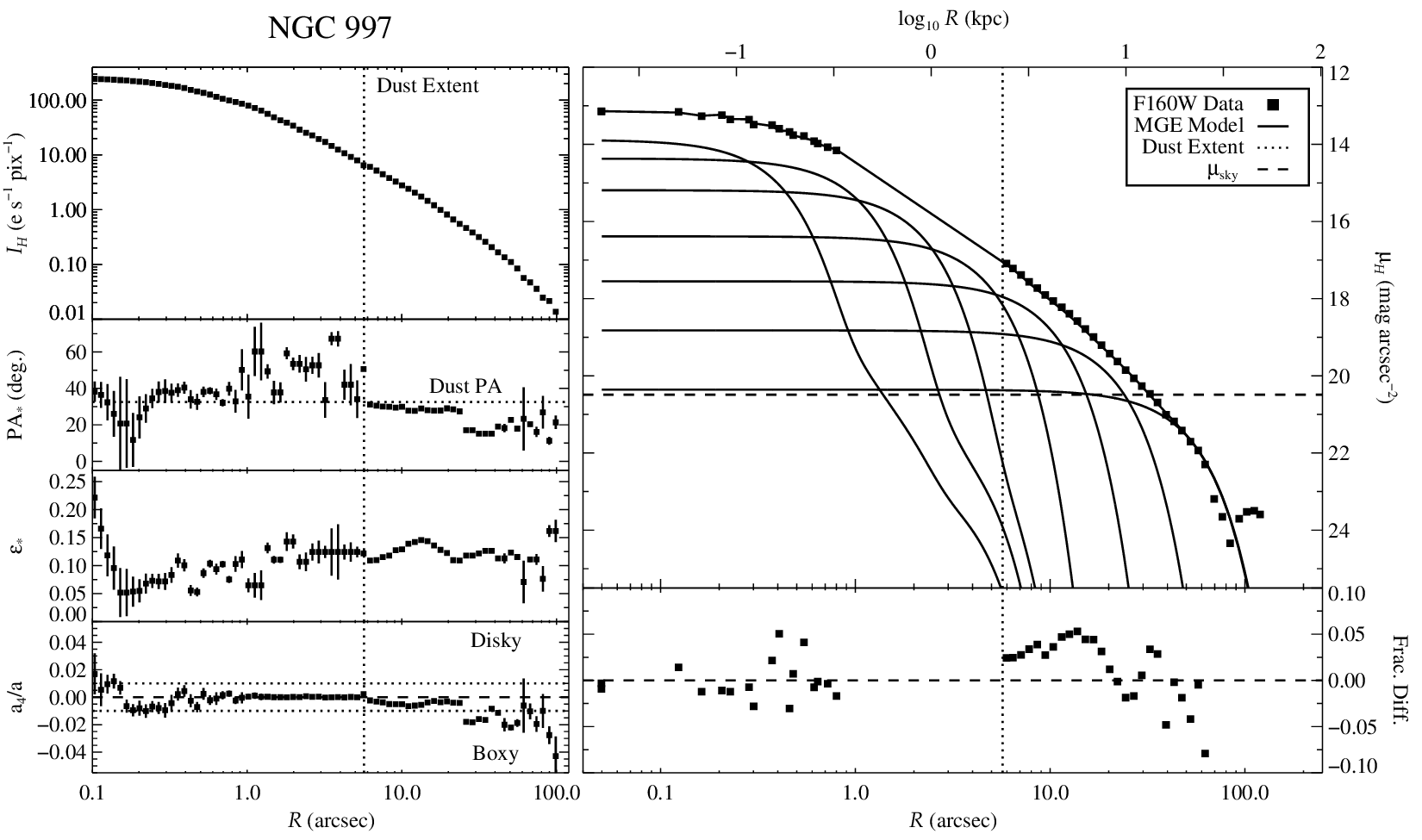}
\centering
\caption{Same as for Figure~\ref{fig:hydraa_ell_mge} but for NGC 997.}
\label{fig:ngc997_ell_mge}
\end{figure*}

\begin{figure*}
\includegraphics[width=0.9\textwidth]{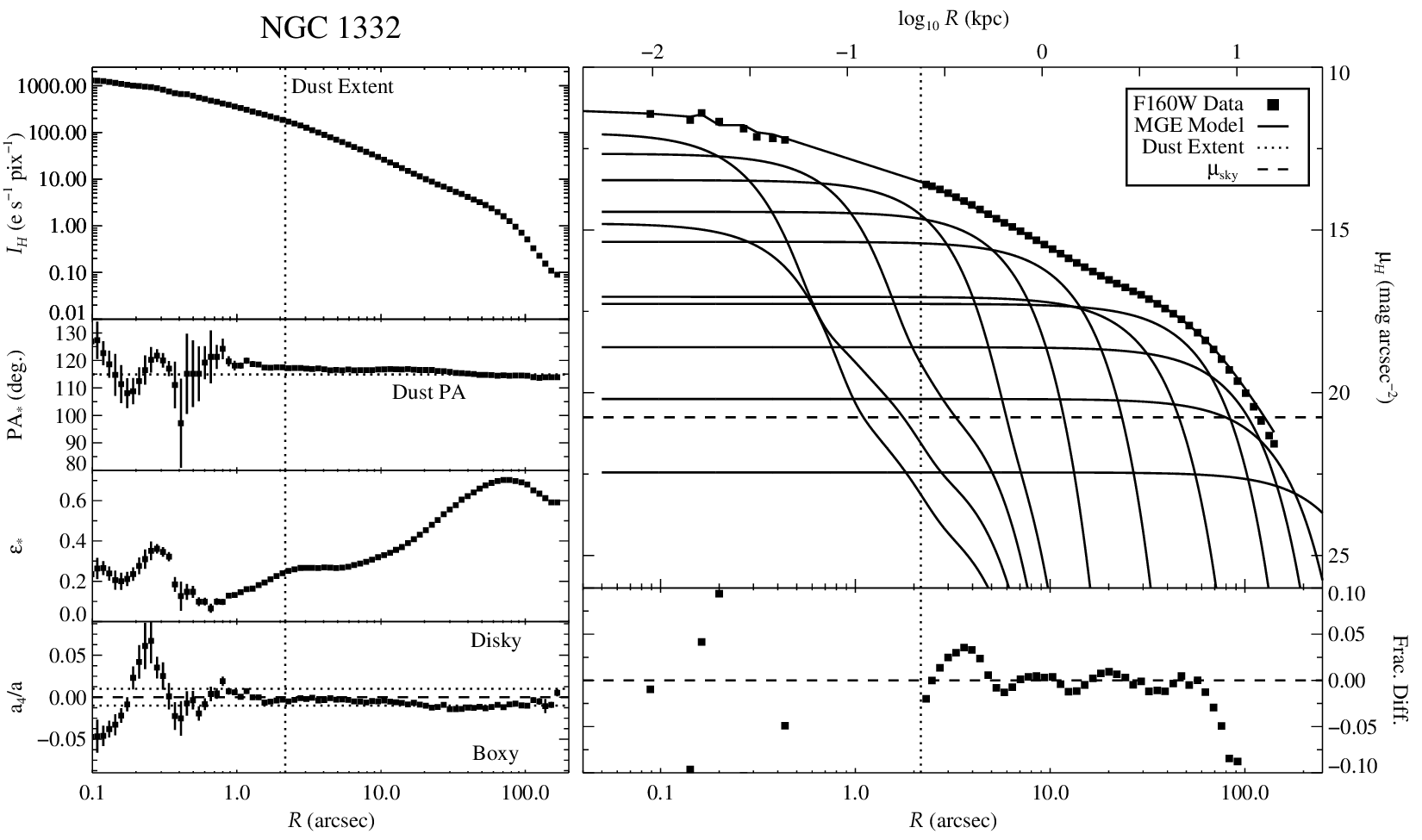}
\centering
\caption{Same as for Figure~\ref{fig:hydraa_ell_mge} but for NGC 1332.}
\label{fig:ngc1332_ell_mge}
\end{figure*}

\begin{figure*}
\includegraphics[width=0.9\textwidth]{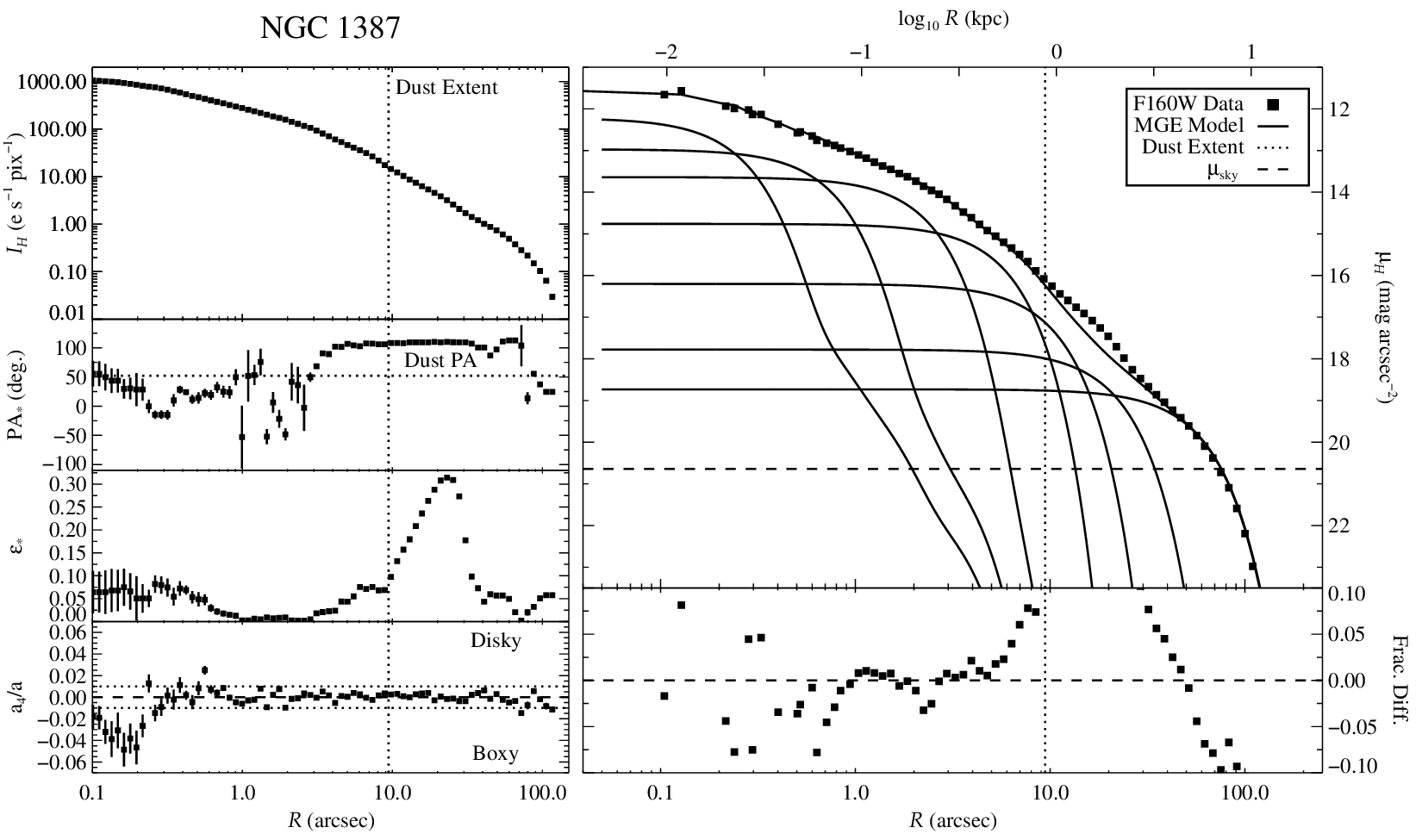}
\centering
\caption{Same as for Figure~\ref{fig:hydraa_ell_mge} but for NGC 1387.}
\label{fig:ngc1387_ell_mge}
\end{figure*}

\begin{figure*}
    \includegraphics[width=0.9\textwidth]{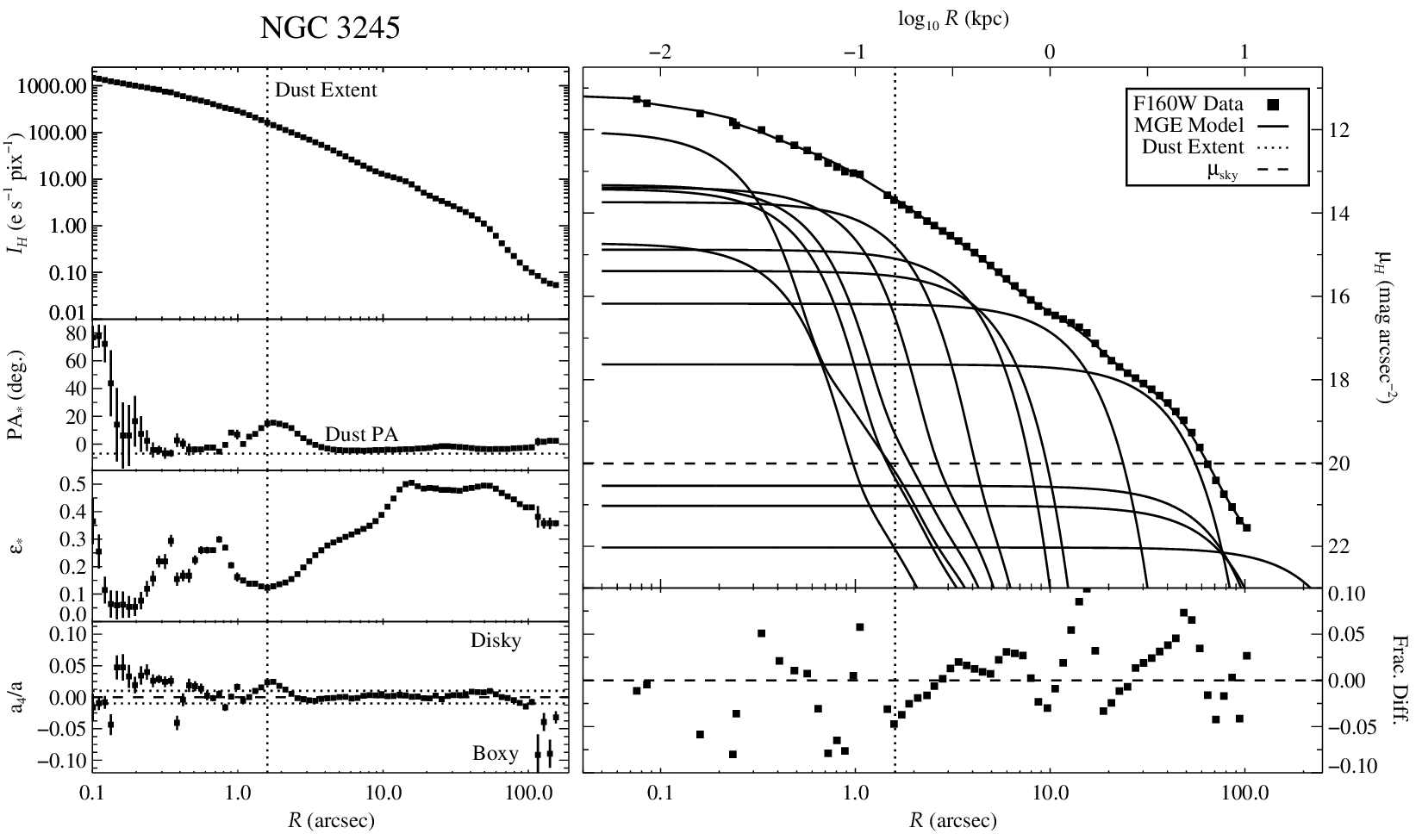}
    \centering
    \caption{Same as for Figure~\ref{fig:hydraa_ell_mge} but for NGC 3245.}
    \label{fig:ngc3245_ell_mge}
\end{figure*}

\begin{figure*}
    \includegraphics[width=0.9\textwidth]{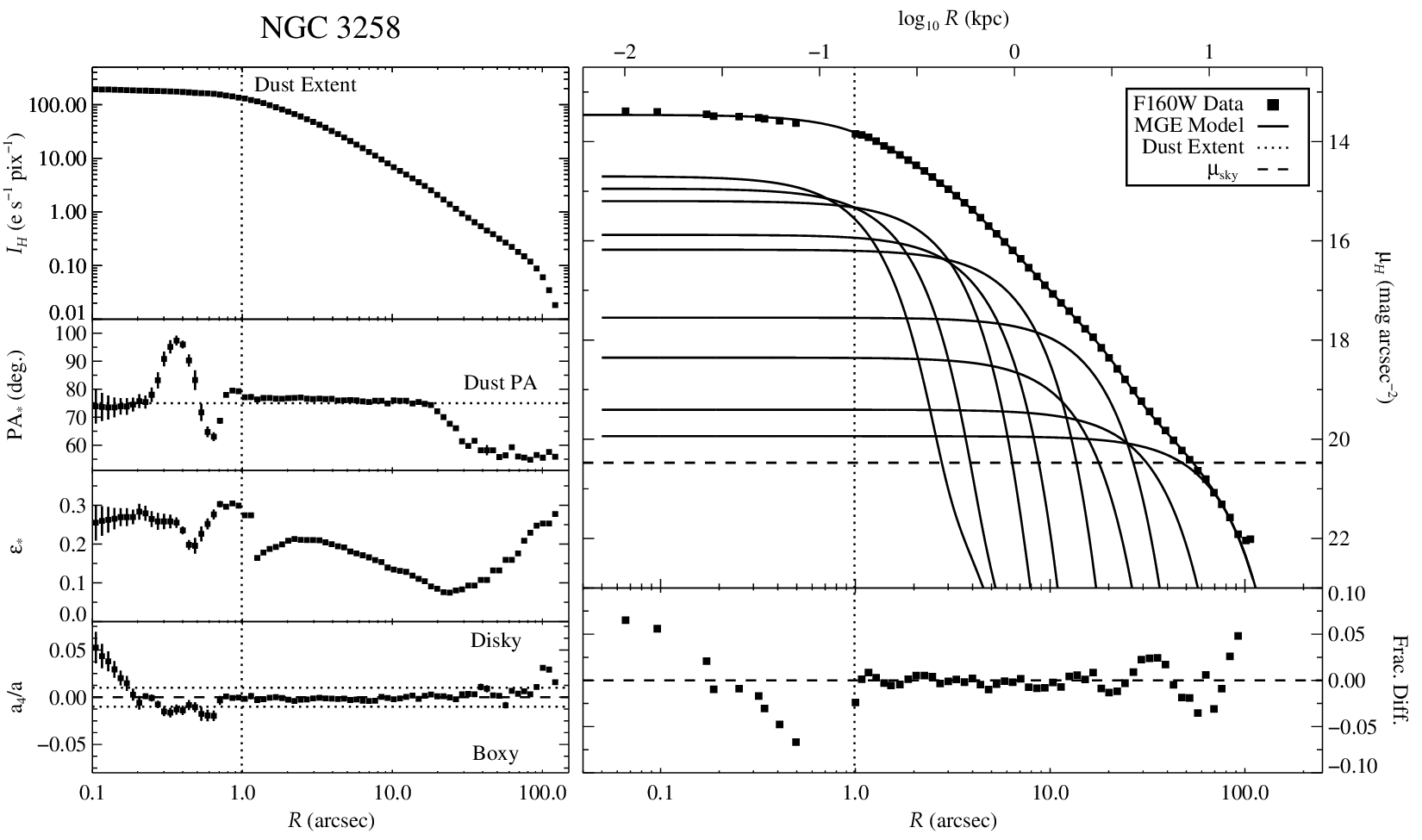}
    \centering
    \caption{Same as for Figure~\ref{fig:hydraa_ell_mge} but for NGC 3258.}
    \label{fig:ngc3258_ell_mge}
\end{figure*}

\begin{figure*}
    \includegraphics[width=0.9\textwidth]{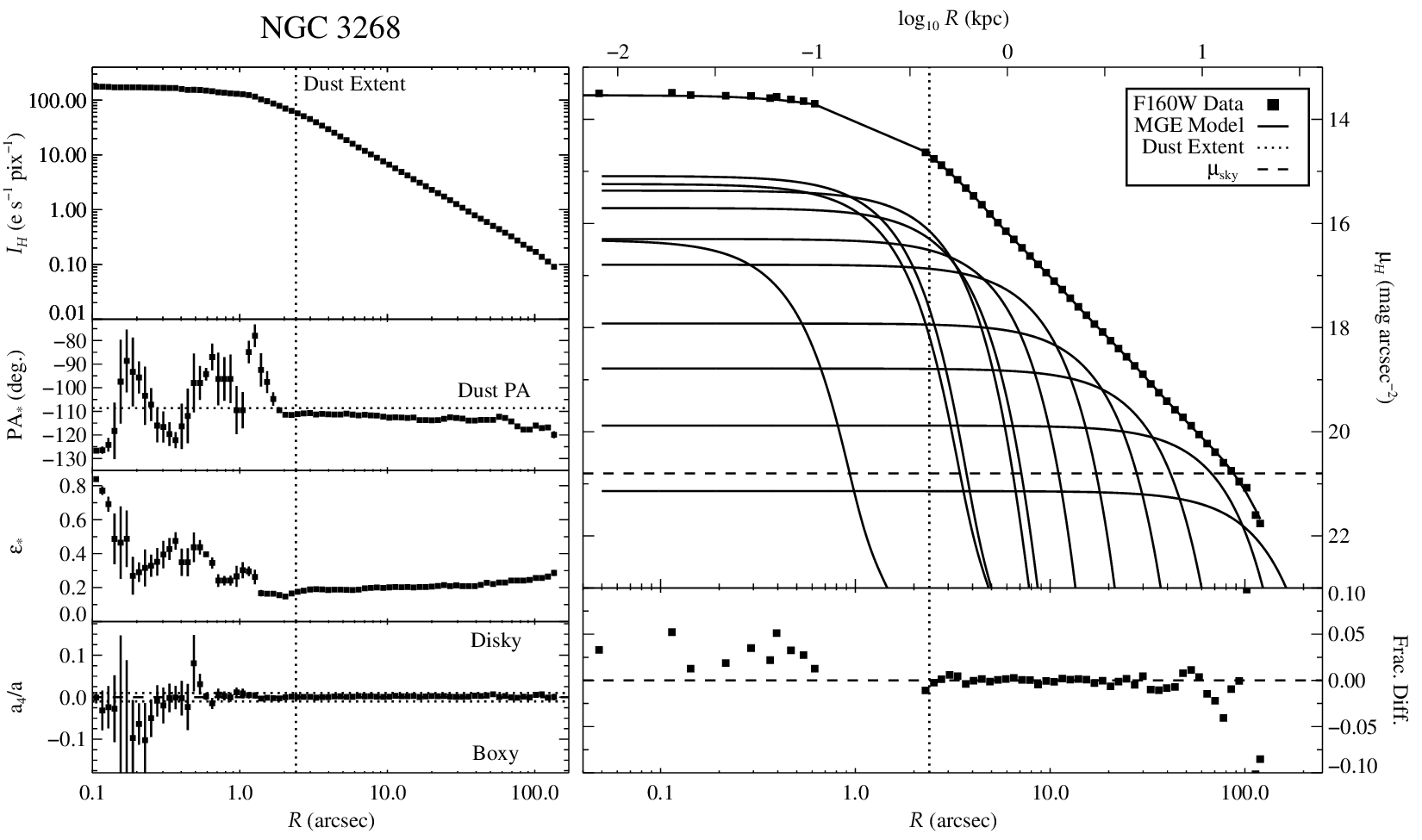}
    \centering
    \caption{Same as for Figure~\ref{fig:hydraa_ell_mge} but for NGC 3268.}
    \label{fig:ngc3268_ell_mge}
\end{figure*}

\begin{figure*}
    \includegraphics[width=0.9\textwidth]{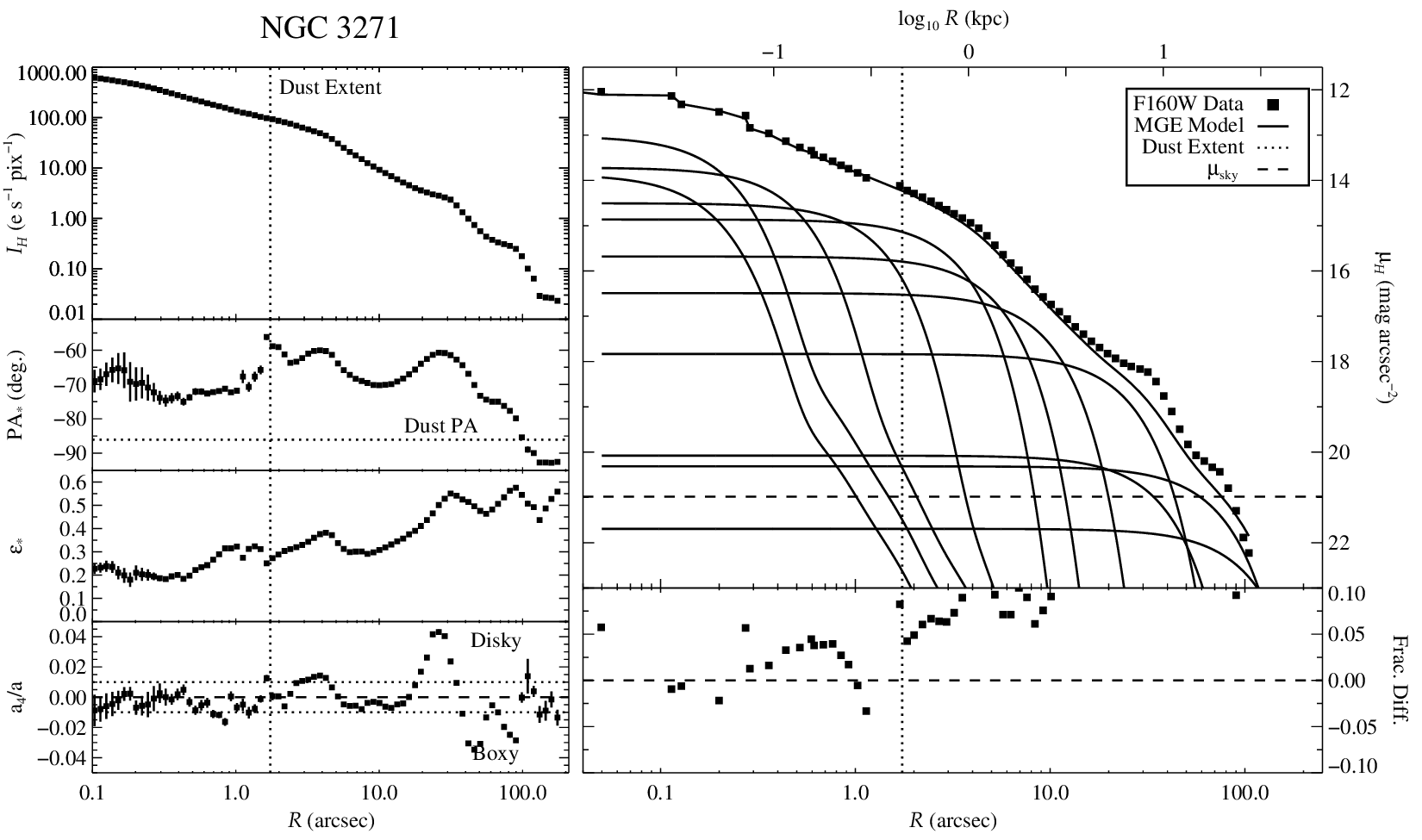}
    \centering
    \caption{Same as for Figure~\ref{fig:hydraa_ell_mge} but for NGC 3271.}
    \label{fig:ngc3271_ell_mge}
\end{figure*}

\begin{figure*}
    \includegraphics[width=0.9\textwidth]{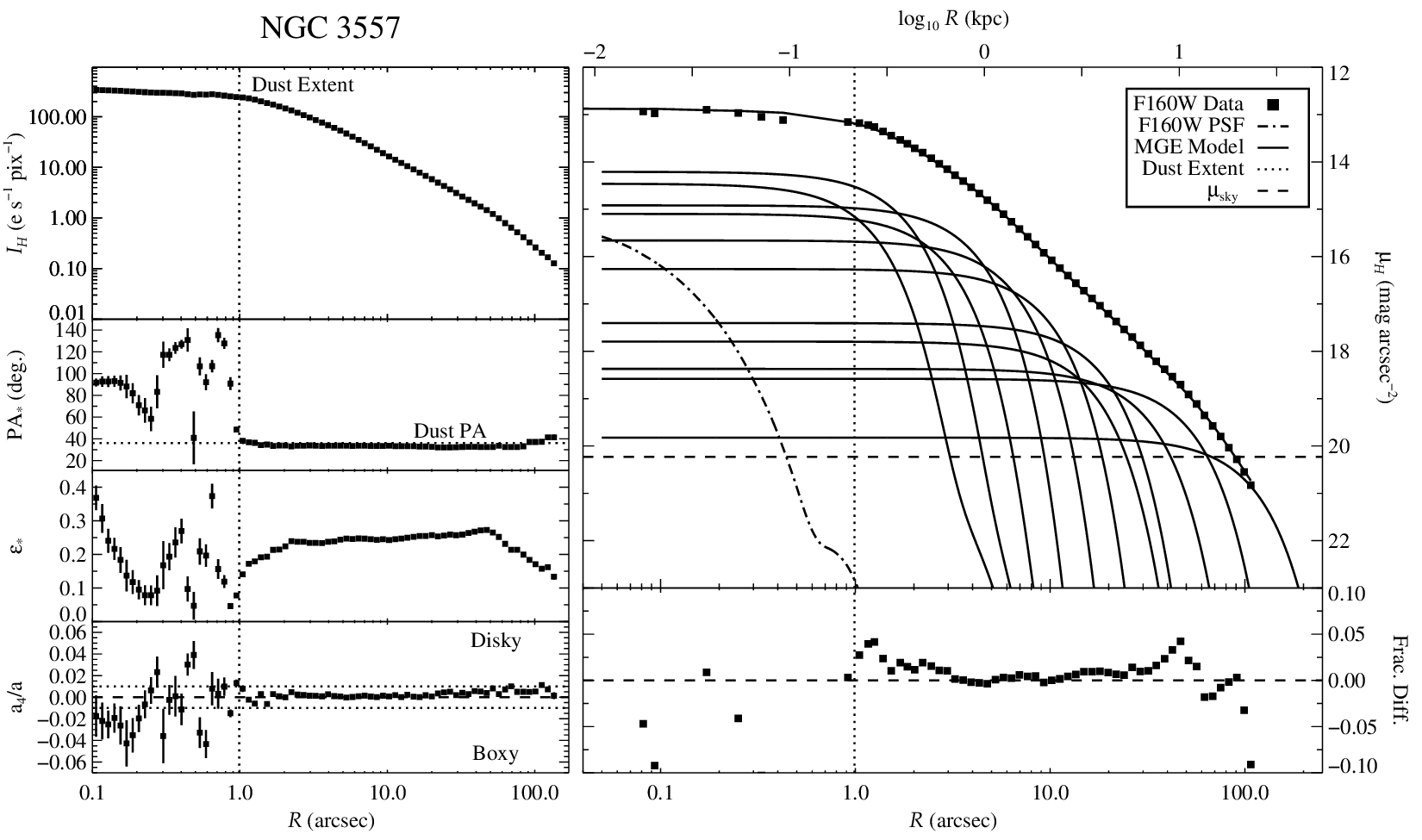}
    \centering
    \caption{Same as for Figure~\ref{fig:hydraa_ell_mge} but for NGC 3557.}
    \label{fig:ngc3557_ell_mge}
\end{figure*}

\begin{figure*}
   \includegraphics[width=0.9\textwidth]{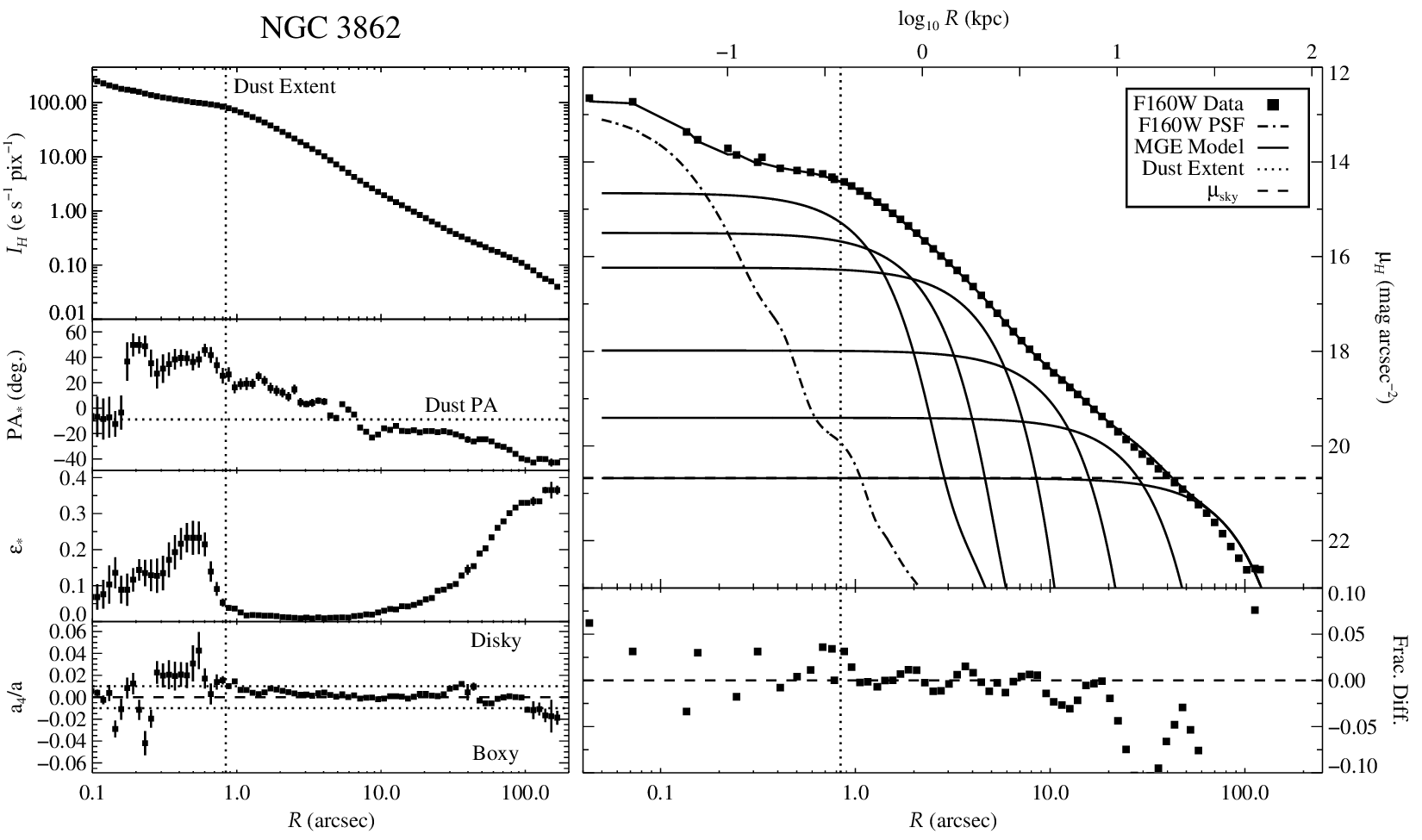}
   \centering
   \caption{Same as for Figure~\ref{fig:hydraa_ell_mge} but for NGC 3862.}
   \label{fig:ngc3862_ell_mge}
\end{figure*}

\begin{figure*}
   \includegraphics[width=0.9\textwidth]{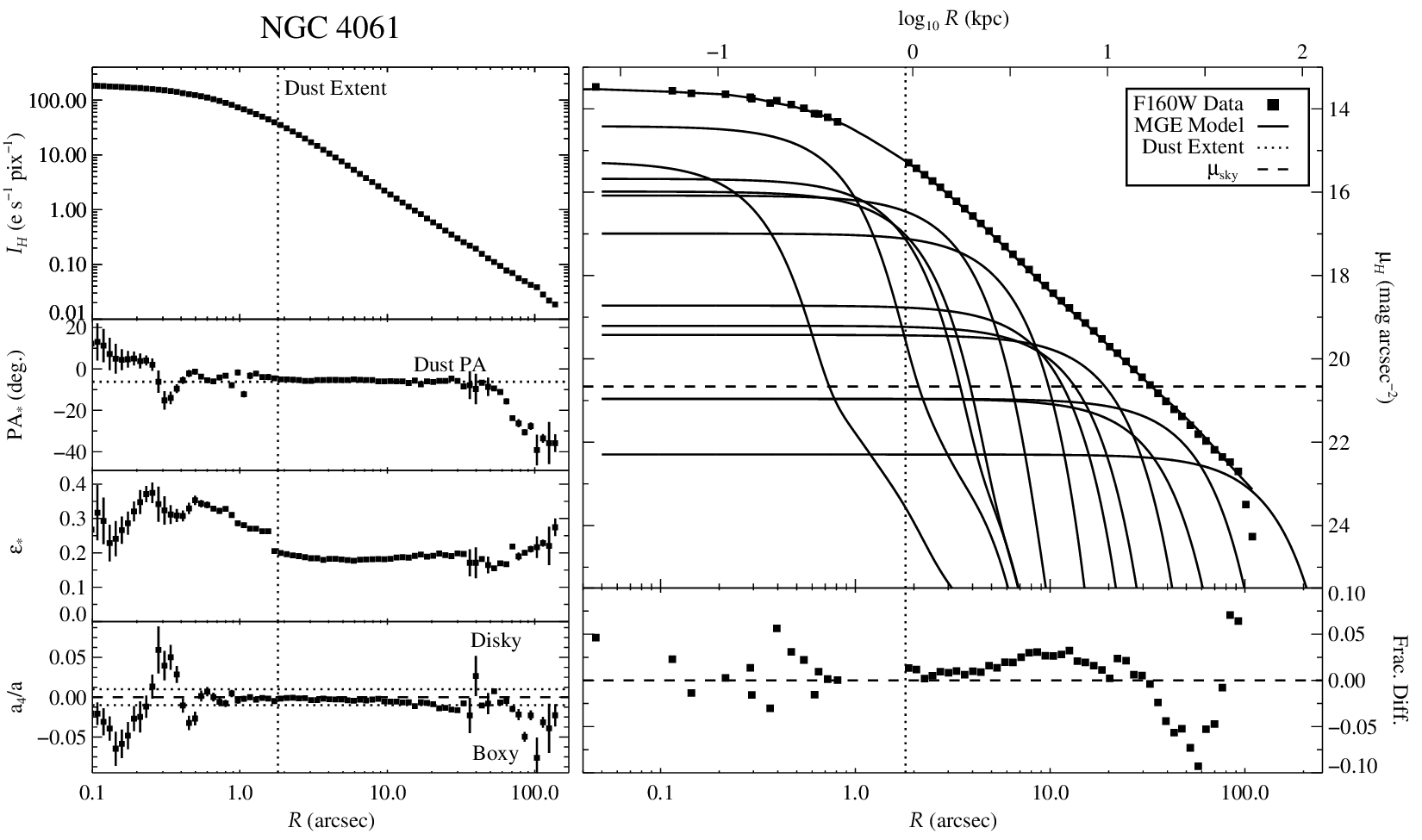}
   \centering
   \caption{Same as for Figure~\ref{fig:hydraa_ell_mge} but for NGC 4061.}
   \label{fig:ngc4061_ell_mge}
\end{figure*}

\begin{figure*}
   \includegraphics[width=0.9\textwidth]{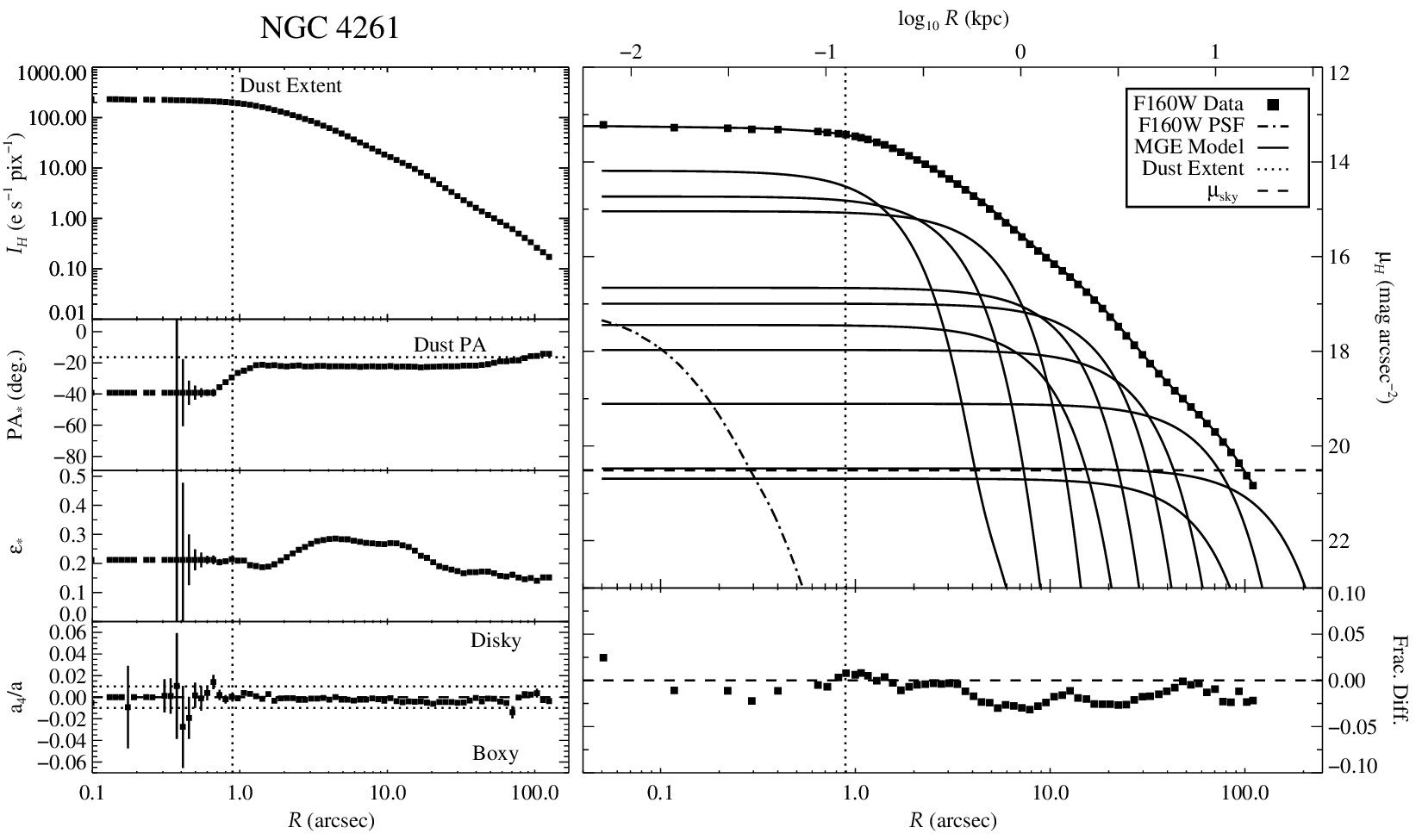}
   \centering
   \caption{Same as for Figure~\ref{fig:hydraa_ell_mge} but for NGC 4261.}
   \label{fig:ngc4261_ell_mge}
\end{figure*}

\begin{figure*}
    \includegraphics[width=0.9\textwidth]{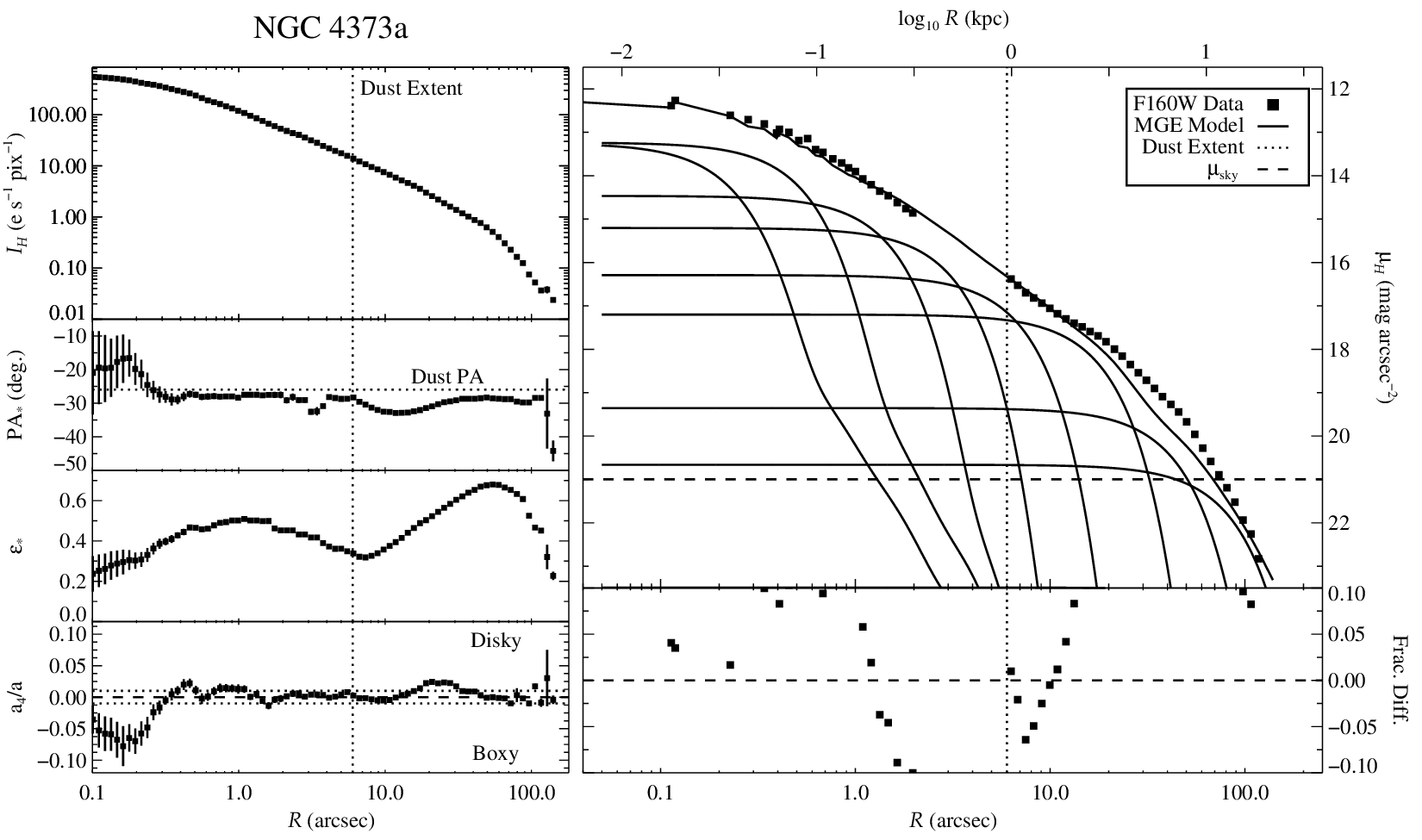}
    \centering
    \caption{Same as for Figure~\ref{fig:hydraa_ell_mge} but for NGC 4373a.}
    \label{fig:ngc4373a_ell_mge}
\end{figure*}

\begin{figure*}
    \includegraphics[width=0.9\textwidth]{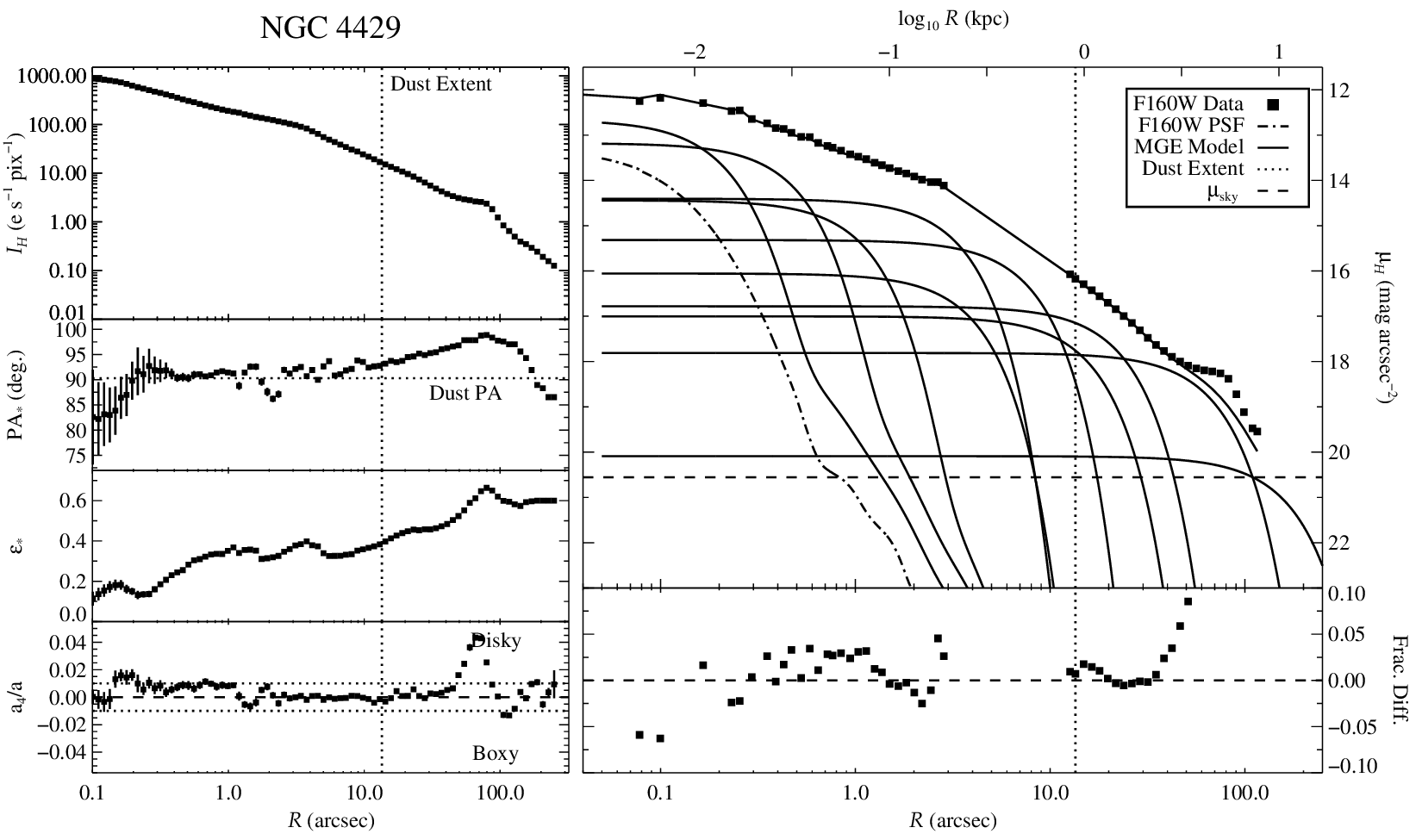}
    \centering
    \caption{Same as for Figure~\ref{fig:hydraa_ell_mge} but for NGC 4429.} 
    \label{fig:ngc4429_ell_mge}
\end{figure*}

\begin{figure*}
    \includegraphics[width=0.9\textwidth]{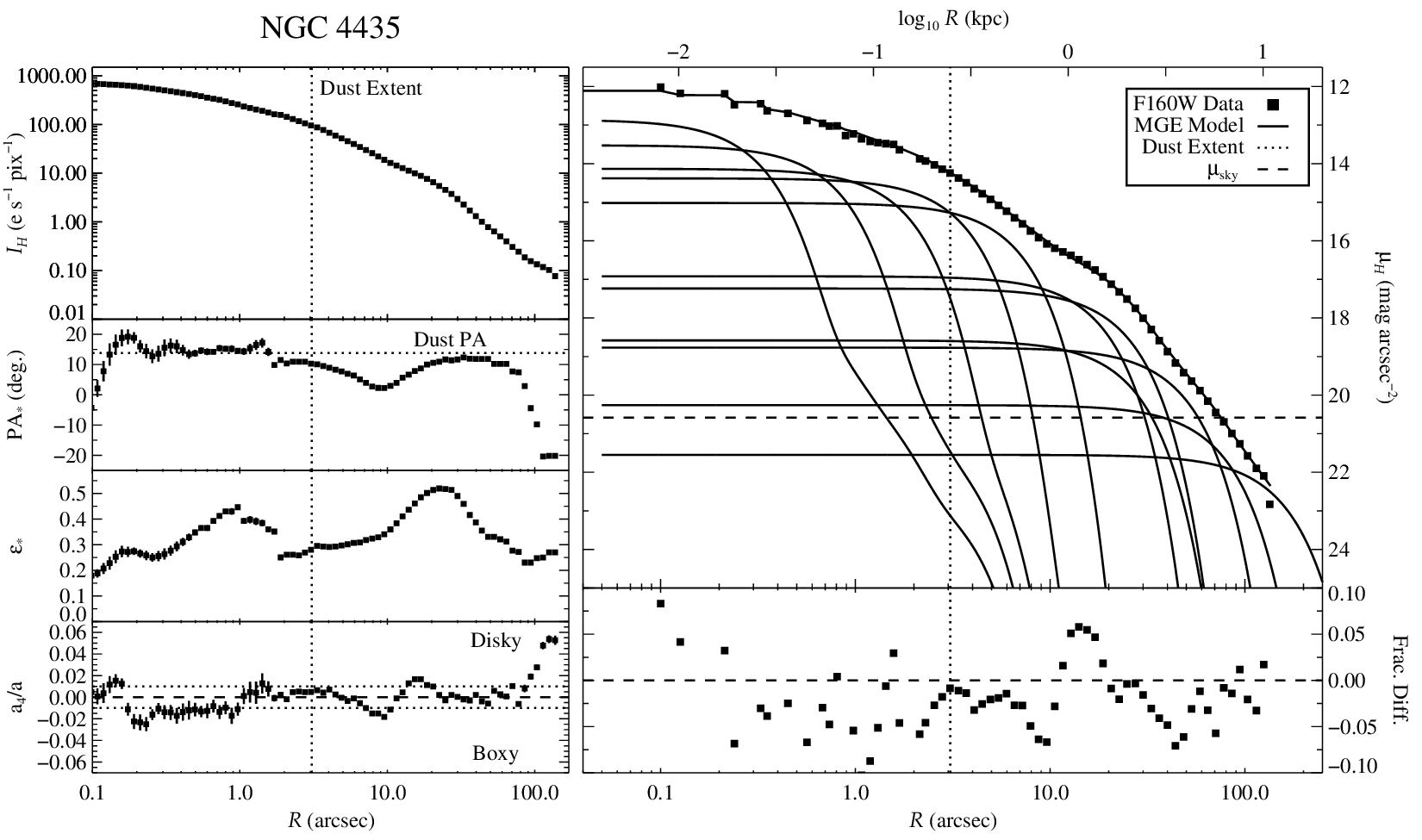}
    \centering
    \caption{Same as for Figure~\ref{fig:hydraa_ell_mge} but for NGC 4435.}
    \label{fig:ngc4435_ell_mge}
\end{figure*}

\begin{figure*}
   \includegraphics[width=0.9\textwidth]{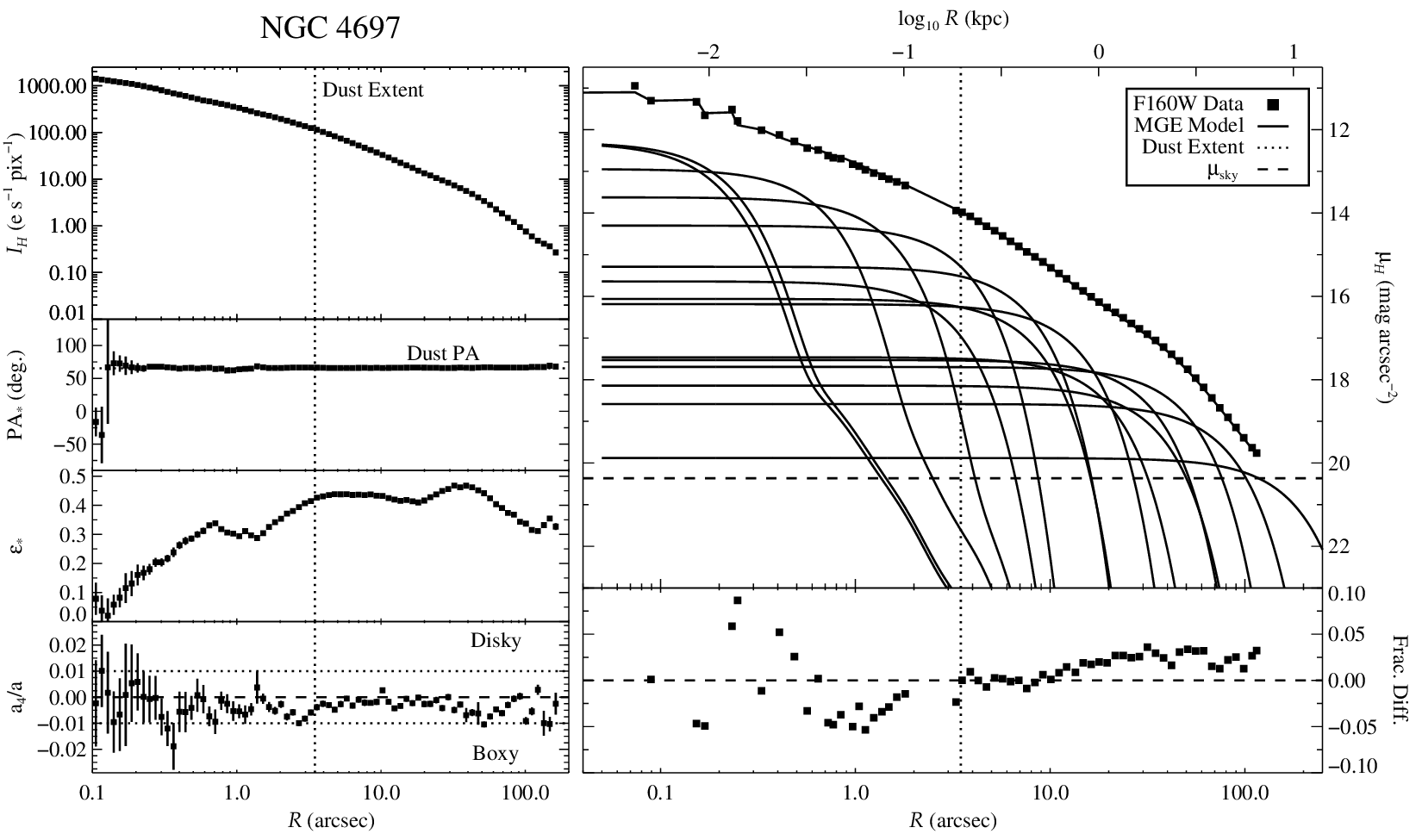}
   \centering
   \caption{Same as for Figure~\ref{fig:hydraa_ell_mge} but for NGC 4697.}
   \label{fig:ngc4697_ell_mge}
\end{figure*}

\begin{figure*}
   \includegraphics[width=0.9\textwidth]{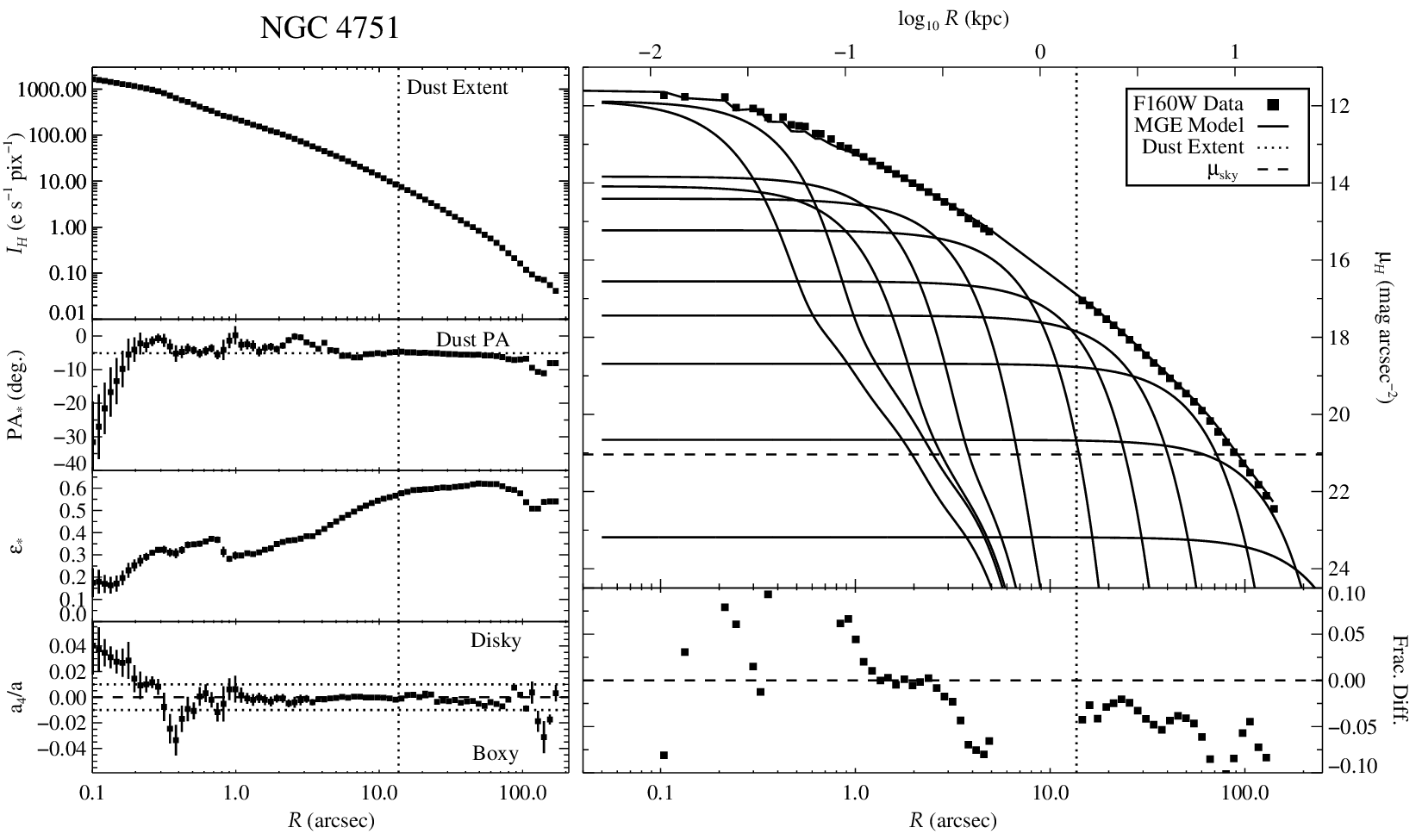}
   \centering
   \caption{Same as for Figure~\ref{fig:hydraa_ell_mge} but for NGC 4751.}
   \label{fig:ngc4751_ell_mge}
\end{figure*}

\begin{figure*}
   \includegraphics[width=0.9\textwidth]{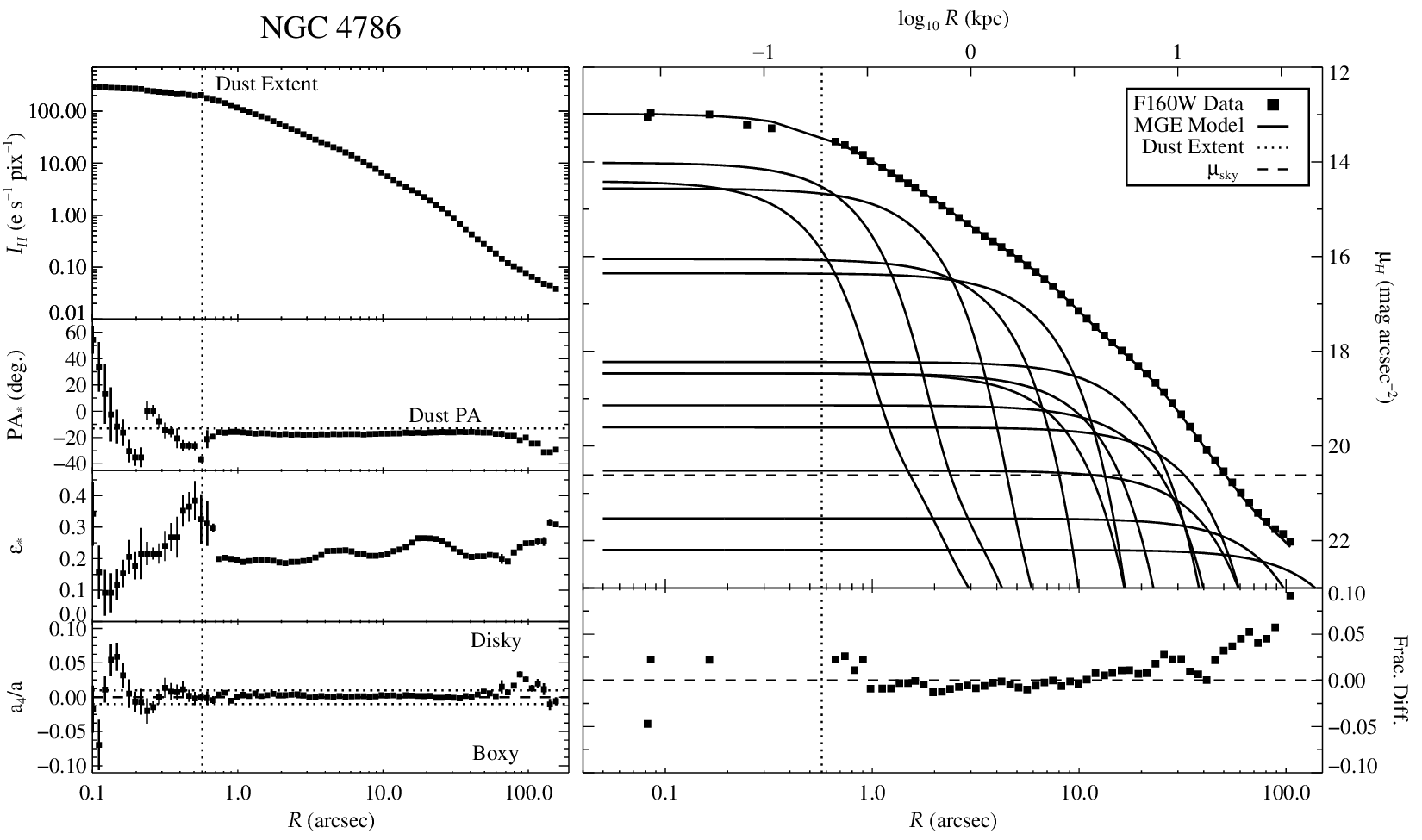}
   \centering
   \caption{Same as for Figure~\ref{fig:hydraa_ell_mge} but for NGC 4786.}
   \label{fig:ngc4786_ell_mge}
\end{figure*}

\begin{figure*}
   \includegraphics[width=0.9\textwidth]{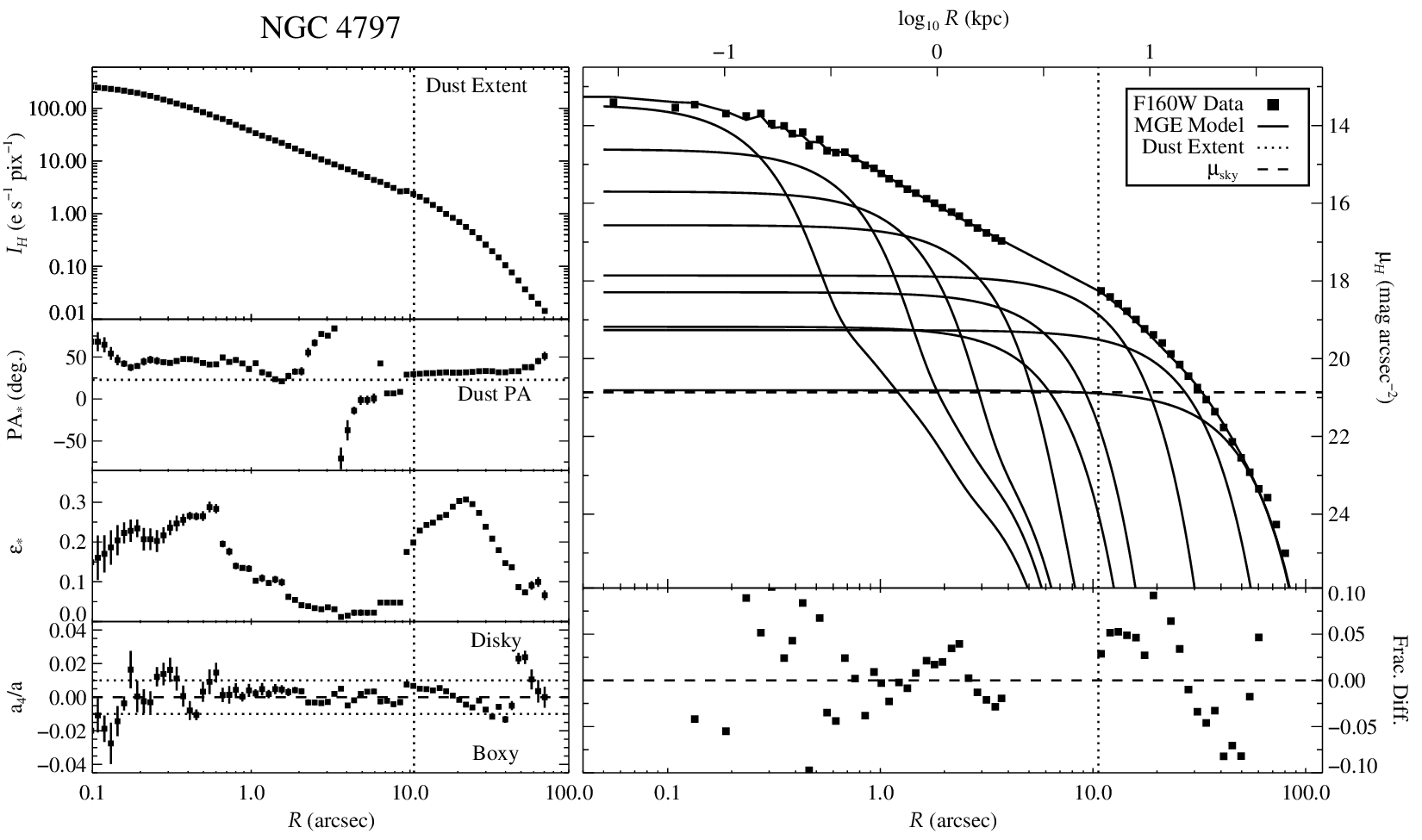}
   \centering
   \caption{Same as for Figure~\ref{fig:hydraa_ell_mge} but for NGC 4797.}
   \label{fig:ngc4797_ell_mge}
\end{figure*}

\begin{figure*}
    \includegraphics[width=0.9\textwidth]{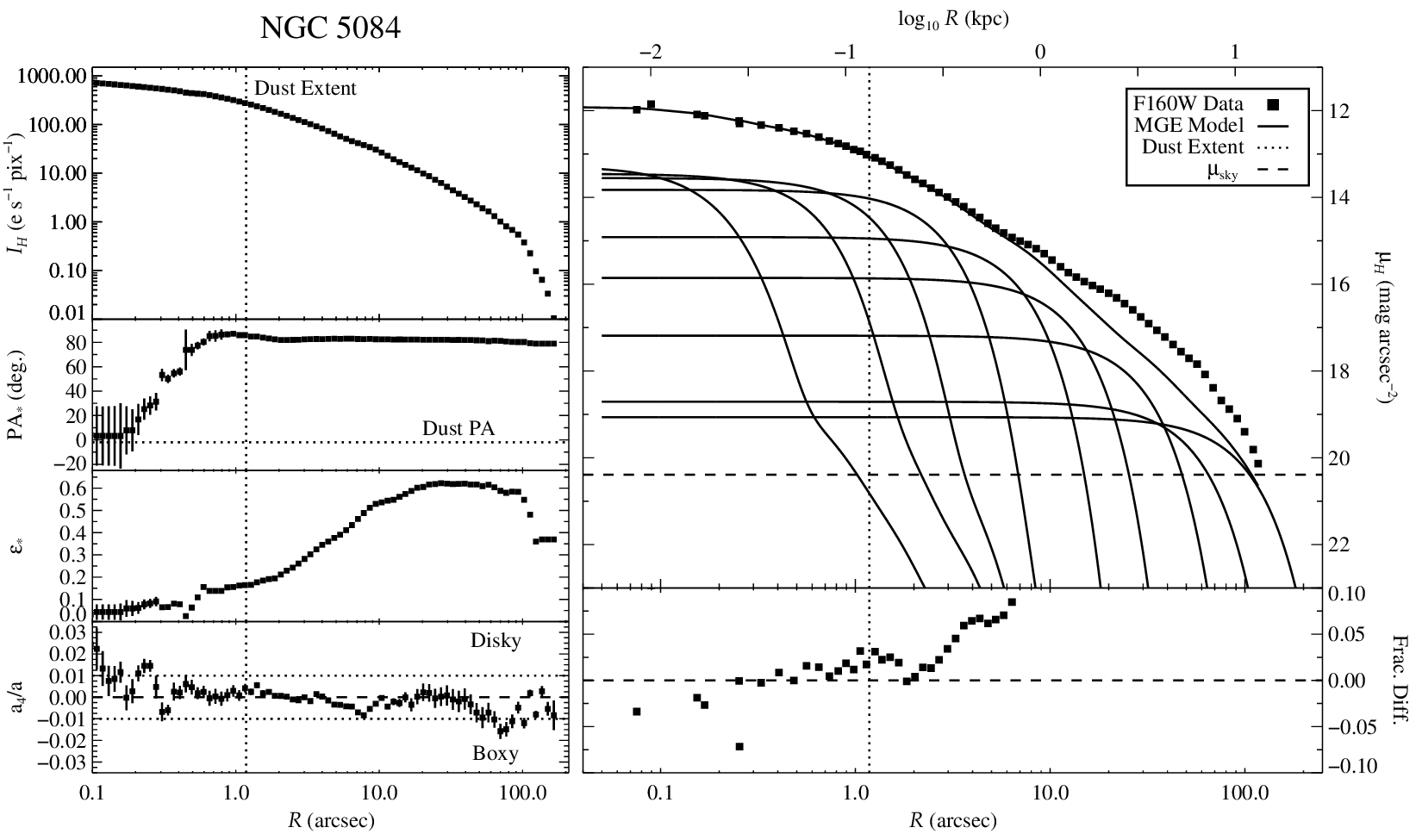}
    \centering
    \caption{Same as for Figure~\ref{fig:hydraa_ell_mge} but for NGC 5084.}
    \label{fig:ngc5084_ell_mge}
\end{figure*}

\begin{figure*}
   \includegraphics[width=0.9\textwidth]{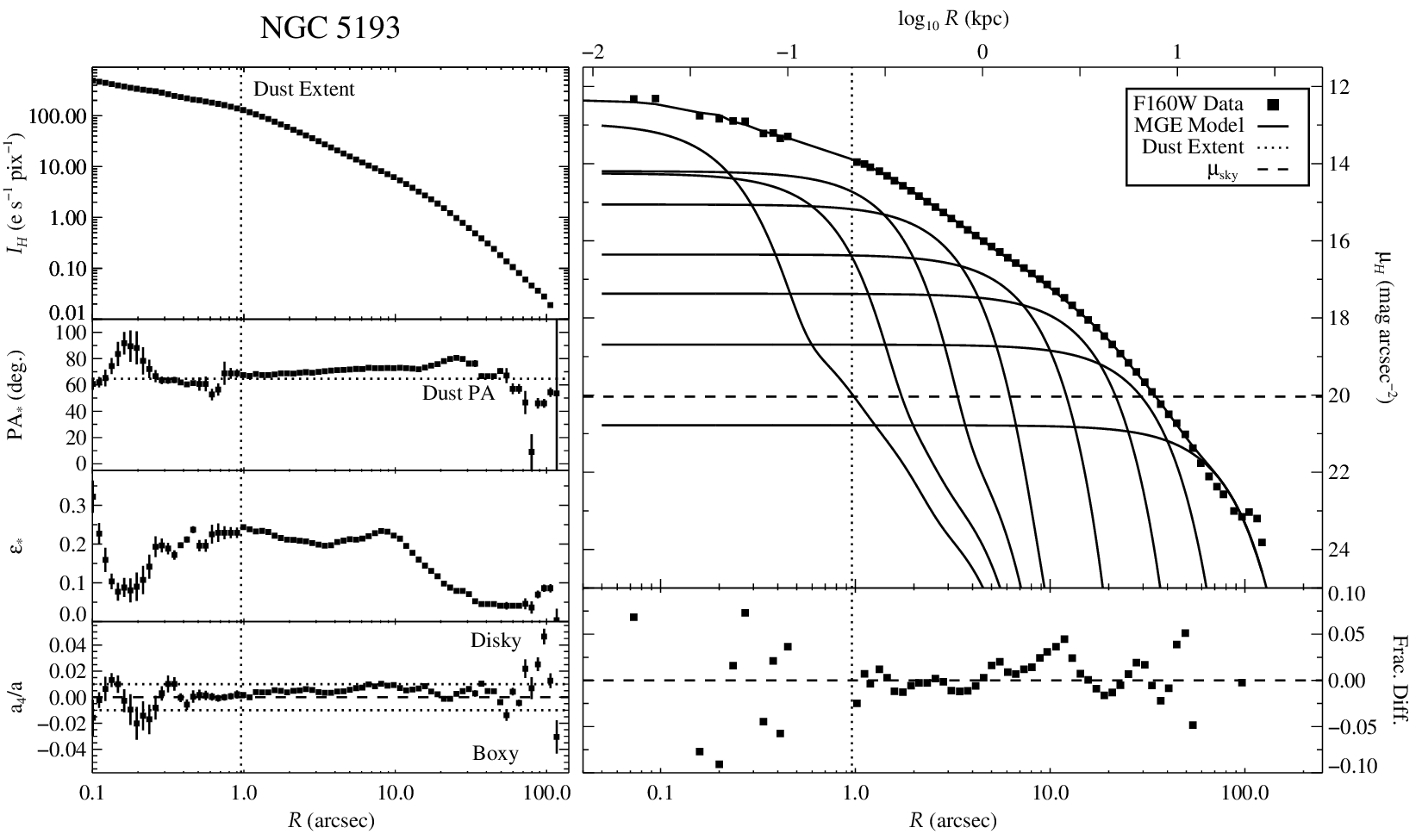}
   \centering
   \caption{Same as for Figure~\ref{fig:hydraa_ell_mge} but for NGC 5193.}
   \label{fig:ngc5193_ell_mge}
\end{figure*}

\begin{figure*}
   \includegraphics[width=0.9\textwidth]{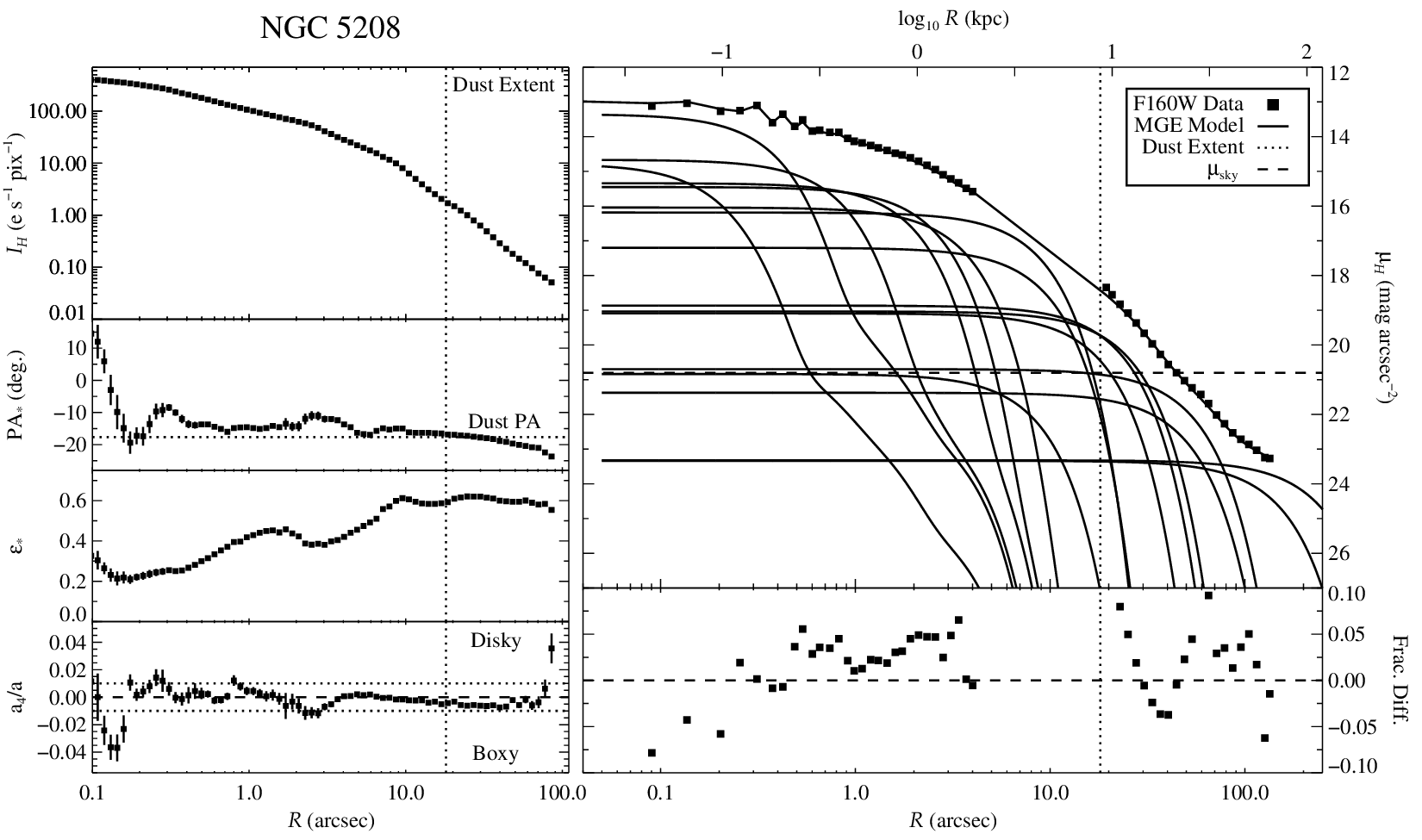}
   \centering
   \caption{Same as for Figure~\ref{fig:hydraa_ell_mge} but for NGC 5208.}
   \label{fig:ngc5208_ell_mge}
\end{figure*}

\begin{figure*}
    \includegraphics[width=0.9\textwidth]{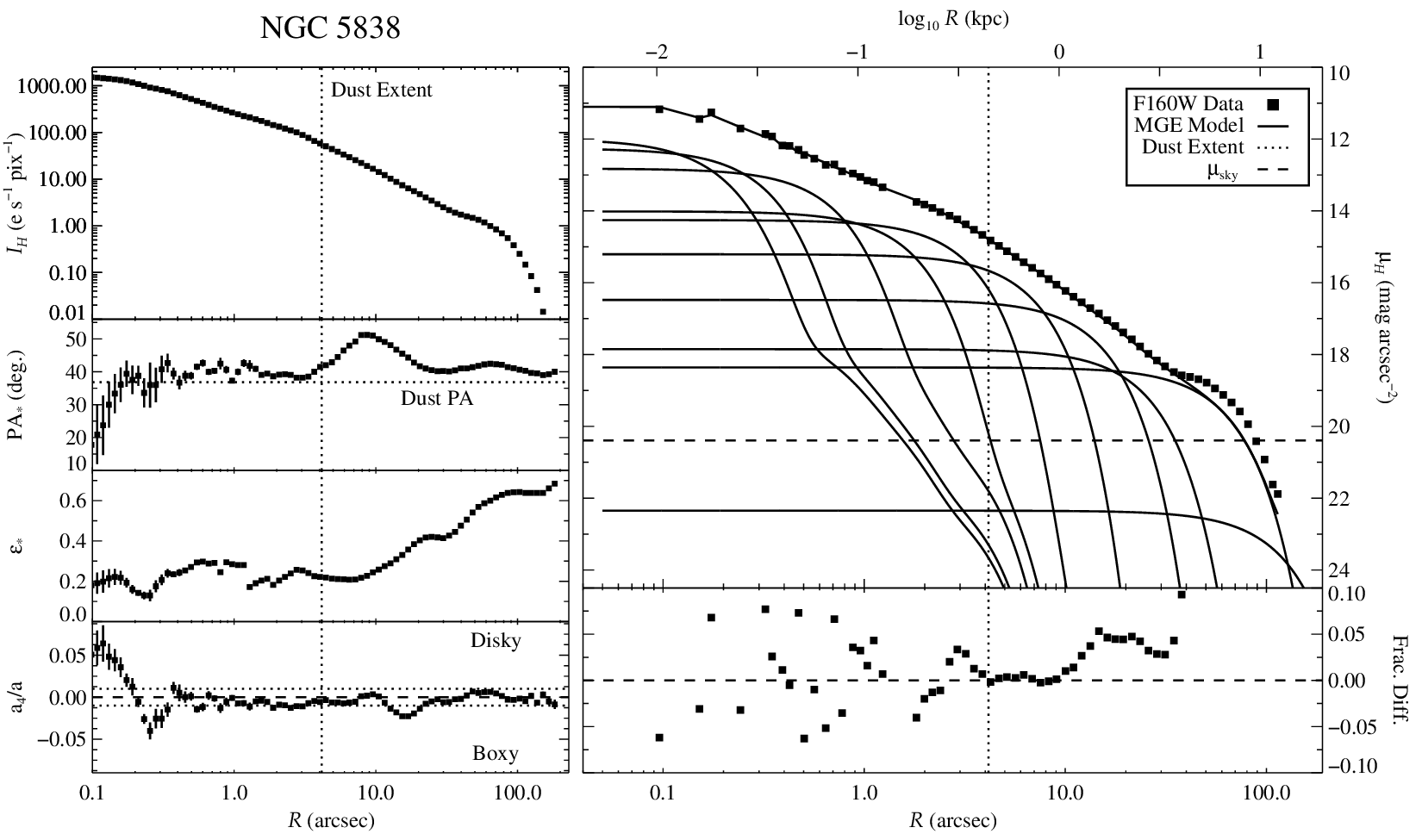}
    \centering
    \caption{Same as for Figure~\ref{fig:hydraa_ell_mge} but for NGC 5838.}
    \label{fig:ngc5838_ell_mge}
\end{figure*}

\begin{figure*}
   \includegraphics[width=0.9\textwidth]{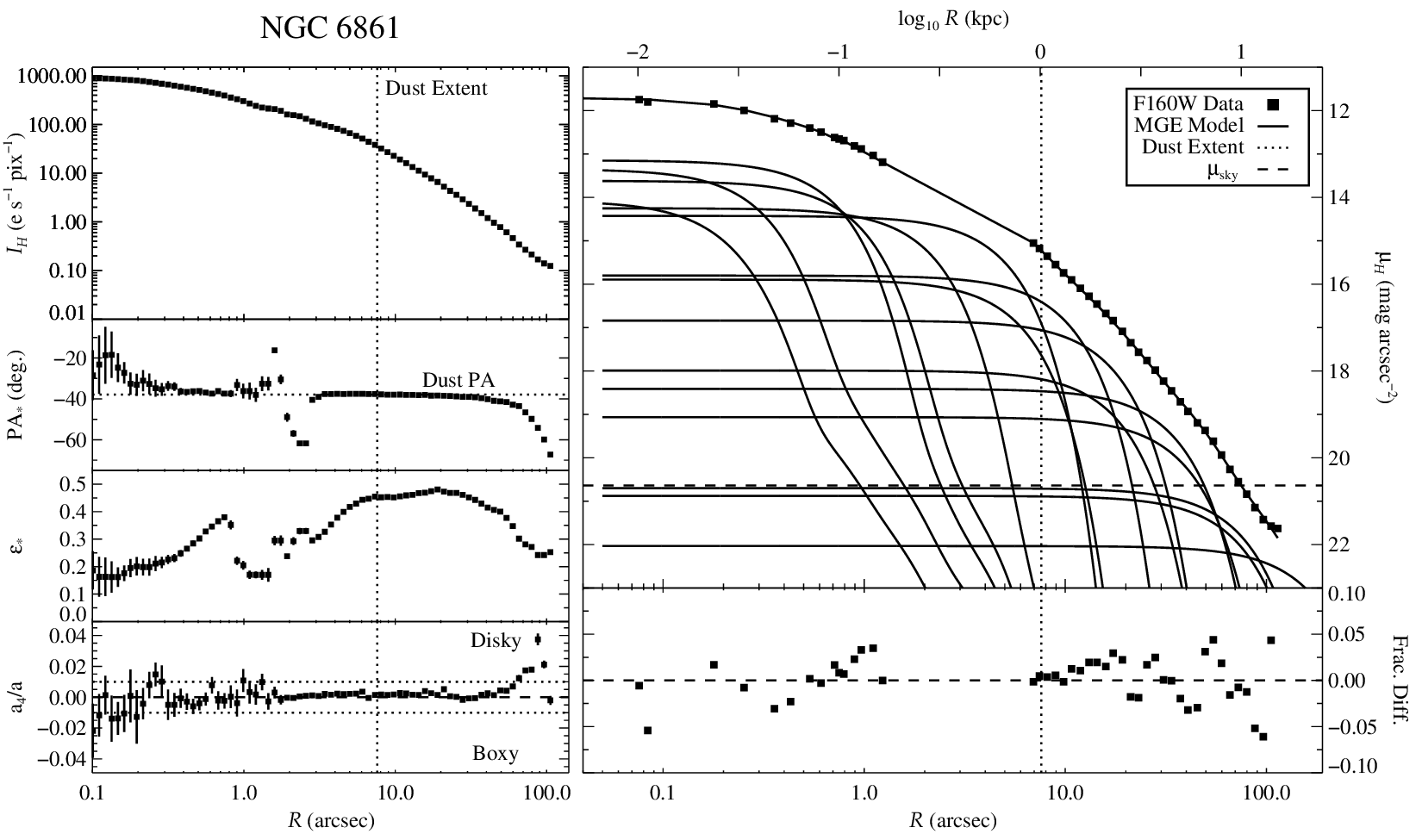}
   \centering
   \caption{Same as for Figure~\ref{fig:hydraa_ell_mge} but for NGC 6861.}
   \label{fig:ngc6861_ell_mge}
\end{figure*}

\begin{figure*}
    \includegraphics[width=0.9\textwidth]{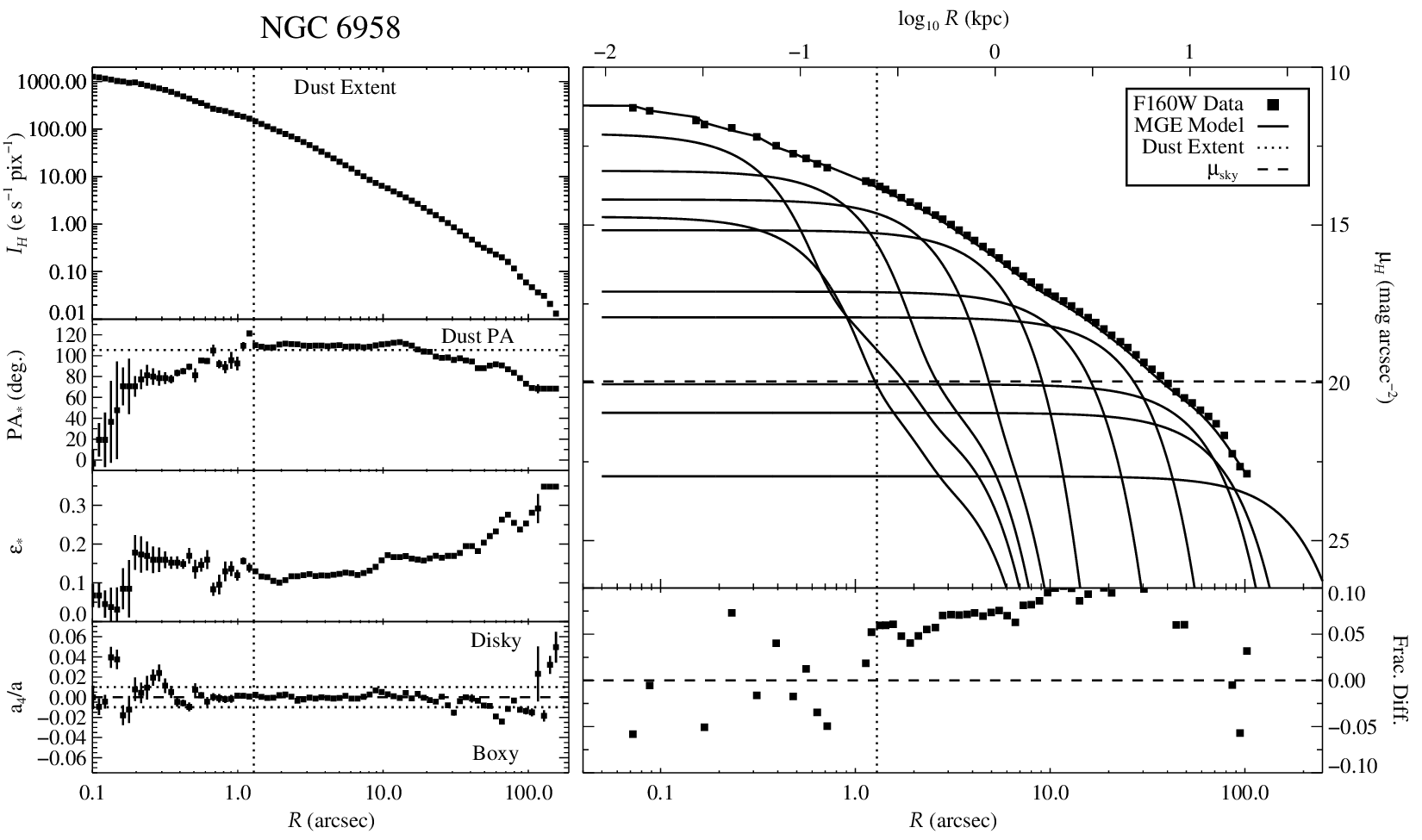}
    \centering
    \caption{Same as for Figure~\ref{fig:hydraa_ell_mge} but for NGC 6958.}
    \label{fig:ngc6958_ell_mge}
\end{figure*}

Here, we provide more detailed and individual analysis of the $H$-band surface brightnesses. In Figures~\ref{fig:hydraa_ell_mge}-\ref{fig:ngc6958_ell_mge}, we provide isophotal analysis results for each ETG as a function of radius nearly out to the edge of the WFC3/IR mosaic FOV. These figures also include 1D surface brightnesses extracted along the major axis from both the $H$-band data and the best-fitting 2D MGE decomposition using \texttt{GALFIT}. 

These surface brightness measurements were made after masking out the most dust-obscured regions with $\Delta (J-H) \gtrsim 0.08$ mag, which leads to (sometimes large) gaps in major-axis plots but only a reduced number of data points for the 2D fits. The majority of our MGE fits show good agreement to the $H$-band data, with fractional differences generally at the 10\% level or less across all (relevant) radii. The major axes of the dust disks tend to be fairly well aligned with the respective stellar photometric axes, although some targets do show significant ($>$20\degr) divergence between PA$_\star$ and PA$_{\mathrm{dust},H}$. The isophotal analyses for these targets are unreliable within the CND radii but reveal smoother variations in the fitted ellipse PA and ellipticity beyond the respective dust extents. The $a_4/a$ deviations are likewise small, with some tending toward boxy or disky isophotes at large radii. In nearly every case, the best-fit MGE reproduces the central stellar light distribution quite well, and should therefore be useful in future/ongoing ALMA CO dynamical modeling efforts. For just under a third of the sample, however, constraints on the MGE components worsen the global goodness-of-fit. Below, we briefly discuss particulars for all fits individually.

\begin{itemize}[leftmargin=*,nolistsep]
\item \textbf{Hydra A} (Figure~\ref{fig:hydraa_ell_mge}): Despite showing close agreement between the stellar and disk orientation at $R\sim R_{\mathrm{dust},H}$, the isophotal analysis shows a rapid shift in PA$_\star$ and a steady increase in $\varepsilon_\star$ starting just afterwards. The $a_4/a$ deviations are generally small but show some preference for boxy isophotes. Such gradients and boxiness are consistent with the dense galactic environment and merger history for BCGs. The best-fit MGE has PAs that align with $\overline{\mathrm{PA}}_\star$, which here hinges on the greater number of measurements at larger radii. Nevertheless, the generally good agreement (with data$-$MGE fractional differences of $\lesssim$5\%) is due to fairly circular isophotes at small $R$.

\item \textbf{NGC 612} (Figure~\ref{fig:ngc612_ell_mge}): The MGE shows passable agreement to the $H$-band data out to $R\sim 70\arcsec$ despite the challenges inherent in fitting such a large dust disk. Of course, the high $\Delta \mathrm{PA}_\star > 45\degr$ cannot be reproduced by the MGE, but the low $\varepsilon_\star$ over most of this $\Delta \mathrm{PA}_\star$ range allows for MGE components with a uniform PA to still fit the data without large discrepancies. The $a_4/a$ values show clear preference for disky isophotes between 40$-$50\arcsec\ with no clear trend beyond, although higher $\varepsilon_\star$ towards the mosaic edge suggests overall greater flattening. Interior to these radii, the outer dust disk features are generally aligned to within $\sim$20\degr\ of PA$_\star$.

\item \textbf{NGC 997} (Figure~\ref{fig:ngc997_ell_mge}): This E galaxy shows close agreement between dust and stellar orientation until $R \sim 20\arcsec$. This radius coincides with the angular distance of a possible companion, PGC 200205. NGC 997 is tidally interacting with NGC 998, and these two other galaxies may both contribute to the apparent shift in PA$_\star$. The MGE shows excellent agreement in the central couple $\times$10 pc but the fit remains good out to $R\sim70\arcsec$. While $\varepsilon_\star$ remains flat, the limiting $q^{\prime}$ still limits most of the MGE components, although the global fit only appears slightly affected. In addition, the $a_4/a$ values shows a clear preference for boxy isophotes with increasing radius. In either case, the global fit appears to be only slightly affected. We note that \citet{dom24} ran an independent MGE model in the \emph{H}-band for this galaxy, finding an additional compact component (with $\sigma^{\prime} < 0\farcs06$) at lower luminosity that does not appear necessary based on the MGE residuals in Figure~\ref{fig:ngc997_ell_mge}. However, these differences can likely be explained by the choice of PSF and lack of masking.

\item \textbf{NGC 1332} (Figure~\ref{fig:ngc1332_ell_mge}): The dust disk orientation remains closely aligned to PA$_\star$ at all radii, even while the isophotes become increasingly flattened. Despite having disk-dominated outskirts, this lenticular galaxy shows slightly boxy isophotes. The MGE reproduces the observed $H$-band stellar light distribution well, although a slight PA$_\star$ shift and flattened isophotes beyond $R \sim 60\arcsec$ accentuate this discrepancy along the major axis.

\item \textbf{NGC 1387} (Figure~\ref{fig:ngc1387_ell_mge}): The measured PA$_\star$ reveals one of the most misaligned ($>$50\degr) dust disks in our sample, with the stellar isophotes lining up well with the stellar bar instead of PA$_{\mathrm{dust},H}$. Given the nearly face-on CND orientation, this 2D MGE cannot match the flattened isophotes around the bar-dominated regions. Because of the CND shape and uniformly low $\Delta (J-H)$ observed for this ETG, we do not create any dust mask. The nearly face-on CND should not affect either the isophotal results or the MGE $\sigma^{\prime}$ or $q^{\prime}$ values; however, MGE central surface brightness $I_H$ may be slightly suppressed, especially for the innermost components that lie almost entirely within the large-scale $R_{\mathrm{dust},H}$.

\item \textbf{NGC 3245} (Figure~\ref{fig:ngc3245_ell_mge}): The dust disk is generally well-aligned with PA$_\star$ except in the first couple arcseconds after $R_{\mathrm{dust},H}$. However, the more circular isophotes at these radii limit the data$-$MGE discrepancies. At larger radii, the more flattened isophotes do result in a couple of MGE component $q^{\prime}$ values settling to $q^{\prime}_\mathrm{min}$, but the overall fit does not seem to be affected.

\item \textbf{NGC 3258} (Figure~\ref{fig:ngc3258_ell_mge}): The best-fit MGE is good at all radii, as is demonstrated along the major axis, although the data do show a moderate PA$_\star$ shift from PA$_{\mathrm{dust},H}$ past $R \sim 20\arcsec$. The data--MGE discrepancy that would otherwise arise from this $\Delta \mathrm{PA}_\star$ is lessened for surface brightness measurements between $\sim$20$-$50\arcsec\ because of more circular isophotes at these radii.

\item \textbf{NGC 3268} (Figure~\ref{fig:ngc3268_ell_mge}): This galaxy shows nearly perfect power-law surface brightness behavior beyond $R_{\mathrm{dust},H}$ with only minor changes in PA$_\star$ and $\varepsilon_\star$ and no significant $a_4/a$ deviations from pure elliptical isophotes. As a result, the MGE provides a very good global fit.

\item \textbf{NGC 3271} (Figure~\ref{fig:ngc3271_ell_mge}): The consistent and moderate-amplitude $\mathrm{PA}_\star - \mathrm{PA}_{\mathrm{dust},H}$ offset arises in part due to the MGE fitting to a prominent, misaligned stellar bar, although the PA$_\star$ of the stellar disk remains persistently high out to $R \sim 50\arcsec$. The generally poorer MGE for this cuspy galaxy also reflects the $a_4/a$ values changing from very disky behavior around the bar to boxy features with increasing radius.

\item \textbf{NGC 3557} (Figure~\ref{fig:ngc3557_ell_mge}): Good alignment between PA$_{\mathrm{dust},H}$ and PA$_\star$ at all radii helps this cored radio galaxy be fit well with an MGE. While we include a central PSF component in the fit, at these wavelengths the AGN contributions are essentially negligible. Beyond $R \sim 50\arcsec$, the $\varepsilon_\star$ decrease is coincident with a gradual increase in $a_4/a$ towards disky behavior.

\item \textbf{NGC 3862} (Figure~\ref{fig:ngc3862_ell_mge}): This galaxy shows a nearly uniform PA$_\star$ gradient while becoming increasingly disk-dominated beyond \re. Despite the large change in PA$_\star$, the MGE does not result in a poorer fit until $R \gtrsim 20\arcsec$, when the isophotes become less circular. The measured $a_4/a$ values show some preference for boxy shapes only near the edge of the WFC3/IR FOV. The central surface brightnesses for this radio galaxy are consistent with a point source, and including a PSF component recovers the expected core behavior.

\item \textbf{NGC 4061} (Figure~\ref{fig:ngc4061_ell_mge}): Out to $\sim$2\re, the stellar isophotal behavior remains consistent and the measured PA$_\star$ aligns with the observed dust disk orientation. At larger radii, an abrupt shift in PA$_\star$ is accompanied by boxy isophotes. The more disturbed features appear to be due to interactions with the nearby NGC 4065. However, the tidal influence is limited to the outer regions of the WFC3/IR FOV, and the \texttt{GALFIT} MGE reproduces the inner 2D stellar surface brightnesses well.

\item \textbf{NGC 4261} (Figure~\ref{fig:ngc4261_ell_mge}): The relatively close agreement between stellar isophotal and outer dust orientations, together with overall low $\varepsilon_\star$, allows for a good global fit without any preference for disky or boxy isophotes. Interestingly, the CND and stellar PAs remain offset by $\sim$5\degr\ out to \re. \citet{boizelle21} show that the outer CO kinematics appear to match PA$_{\mathrm{dust},H}$ reported in Table~\ref{tbl:ell_results}. However, the CO kinematic line-of-nodes PA approaches $\sim$0\degr\ towards the disk center, which only increases the $\mathrm{PA}_\star - \mathrm{PA}_{\mathrm{dust},H}$ offset. We note that the MGE includes a central PSF component fitted simultaneously in \texttt{GALFIT}, but this point source is negligible at these wavelengths.

\item \textbf{NGC 4373a} (Figure~\ref{fig:ngc4373a_ell_mge}): The MGE fit shows generally good agreement out to $R\sim13\arcsec$. At larger radii, high $\varepsilon_\star > 1-q_\mathrm{min}^{\prime}$ for $R\gtrsim 15\arcsec$ and isophotes with $a_4/a > 0.01$ for $15\arcsec \lesssim R \lesssim 40\arcsec$ lead to a poorer overall fit. Indeed, most of the MGE components have $q^{\prime}$ that are constrained by $q_\mathrm{min}^{\prime}$. Inside the disky isophotes, the stellar isophotal analysis returns $a_4/a \approx -0.01$ and the inner MGE fit around $R\sim 10\arcsec$ struggles to match the more boxy features. The dust disk is aligned to within $\sim$10\degr\ of the stellar photometric axis at most radii; at large $R$, PA$_\star$ may begin to diverge significantly from PA$_{\mathrm{dust},H}$, although the certainty of these final measurements is not high.

\item \textbf{NGC 4429} (Figure~\ref{fig:ngc4429_ell_mge}): Overall, the MGE shows generally good agreement out to $R\sim45\arcsec$. $\mathrm{PA}_\star \sim \mathrm{PA}_{\mathrm{dust},H}$ just beyond $R_{\mathrm{dust},H} \sim 13\arcsec$, although PA$_\star$ increases thereafter to a maximum $10\degr$ offset. The isophotal analysis shows an $I(R)$ bump for $50\arcsec \lesssim R \lesssim 90\arcsec$ that corresponds to increased PA$_\star$, $\varepsilon_\star$, and $a_4/a$ values. Figure~\ref{fig:contours2} reveals this feature arises from the (inner) stellar ring. All of the outer MGE components have $q^{\prime} \rightarrow q_\mathrm{min}^{\prime}$, and the global MGE does not reproduce the outer stellar light distribution as well as it does the inner regions. This \texttt{GALFIT} fit includes a PSF component to account for moderate AGN contamination, although the central MGE component(s) exceed(s) the central surface brightness  of this unresolved source.

\item \textbf{NGC 4435} (Figure~\ref{fig:ngc4435_ell_mge}): This barred lenticular galaxy shows only modest PA$_\star$ changes and moderate $\varepsilon_\star$ fluctuations that are below $1-q_\mathrm{min}^{\prime}$ for $R < 70\arcsec$, allowing the MGE to reproduce the inner stellar light distribution well. In this region, fluctuating $a_4/a$ indicates both disky and boxy isophotal behavior at different radii. In the outer regions (at $R \gtrsim 80\arcsec$), the PA$_\star$ abruptly shifts while the isophotes become very disky. Some portion of this shift may be due to the stellar light distribution of (or perhaps interactions with) neighboring NGC 4438.

\item \textbf{NGC 4697} (Figure~\ref{fig:ngc4697_ell_mge}): The MGE provides very good fits to both inner and global stellar light distributions, with $\mathrm{PA}_\star \sim \mathrm{PA}_{\mathrm{dust},H}$ at all $R$. The measured $a_4/a$ values approach boxy behavior towards the outer parts of this galaxy without a noticeable impact on the MGE goodness-of-fit. Positive data$-$MGE residuals along the major axis for $R\gtrsim 15\arcsec$ are evidence of a disk-like stellar component ($\varepsilon_{\mathrm{disk},\star} > 0.7$) for this fast-rotating ETG \citep{spiniello15} that is not fully fit with this MGE approach.

\item \textbf{NGC 4751} (Figure~\ref{fig:ngc4751_ell_mge}): This cuspy lenticular galaxy is moderately well-fit by an MGE at all radii, with $\mathrm{PA}_\star \approx \mathrm{PA}_{\mathrm{dust},H}$ across most of the WFC3/IR FOV. At larger radii, the $a_4/a$ values hint at slightly more boxy features, which result in negative data$-$MGE fractional residuals along the major axis.

\item \textbf{NGC 4786} (Figure~\ref{fig:ngc4786_ell_mge}): For this more cored galaxy, the $\mathrm{PA}_\star \sim \mathrm{PA}_{\mathrm{dust},H}$ and $\varepsilon_\star$ values are consistent across nearly the entire FOV, only showing discrepancies for $R\gtrsim 90\arcsec$ coincident with the start of disky isophotes. The MGE reproduces the stellar light distributions well on both inner and global scales.

\item \textbf{NGC 4797} (Figure~\ref{fig:ngc4797_ell_mge}): This fairly compact lenticular galaxy shows consistent PA$_\star$ out to $>$4\re\ even as the stellar ellipticity fluctuates. Beyond that radius, PA$_\star$ and $a_4/a$ both increase, with the isophotes transitioning from boxy to disky behavior. The data--MGE residuals are around the 10\% level for many of the major-axis measurements, although the 2D fits show generally good agreement throughout the WFC3/IR FOV.

\item \textbf{NGC 5084} (Figure~\ref{fig:ngc5084_ell_mge}): The MGE shows good agreement to the data only out to $\sim$7\arcsec, beyond which the highly elliptical stellar light distribution and constrained $q^{\prime}$ lead to a poor fit, especially along the major-axis direction. Beyond the dust extent, the isophotal analysis shows fairly constant PA$_\star$. However, PA$_{\mathrm{dust},H}$ is nearly orthogonal to $\overline{\mathrm{PA}}_\star$, indicating a polar-ring CND. In such a case, the MGE may still be useful but one cannot simply assume the CND lies in the galaxy's midplane when calculating $v_{\mathrm{c},\star}$. At the transition between disk-dominated regions (out to $R \sim 70\arcsec$) and the stellar halo, the stellar isophotes show a clear trend towards boxy behavior.

\item \textbf{NGC 5193} (Figure~\ref{fig:ngc5193_ell_mge}): This peculiar E galaxy is well-fit by the standard MGE. The stellar and dust-disk orientation angles agree just beyond $R_{\mathrm{dust},H}$, although an increasing PA$_\star$ leads to a $\sim$15\degr\ offset between $R\sim 20-30\arcsec$. This discrepancy may be due to a companion galaxy to the southwest whose light may not be fully masked. This PA$_\star$ shift does not result in a noticeably worse MGE goodness-of-fit, due to a coincident drop in $\varepsilon_\star$ to nearly circular isophotes.

\item \textbf{NGC 5208} (Figure~\ref{fig:ngc5208_ell_mge}): For this cuspy galaxy with a large (projected) disk size, the remaining unmasked $H$-band data show only mild gradients in PA$_\star$ and $\varepsilon_\star$. Despite slightly larger data--MGE residuals along the major axis, the overall MGE is not a poor representation of the $H$-band mosaic.

\item \textbf{NGC 5838} (Figure~\ref{fig:ngc5838_ell_mge}): The MGE shows generally good agreement with the data out to $R \sim 35\arcsec$, beyond which highly elliptical isophotes and constrained $q^{\prime} \rightarrow q_\mathrm{min}^{\prime}$ lead to a poorer fit, especially along the major axis. Modest changes in PA$_\star$ for $1< R/R_{\mathrm{dust},H} \lesssim 3$ exacerbate the discrepancies. However, $\mathrm{PA} \sim \mathrm{PA}_{\mathrm{dust},H}$ both near the CND edge and for $R\gtrsim 5 R_{\mathrm{dust},H}$. At more intermediate radii, the isophotes show boxy behavior.

\item \textbf{NGC 6861} (Figure~\ref{fig:ngc6861_ell_mge}): The MGE for this BGG is able to reproduce the observed surface brightness distribution fairly well. Despite close agreement between PA$_\star$ and PA$_{\mathrm{dust},H}$ in the inner regions, PA$_\star$ begins to noticeably decrease for $R\gtrsim 50\arcsec$, leading to large-scale data$-$MGE discrepancies. The disky isophotes at large $R$ complicate the global fit, but not more than the observed $\Delta\mathrm{PA}_\star$.

\item \textbf{NGC 6958} (Figure~\ref{fig:ngc6958_ell_mge}): In this cD galaxy, the very face-on [$(b/a)_{\mathrm{dust},H}\sim 1$] disk orientation severely limits the $q^{\prime}$ values that the MGE components can assume at all radii, leading to high major-axis data$-$MGE residuals. As a result of a large PA$_\star$ decrease and a moderate $\varepsilon_\star$ increase for $R\gtrsim 40\arcsec$, the MGE solution beyond this radius becomes less reliable.

\end{itemize}

\pagebreak
\section{Multi-Gaussian Expansions with a Free PA}
\label{app:free_pa}
\begin{figure*}[!ht]
    \includegraphics[width=\textwidth]{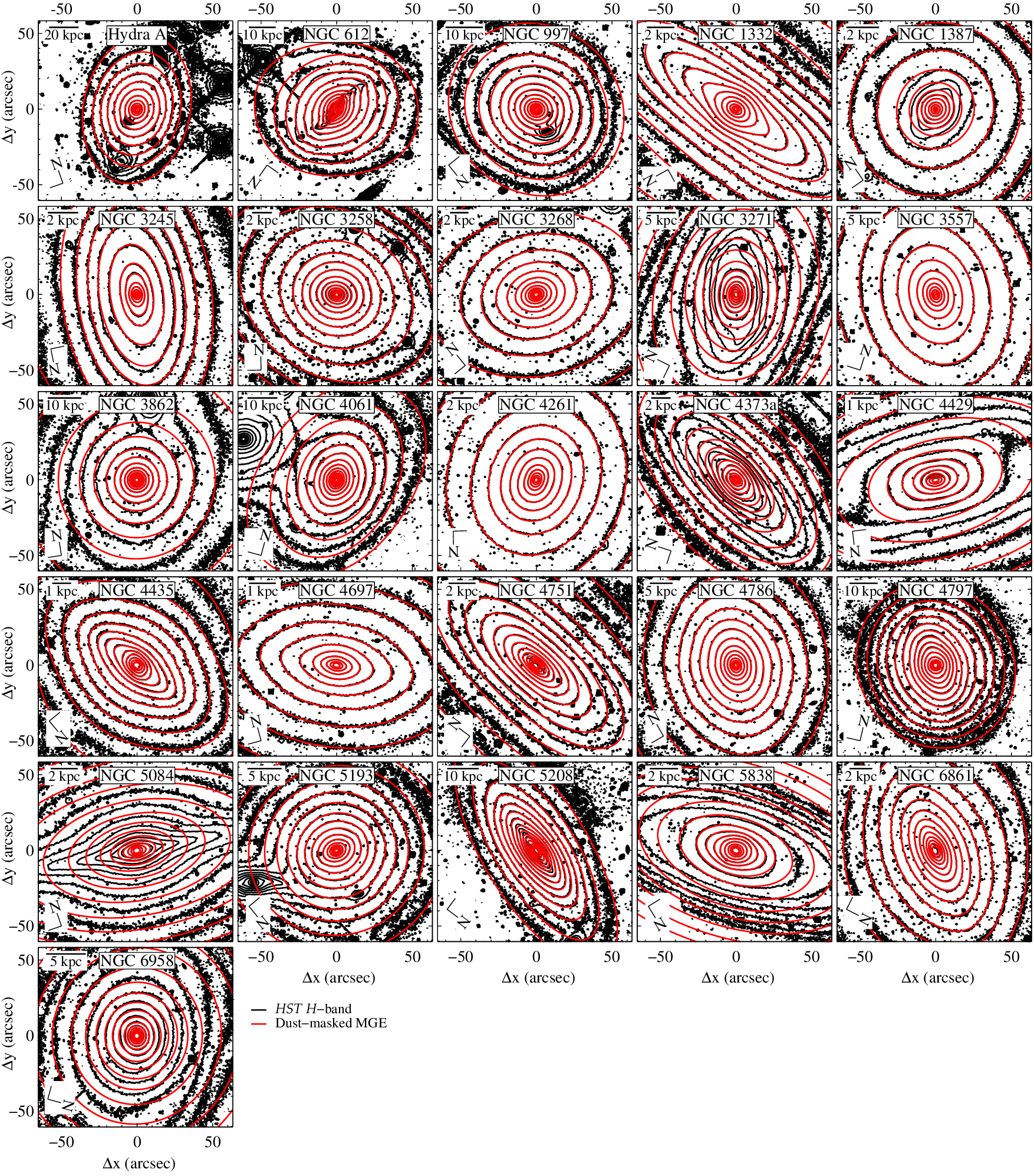}
    \centering
    \caption{Nearly full-frame HST WFC3/F160W mosaics, together with the (dust-masked) \texttt{GALFIT} MGE solutions (overplotted in red) that fits a variable PA for all components. At larger radii, some galaxies exhibit highly flattened stellar isophotes, resulting in unavoidable discrepancies. Contours are shown at logarithmic intensity intervals.}
    \label{fig:contours_free_pa}
\end{figure*}

We report \texttt{GALFIT} MGE solutions that allow for PA to vary between components in Table~\ref{tbl:mge_free_pa} and show the resulting contour plots in Figure~\ref{fig:contours_free_pa}. Photometric PA twists are common in this sample, and accounting for these PA twists results in better overall fits. For uniformity with our previous MGE solutions, we restrict these models to an appropriate range of inclination angles, which can still result in some discrepancies. Accounting for PA twists can also be useful for stellar-dynamical efforts that explore triaxality \citep[e.g.,][]{vandb08,krajnovic11,liepold23}. However, allowing for a variable PA between Gaussian components prevents a simple deprojection. Isophotal centroids are not always consistent for different radii \citep[e.g.,][]{goullaud18}, likely due to ongoing settling of a recent merger or tidal interaction. However, non-concentric series expansions are not currently viable for standard gas-dynamical modeling efforts, so we opt to force a concentric MGE solution. Compared to the MGE solutions with a fixed PA, changes in \chisq\ are minimal in most cases, reaching up to $\sim4\%$ improvement at best.

\begin{deluxetable}{ccccc@{\hspace{5mm}}cccc@{\hspace{5mm}}cccc}[!ht]
\tabletypesize{\scriptsize}
\tablecaption{MGE Parameters (with a Free PA)}
\tablewidth{0pt}
\tablehead{\\[-5mm]
\colhead{\emph{j}} & \colhead{$\log_{10}~I_{H,j}$} & \colhead{$\sigma^{\prime}_{j}$} & \colhead{$q^{\prime}_{j}$} & \colhead{PA}  & \colhead{$\log_{10}~I_{H,j}$} & \colhead{$\sigma^{\prime}_{j}$} & \colhead{$q^{\prime}_{j}$} & \colhead{PA}  & \colhead{$\log_{10}~I_{H,j}$} & \colhead{$\sigma^{\prime}_{j}$} & \colhead{$q^{\prime}_{j}$} & \colhead{PA} \\[-2.5mm]
\colhead{} & \colhead{$(L_{\sun}~\rm{pc}^{-2})$} & \colhead{(arcsec)} & \colhead{} & \colhead{(deg.)} & \colhead{$(L_{\sun}~\rm{pc}^{-2})$} & \colhead{(arcsec)} & \colhead{} & \colhead{(deg.)} & \colhead{$(L_{\sun}~\rm{pc}^{-2})$} & \colhead{(arcsec)} & \colhead{} & \colhead{(deg.)} \\[-2mm]
\colhead{(1)} & \colhead{(2)} & \colhead{(3)} & \colhead{(4)} & \colhead{(5)}  & \colhead{(2)} & \colhead{(3)} & \colhead{(4)} & \colhead{(5)} & \colhead{(2)} & \colhead{(3)} & \colhead{(4)} & \colhead{(5)}
}
\startdata
& & & $q^{\prime}_{min} = 0.23$ & & & & $q^{\prime}_{min} = 0.24$ & & & & $q^{\prime}_{min} = 0.91$ & \\[-2mm]
 & \multicolumn{3}{c}{\hspace{-1mm}\textbf{Hydra A}} & & \multicolumn{3}{c}{\hspace{-1mm}\textbf{NGC 612}} & & \multicolumn{3}{c}{\hspace{-1mm}\textbf{NGC 997}} & \\
\cline{2-5} \cline{6-9} \cline{10-13}
1 & 3.4568 & 1.2335 & 0.8913 & $-$79.83 & 4.3281 & 0.1903 & 0.7587 & $-$39.44 & 4.5662 & 0.2499 & 0.9100 & 36.57 \\
2 & 2.9176 & 2.9311 & 0.9040 & $-$30.96 & 3.8095 & 0.9655 & 0.2893 & $-$23.97 & 4.2771 & 0.7129 & 0.9127 & 34.68 \\
3 & 2.1985 & 6.6571 & 0.9179 & \phantom{$-$}6.58 & 3.6600 & 1.0974 & 0.2400 & \phantom{$-$}40.32 & 3.9215 & 1.4906 & 0.9309 & 35.48 \\
4 & 2.0283 & 7.7789 & 0.6524 & $-$44.11 & 3.7754 & 0.8971 & 0.5199 & $-$87.74 & 3.4213 & 3.2687 & 0.9100 & 30.44 \\
5 & 2.1393 & 13.591 & 0.7972 & $-$34.43 & 3.1582 & 2.3074 & 0.8134 & \phantom{$-$}84.61 & 2.9231 & 7.1770 & 0.9100 & 29.83 \\
6 & 1.3498 & 27.250 & 0.7011 & $-$29.73 & 3.8082 & 3.8546 & 0.2658 & \phantom{$-$}4.70 & 2.3616 & 15.141 & 0.9100 & 21.84 \\
7 & 0.9367 & 59.514 & 0.6606 & $-$33.63 & 3.0907 & 4.7088 & 0.6825 & \phantom{$-$}9.43 & 1.7550 & 36.268 & 0.9100 & 10.67 \\
8 & \nodata & \nodata & \nodata & \nodata & 2.5885 & 5.9420 & 0.9211 & \phantom{$-$}86.23 & \nodata & \nodata & \nodata & \nodata \\
9 & \nodata & \nodata & \nodata & \nodata & 2.7923 & 7.1097 & 0.2400 & $-$8.91 & \nodata & \nodata & \nodata & \nodata \\
10 & \nodata & \nodata & \nodata & \nodata & 2.3449 & 12.666 & 0.4365 & $-$9.20 & \nodata & \nodata & \nodata & \nodata \\
11 & \nodata & \nodata & \nodata & \nodata & 2.2925 & 16.064 & 0.8224 & $-$22.72 & \nodata & \nodata & \nodata & \nodata \\
12 & \nodata & \nodata & \nodata & \nodata & 1.1335 & 52.983 & 0.6825 & $-$31.82 & \nodata & \nodata & \nodata & \nodata \\
13 & \nodata & \nodata & \nodata & \nodata & 1.1010 & 53.895 & 0.5797 & $-$84.96 & \nodata & \nodata & \nodata & \nodata \\
\hline
& & & $q^{\prime}_{min} = 0.17$ & & & & $q^{\prime}_{min} = 0.95$ & & & & $q^{\prime}_{min} = 0.52$ & \\[-2mm]
 & \multicolumn{3}{c}{\hspace{-1.5mm}\textbf{NGC 1332}} & & \multicolumn{3}{c}{\hspace{-1.5mm}\textbf{NGC 1387}} & & \multicolumn{3}{c}{\hspace{-1.5mm}\textbf{NGC 3245}} \\
\cline{2-5} \cline{6-9} \cline{10-13}
1 & 5.8054 & 0.1279 & 0.3348 & $-$63.74 & 5.3558 & 0.1549 & 0.9500 & \phantom{$-$}7.45 & 5.6183 & 0.1000 & 0.7331 & $-$39.10 \\
2 & 4.9808 & 0.4945 & 0.9725 & $-$49.77 & 4.8527 & 0.5285 & 0.9744 & \phantom{$-$}12.38 & 5.1971 & 0.1626 & 0.5200 & \phantom{$-$}13.55 \\
3 & 4.6118 & 1.5343 & 0.7284 & $-$62.34 & 4.5401 & 1.6483 & 0.9922 & \phantom{$-$}42.47 & 4.9520 & 0.4882 & 0.8925 & \phantom{$-$}3.04 \\
4 & 4.2028 & 3.4563 & 0.7279 & $-$64.51 & 4.0631 & 4.0942 & 0.9500 & $-$73.88 & 4.3391 & 1.0742 & 0.8129 & \phantom{$-$}36.30 \\
5 & 3.8253 & 7.4695 & 0.7682 & $-$60.44 & 3.5116 & 6.8412 & 0.9500 & $-$71.07 & 4.2465 & 1.1928 & 0.9226 & \phantom{$-$}19.57 \\
6 & 3.1793 & 18.073 & 0.3133 & $-$64.62 & 2.9294 & 14.219 & 0.9500 & $-$70.62 & 4.0351 & 2.5475 & 0.6392 & $-$3.27 \\
7 & 3.0388 & 34.972 & 0.2839 & $-$66.38 & 2.4982 & 39.993 & 0.9599 & $-$73.20 & 3.8187 & 3.2562 & 0.7724 & $-$7.94 \\
8 & 2.1287 & 55.919 & 0.2547 & $-$52.75 & \nodata & \nodata & \nodata & \nodata & 3.5250 & 8.6738 & 0.5200 & $-$3.53 \\
9 & 2.3712 & 60.489 & 0.3236 & $-$71.54 & \nodata & \nodata & \nodata & \nodata & 2.8619 & 26.303 & 0.5200 & $-$2.97 \\
10 & 1.5591 & 86.725 & 0.6815 & $-$55.92 & \nodata & \nodata & \nodata & \nodata & 2.1463 & 42.513 & 0.5269 & $-$4.76 \\
11 & \nodata & \nodata & \nodata & \nodata & \nodata & \nodata & \nodata & \nodata & 2.0451 & 18.731 & 0.5200 & \phantom{$-$}3.92 \\
12 & \nodata & \nodata & \nodata & \nodata & \nodata & \nodata & \nodata & \nodata & 1.3133 & 109.00 & 0.9668 & \phantom{$-$}15.42 \\
\hline
& & & $q^{\prime}_{min} = 0.72$ & & & & $q^{\prime}_{min} = 0.67$ & & & & $q^{\prime}_{min} = 0.73$ & \\[-2mm]
 & \multicolumn{3}{c}{\hspace{-1.5mm}\textbf{NGC 3258}} & & \multicolumn{3}{c}{\hspace{-1.5mm}\textbf{NGC 3268}} & & \multicolumn{3}{c}{\hspace{-1.5mm}\textbf{NGC 3271}} \\
\cline{2-5} \cline{6-9} \cline{10-13}
1 & 4.1544 & 0.7817 & 0.9534 & 81.66 & 3.8018 & 0.2295 & 0.6700 & \phantom{$-$}64.47 & 4.9283 & 0.1000 & 0.7300 & $-$56.05 \\
2 & 3.9772 & 1.1312 & 0.7200 & 75.57 & 3.9040 & 1.0116 & 0.9786 & $-$44.81 & 5.1004 & 0.1420 & 0.7300 & $-$71.75 \\
3 & 3.9493 & 1.9658 & 0.7573 & 77.43 & 3.9745 & 1.0911 & 0.7336 & \phantom{$-$}70.80 & 4.5431 & 0.3628 & 0.8602 & $-$85.05 \\
4 & 3.6280 & 3.0048 & 0.8116 & 75.10 & 3.8481 & 1.9805 & 0.7185 & \phantom{$-$}66.30 & 4.2114 & 0.9708 & 0.7300 & $-$64.29 \\
5 & 3.4937 & 4.7532 & 0.8475 & 74.75 & 3.7059 & 2.2801 & 0.8904 & \phantom{$-$}73.63 & 4.0583 & 2.4108 & 0.7300 & $-$60.23 \\
6 & 2.7343 & 8.3235 & 0.8346 & 84.57 & 3.4643 & 3.8046 & 0.7788 & \phantom{$-$}68.70 & 3.7306 & 3.7406 & 0.7300 & $-$66.85 \\
7 & 2.9480 & 11.477 & 0.9222 & 67.16 & 3.2627 & 6.3286 & 0.8091 & \phantom{$-$}67.99 & 3.3797 & 7.0013 & 0.7300 & $-$69.50 \\
8 & 2.2619 & 20.600 & 0.9724 & 61.57 & 2.8096 & 12.068 & 0.7836 & \phantom{$-$}65.37 & 2.9196 & 17.385 & 0.7300 & $-$63.39 \\
9 & 2.0426 & 47.745 & 0.7563 & 56.47 & 2.4636 & 21.539 & 0.8038 & \phantom{$-$}69.14 & 1.7699 & 33.410 & 0.7300 & $-$73.50 \\
10 & \nodata & \nodata & \nodata & \nodata & 2.0176 & 51.615 & 0.7267 & \phantom{$-$}64.49 & 1.9335 & 50.589 & 0.7300 & $-$81.49 \\
11 & \nodata & \nodata & \nodata & \nodata & 1.5354 & 89.674 & 0.8499 & \phantom{$-$}45.60 & 0.72101 & 119.88 & 0.7390 & $-$86.53 \\
\hline
& & & $q^{\prime}_{min} = 0.75$ & & & & $q^{\prime}_{min} = 0.99$ & & & & $q^{\prime}_{min} = 0.66$ & \\[-2mm]
 & \multicolumn{3}{c}{\hspace{-1.5mm}\textbf{NGC 3557}} & & \multicolumn{3}{c}{\hspace{-1.5mm}\textbf{NGC 3862}} & & \multicolumn{3}{c}{\hspace{-1.5mm}\textbf{NGC 4061}} \\
\cline{2-5} \cline{6-9} \cline{10-13}
1 & 4.3101 & 0.8964 & 0.8939 & 40.58 & 4.1623 & 0.7526 & 0.9900 & \phantom{$-$}24.23 & 4.2038 & 0.1643 & 0.6600 & $-$7.04 \\
2 & 4.2852 & 1.3284 & 0.7500 & 32.82 & 3.8002 & 1.4249 & 0.9913 & \phantom{$-$}16.53 & 4.2872 & 0.5382 & 0.6600 & $-$4.16 \\
3 & 3.9338 & 2.0829 & 0.7500 & 33.04 & 3.4945 & 2.9143 & 0.9900 & \phantom{$-$}1.82 & 3.7492 & 1.1051 & 0.9019 & $-$5.77 \\
4 & 4.0034 & 3.0179 & 0.7500 & 34.30 & 2.7925 & 7.0073 & 0.9900 & $-$16.24 & 3.6006 & 1.2674 & 0.6600 & $-$3.32 \\
5 & 3.7192 & 4.5527 & 0.7606 & 32.91 & 2.2356 & 17.893 & 0.9900 & $-$21.13 & 3.5601 & 2.1423 & 0.8069 & $-$6.01 \\
6 & 3.4630 & 7.0489 & 0.7509 & 34.41 & 1.6486 & 63.145 & 0.9900 & $-$47.07 & 3.1868 & 3.7987 & 0.8128 & $-$5.24 \\
7 & 2.6865 & 17.322 & 0.7500 & 32.86 & \nodata & \nodata & \nodata & \nodata & 2.4864 & 6.3725 & 0.9544 & \phantom{$-$}11.23 \\
8 & 3.1593 & 12.247 & 0.7500 & 31.70 & \nodata & \nodata & \nodata & \nodata & 2.3310 & 8.5164 & 0.6600 & $-$9.11 \\
9 & 2.6519 & 31.261 & 0.7500 & 32.73 & \nodata & \nodata & \nodata & \nodata & 2.1361 & 13.776 & 0.8540 & $-$6.35 \\
10 & 2.0201 & 47.821 & 0.8375 & 33.35 & \nodata & \nodata & \nodata & \nodata & 1.6802 & 24.945 & 0.6938 & $-$2.47 \\
11 & 1.9106 & 91.520 & 0.7986 & 38.25 & \nodata & \nodata & \nodata & \nodata & 1.4371 & 50.149 & 0.6600 & $-$26.66 \\
12 & \nodata & \nodata & \nodata & \nodata & \nodata & \nodata & \nodata & \nodata & 0.55051 & 144.60 & 0.7209 & $-$50.73 \\
\enddata
   \tablecomments{Individual Gaussian components from the best-fitting MGE for each galaxy in this \emph{H}-band sample, after masking out neighboring galaxies, foreground stars, and the most dust-obscured regions of the CND. The PA for each component was allowed to vary freely. Projected terms are indicated by a $^\prime{}$. During these fits, the individual $q^\prime$ values were constrained to be equal to or greater than the listed limit $q^{\prime{}}_\mathrm{min}$.}
   \label{tbl:mge_free_pa}
\end{deluxetable}

\setcounter{table}{5}
\begin{deluxetable}{ccccc@{\hspace{5mm}}cccc@{\hspace{5mm}}cccc}[!ht]
\tabletypesize{\scriptsize}
\tablecaption{MGE Parameters (with a Free PA)}
\tablewidth{0pt}
\tablehead{\\[-5mm]
\colhead{\emph{j}} & \colhead{$\log_{10}~I_{H,j}$} & \colhead{$\sigma^{\prime}_{j}$} & \colhead{$q^{\prime}_{j}$} & \colhead{PA}  & \colhead{$\log_{10}~I_{H,j}$} & \colhead{$\sigma^{\prime}_{j}$} & \colhead{$q^{\prime}_{j}$} & \colhead{PA}  & \colhead{$\log_{10}~I_{H,j}$} & \colhead{$\sigma^{\prime}_{j}$} & \colhead{$q^{\prime}_{j}$} & \colhead{PA} \\[-2.5mm]
\colhead{} & \colhead{$(L_{\sun}~\rm{pc}^{-2})$} & \colhead{(arcsec)} & \colhead{} & \colhead{(deg.)} & \colhead{$(L_{\sun}~\rm{pc}^{-2})$} & \colhead{(arcsec)} & \colhead{} & \colhead{(deg.)} & \colhead{$(L_{\sun}~\rm{pc}^{-2})$} & \colhead{(arcsec)} & \colhead{} & \colhead{(deg.)} \\[-2mm]
\colhead{(1)} & \colhead{(2)} & \colhead{(3)} & \colhead{(4)} & \colhead{(5)}  & \colhead{(2)} & \colhead{(3)} & \colhead{(4)} & \colhead{(5)} & \colhead{(2)} & \colhead{(3)} & \colhead{(4)} & \colhead{(5)}
}
\startdata
& & & $q^{\prime}_{min} = 0.71$ & & & & $q^{\prime}_{min} = 0.52$ & & & & $q^{\prime}_{min} = 0.57$ & \\[-2mm]
 & \multicolumn{3}{c}{\hspace{-1.5mm}\textbf{NGC 4261}} & & \multicolumn{3}{c}{\hspace{-1.5mm}\textbf{NGC 4373a}} & & \multicolumn{3}{c}{\hspace{-1.5mm}\textbf{NGC 4429}} \\
\cline{2-5} \cline{6-9} \cline{10-13}
1 & 4.3300 & 1.1171 & 0.8200 & $-$22.92 & 5.1906 & 0.1133 & 0.5200 & $-$18.92 & 5.2744 & 0.1133 & 0.8697 & \phantom{$-$}89.55 \\
2 & 4.0998 & 2.1901 & 0.7100 & $-$20.82 & 4.8712 & 0.3296 & 0.5200 & $-$27.01 & 4.8345 & 0.3249 & 0.5700 & $-$89.49 \\
3 & 3.9659 & 3.7306 & 0.7257 & $-$24.13 & 4.2773 & 0.9670 & 0.5505 & $-$21.93 & 4.3016 & 0.7510 & 0.7012 & $-$85.52 \\
4 & 2.9954 & 6.5190 & 0.7100 & $-$2.520 & 3.9327 & 2.1215 & 0.5788 & $-$27.10 & 4.2403 & 2.4643 & 0.5700 & \phantom{$-$}89.96 \\
5 & 3.3170 & 8.3750 & 0.7100 & $-$29.93 & 3.4518 & 4.6950 & 0.7496 & $-$35.00 & 3.6811 & 3.1301 & 0.8392 & \phantom{$-$}85.91 \\
6 & 3.1818 & 12.620 & 0.8165 & $-$17.42 & 3.1189 & 11.913 & 0.5200 & $-$29.96 & 3.8262 & 6.6209 & 0.6155 & $-$87.26 \\
7 & 2.7878 & 19.976 & 0.8322 & $-$28.76 & 2.2540 & 28.705 & 0.5200 & $-$28.46 & 2.9329 & 19.703 & 0.5700 & $-$75.08 \\
8 & 1.7203 & 41.033 & 0.7560 & \phantom{$-$}7.40 & 1.7247 & 54.713 & 0.5200 & $-$29.52 & 3.2075 & 15.868 & 0.5700 & \phantom{$-$}88.27 \\
9 & 2.3330 & 45.969 & 0.8291 & $-$20.43 & 0.6228 & 137.36 & 0.9964 & $-$38.48 & 2.4565 & 52.720 & 0.5700 & $-$81.74 \\
10 & 1.7945 & 95.693 & 0.9336 & $-$14.80 & \nodata & \nodata & \nodata & \nodata & 2.2329 & 42.422 & 0.5700 & $-$81.01 \\
11 & \nodata & \nodata & \nodata & \nodata & \nodata & \nodata & \nodata & \nodata & 2.4152 & 71.394 & 0.5700 & $-$83.12 \\
\hline
& & & $q^{\prime}_{min} = 0.41$ & & & & $q^{\prime}_{min} = 0.40$ & & & & $q^{\prime}_{min} = 0.35$ & \\[-2mm]
 & \multicolumn{3}{c}{\hspace{-1.5mm}\textbf{NGC 4435}} & & \multicolumn{3}{c}{\hspace{-1.5mm}\textbf{NGC 4697}} & & \multicolumn{3}{c}{\hspace{-1.5mm}\textbf{NGC 4751}} \\
\cline{2-5} \cline{6-9} \cline{10-13}
1 & 5.1424 & 0.1773 & 0.4820 & \phantom{$-$}9.47 & 5.7171 & 0.1000 & 0.4000 & $-$32.04 & 6.1303 & 0.1136 & 0.3500 & $-$1.66 \\
2 & 4.6677 & 0.5383 & 0.6624 & \phantom{$-$}32.01 & 5.4697 & 0.1805 & 0.4117 & \phantom{$-$}63.62 & 5.2241 & 0.3007 & 0.3500 & $-$4.49 \\
3 & 4.3822 & 1.1710 & 0.8042 & \phantom{$-$}4.91 & 4.8204 & 0.5119 & 0.7567 & \phantom{$-$}65.81 & 4.4520 & 0.5333 & 0.9973 & $-$27.53 \\
4 & 4.2347 & 2.4201 & 0.6675 & \phantom{$-$}16.38 & 4.5250 & 1.0984 & 0.7124 & \phantom{$-$}66.88 & 4.4722 & 0.9159 & 0.6330 & $-$3.20 \\
5 & 3.9720 & 4.5892 & 0.7299 & $-$6.98 & 3.7593 & 2.2197 & 0.9620 & \phantom{$-$}63.31 & 4.2449 & 1.7674 & 0.6551 & $-$5.38 \\
6 & 3.2195 & 12.116 & 0.4100 & \phantom{$-$}10.66 & 4.2598 & 2.5815 & 0.4380 & \phantom{$-$}65.90 & 3.9386 & 4.1249 & 0.4735 & $-$5.52 \\
7 & 3.0735 & 16.582 & 0.4100 & \phantom{$-$}11.94 & 3.8533 & 5.3273 & 0.4277 & \phantom{$-$}65.67 & 3.3793 & 8.3846 & 0.4188 & $-$4.49 \\
8 & 2.5324 & 16.508 & 0.9569 & \phantom{$-$}28.45 & 3.5599 & 5.5314 & 0.6935 & \phantom{$-$}66.59 & 2.9959 & 15.765 & 0.4111 & $-$5.25 \\
9 & 2.4556 & 31.778 & 0.5339 & \phantom{$-$}11.20 & 3.4864 & 9.5203 & 0.7144 & \phantom{$-$}67.14 & 2.4976 & 34.652 & 0.3970 & $-$6.04 \\
10 & 1.7724 & 64.592 & 0.7168 & $-$70.50 & 3.0765 & 12.411 & 0.4000 & \phantom{$-$}65.41 & 1.7031 & 75.996 & 0.4970 & $-$9.94 \\
11 & 1.5648 & 88.297 & 0.4140 & \phantom{$-$}9.91 & 2.8323 & 22.877 & 0.4000 & \phantom{$-$}65.10 & 0.60477 & 181.39 & 0.8516 & $-$19.76 \\
12 & \nodata & \nodata & \nodata & \nodata & 2.9443 & 23.326 & 0.5089 & \phantom{$-$}66.05 & \nodata & \nodata & \nodata & \nodata \\
13 & \nodata & \nodata & \nodata & \nodata & 2.9126 & 34.276 & 0.6047 & \phantom{$-$}66.95 & \nodata & \nodata & \nodata & \nodata \\
14 & \nodata & \nodata & \nodata & \nodata & 2.5195 & 58.878 & 0.6590 & \phantom{$-$}66.11 & \nodata & \nodata & \nodata & \nodata \\
15 & \nodata & \nodata & \nodata & \nodata & 1.9977 & 118.41 & 0.8048 & \phantom{$-$}69.42 & \nodata & \nodata & \nodata & \nodata \\
\hline
& & & $q^{\prime}_{min} = 0.69$ & & & & $q^{\prime}_{min} = 0.59$ & & & & $q^{\prime}_{min} = 0.53$ & \\[-2mm]
 & \multicolumn{3}{c}{\hspace{-1.5mm}\textbf{NGC 4786}} & & \multicolumn{3}{c}{\hspace{-1.5mm}\textbf{NGC 4797}} & & \multicolumn{3}{c}{\hspace{-1.5mm}\textbf{NGC 5084}} \\
\cline{2-5} \cline{6-9} \cline{10-13}
1 & 4.3086 & 0.3298 & 0.9381 & \phantom{$-$}75.16 & 4.9673 & 0.1404 & 0.5900 & \phantom{$-$}43.03 & 5.2766 & 0.1000 & 0.5300 & 21.32 \\
2 & 4.4393 & 0.5740 & 0.7206 & $-$15.72 & 4.2167 & 0.4608 & 0.8116 & \phantom{$-$}36.34 & 4.7053 & 0.4536 & 0.8297 & 87.62 \\
3 & 4.1755 & 1.2669 & 0.8197 & $-$18.02 & 3.6904 & 0.9806 & 0.9678 & \phantom{$-$}43.52 & 4.5468 & 0.8996 & 0.8877 & 83.20 \\
4 & 3.5620 & 2.7412 & 0.7348 & $-$17.26 & 3.2996 & 2.0556 & 0.9337 & $-$79.20 & 4.4677 & 1.8366 & 0.7501 & 82.48 \\
5 & 3.4411 & 4.7184 & 0.8148 & $-$17.94 & 2.7301 & 5.4252 & 0.8406 & $-$50.53 & 4.0750 & 4.5063 & 0.5300 & 82.62 \\
6 & 2.5653 & 5.9174 & 0.8593 & $-$22.41 & 2.7371 & 7.4784 & 0.6212 & \phantom{$-$}31.61 & 3.6607 & 8.4099 & 0.5300 & 82.35 \\
7 & 2.5992 & 7.7617 & 0.6900 & $-$14.24 & 2.3562 & 12.931 & 0.6168 & \phantom{$-$}32.26 & 3.2294 & 18.160 & 0.5300 & 82.10 \\
8 & 2.6956 & 12.806 & 0.6900 & $-$16.27 & 1.9419 & 18.686 & 0.8588 & \phantom{$-$}29.80 & 2.6726 & 47.305 & 0.5300 & 80.74 \\
9 & 2.3145 & 14.855 & 0.9089 & $-$16.79 & 1.3603 & 33.385 & 0.9183 & \phantom{$-$}58.23 & 1.5243 & 105.20 & 0.5300 & 80.05 \\
10 & 2.1400 & 23.637 & 0.6900 & $-$14.79 & \nodata & \nodata & \nodata & \nodata & \nodata & \nodata & \nodata & \nodata \\
11 & 1.7393 & 28.847 & 0.9223 & $-$20.55 & \nodata & \nodata & \nodata & \nodata & \nodata & \nodata & \nodata & \nodata \\
12 & 1.3589 & 55.304 & 0.6900 & $-$17.07 & \nodata & \nodata & \nodata & \nodata & \nodata & \nodata & \nodata & \nodata \\
13 & 1.1908 & 109.29 & 0.8174 & $-$39.18 & \nodata & \nodata & \nodata & \nodata & \nodata & \nodata & \nodata & \nodata \\
\hline
& & & $q^{\prime}_{min} = 0.75$ & & & & $q^{\prime}_{min} = 0.31$ & & & & $q^{\prime}_{min} = 0.56$ & \\[-2mm]
 & \multicolumn{3}{c}{\hspace{-1.5mm}\textbf{NGC 5193}} & & \multicolumn{3}{c}{\hspace{-1.5mm}\textbf{NGC 5208}} & & \multicolumn{3}{c}{\hspace{-1.5mm}\textbf{NGC 5838}} \\
\cline{2-5} \cline{6-9} \cline{10-13}
1 & 5.2531 & 0.1000 & 0.7500 & 61.39 & 4.7132 & 0.1000 & 0.6486 & \phantom{$-$}3.62 & 5.3228 & 0.1000 & 0.7369 & $-$30.22 \\
2 & 4.3480 & 0.4486 & 0.7500 & 62.53 & 4.7639 & 0.2374 & 0.6454 & $-$13.42 & 5.5671 & 0.1567 & 0.5600 & \phantom{$-$}38.57 \\
3 & 4.3466 & 0.9357 & 0.7500 & 67.88 & 4.2336 & 0.5364 & 0.6789 & $-$13.20 & 4.9070 & 0.3916 & 0.9063 & \phantom{$-$}73.40 \\
4 & 3.9765 & 1.9775 & 0.8096 & 69.80 & 3.7478 & 1.3023 & 0.6911 & $-$14.50 & 4.5814 & 0.7877 & 0.8646 & \phantom{$-$}37.26 \\
5 & 3.4464 & 4.6026 & 0.7500 & 73.33 & 3.9250 & 1.6661 & 0.3100 & $-$16.77 & 4.4185 & 1.9833 & 0.7133 & \phantom{$-$}36.51 \\
6 & 3.0327 & 9.7373 & 0.8422 & 74.22 & 3.4542 & 2.3510 & 0.5494 & $-$15.21 & 3.9404 & 4.5365 & 0.8312 & \phantom{$-$}58.05 \\
7 & 2.5116 & 18.578 & 0.9809 & 83.30 & 3.5274 & 5.4322 & 0.3100 & $-$16.49 & 3.4456 & 10.621 & 0.6015 & \phantom{$-$}41.87 \\
8 & 1.6530 & 47.070 & 0.9401 & 64.60 & 3.0443 & 6.0382 & 0.5683 & $-$14.00 & 2.7230 & 32.183 & 0.5600 & \phantom{$-$}41.58 \\
9 & \nodata & \nodata & \nodata & \nodata & 2.5614 & 9.2453 & 0.3100 & $-$17.57 & 2.0282 & 57.223 & 0.5600 & \phantom{$-$}41.57 \\
10 & \nodata & \nodata & \nodata & \nodata & 2.5448 & 13.568 & 0.3100 & $-$17.33 & \nodata & \nodata & \nodata & \nodata \\
11 & \nodata & \nodata & \nodata & \nodata & 2.2904 & 17.293 & 0.4537 & $-$16.36 & \nodata & \nodata & \nodata & \nodata \\
12 & \nodata & \nodata & \nodata & \nodata & 1.8148 & 22.700 & 0.3100 & $-$21.79 & \nodata & \nodata & \nodata & \nodata \\
13 & \nodata & \nodata & \nodata & \nodata & 1.5476 & 44.782 & 0.3775 & $-$21.63 & \nodata & \nodata & \nodata & \nodata \\
14 & \nodata & \nodata & \nodata & \nodata & 0.8870 & 129.08 & 0.8838 & $-$29.79 & \nodata & \nodata & \nodata & \nodata \\
\enddata
   \tablecomments{cont.}
   \label{tbl:mge_free_pa2}
\end{deluxetable}

\setcounter{table}{5}
\begin{deluxetable}{ccccc@{\hspace{5mm}}cccc@{\hspace{5mm}}cccc}[!ht]
\tabletypesize{\scriptsize}
\tablecaption{MGE Parameters (with a Free PA)}
\tablewidth{0pt}
\tablehead{\\[-5mm]
\colhead{\emph{j}} & \colhead{$\log_{10}~I_{H,j}$} & \colhead{$\sigma^{\prime}_{j}$} & \colhead{$q^{\prime}_{j}$} & \colhead{PA}  & \colhead{$\log_{10}~I_{H,j}$} & \colhead{$\sigma^{\prime}_{j}$} & \colhead{$q^{\prime}_{j}$} & \colhead{PA}  & \colhead{$\log_{10}~I_{H,j}$} & \colhead{$\sigma^{\prime}_{j}$} & \colhead{$q^{\prime}_{j}$} & \colhead{PA} \\[-2.5mm]
\colhead{} & \colhead{$(L_{\sun}~\rm{pc}^{-2})$} & \colhead{(arcsec)} & \colhead{} & \colhead{(deg.)} & \colhead{$(L_{\sun}~\rm{pc}^{-2})$} & \colhead{(arcsec)} & \colhead{} & \colhead{(deg.)} & \colhead{$(L_{\sun}~\rm{pc}^{-2})$} & \colhead{(arcsec)} & \colhead{} & \colhead{(deg.)} \\[-2mm]
\colhead{(1)} & \colhead{(2)} & \colhead{(3)} & \colhead{(4)} & \colhead{(5)}  & \colhead{(2)} & \colhead{(3)} & \colhead{(4)} & \colhead{(5)} & \colhead{(2)} & \colhead{(3)} & \colhead{(4)} & \colhead{(5)}
}
\startdata
& & & $q^{\prime}_{min} = 0.38$ & & & & $q^{\prime}_{min} = 0.95$ & \\[-2mm]
 & \multicolumn{3}{c}{\hspace{-1.5mm}\textbf{NGC 6861}} & & \multicolumn{3}{c}{\hspace{-1.5mm}\textbf{NGC 6958}} \\
\cline{2-5} \cline{6-9}
1 & 5.0115 & 0.1240 & 0.5642 & \phantom{$-$}19.16 & 5.3939 & 0.1000 & 0.9500 & \phantom{$-$}61.22 \\
2 & 4.8017 & 0.2137 & 0.8592 & $-$24.84 & 5.3091 & 0.1893 & 0.9500 & \phantom{$-$}67.25 \\
3 & 4.8535 & 0.5023 & 0.3800 & $-$35.82 & 4.6775 & 0.6404 & 0.9500 & $-$70.95 \\
4 & 4.5926 & 0.6724 & 0.9826 & $-$25.22 & 4.2927 & 1.4673 & 0.9500 & $-$69.61 \\
5 & 4.2607 & 1.5157 & 0.8176 & $-$36.10 & 3.8961 & 3.0894 & 0.9500 & $-$70.56 \\
6 & 4.2174 & 3.4836 & 0.4681 & $-$37.45 & 3.1265 & 7.1427 & 0.9500 & $-$70.24 \\
7 & 3.7291 & 4.4471 & 0.7497 & $-$38.68 & 2.7974 & 14.133 & 0.9500 & $-$77.93 \\
8 & 3.5860 & 7.5404 & 0.4153 & $-$37.57 & 1.9274 & 33.343 & 0.9500 & \phantom{$-$}89.82 \\
9 & 3.2756 & 11.558 & 0.5075 & $-$38.29 & 1.5789 & 42.066 & 0.9500 & \phantom{$-$}85.45 \\
10 & 2.6651 & 13.092 & 0.6773 & $-$37.38 & 0.87033 & 90.065 & 0.9942 & $-$73.57 \\
11 & 2.6056 & 23.592 & 0.4402 & $-$39.64 & \nodata & \nodata & \nodata & \nodata \\
12 & 2.3448 & 26.382 & 0.7077 & $-$42.19 & \nodata & \nodata & \nodata & \nodata \\
13 & 1.8380 & 43.342 & 0.5320 & $-$39.05 & \nodata & \nodata & \nodata & \nodata \\
14 & 1.6966 & 46.380 & 0.8709 & \phantom{$-$}84.57 & \nodata & \nodata & \nodata & \nodata \\
15 & 1.4120 & 153.99 & 0.5012 & $-$70.61 & \nodata & \nodata & \nodata & \nodata \\
\enddata
   \tablecomments{cont.}
   \label{tbl:mge_free_pa3}
\end{deluxetable}

\end{document}